\documentclass[12pt]{iopart}

\usepackage{graphicx}
\graphicspath{{./Fig/}}
\usepackage{amssymb}
\usepackage{ulem}

\newcommand{\vcr}{({\bf r})}
\newcommand{\paDir}[2]{\frac{\partial #1}{\partial #2}}
\newcommand{\paDirH}[3]{\frac{\partial^{#3} #1}{\partial #2^{#3}}}
\newcommand{\paDirM}[3]{\frac{\partial^2 #1}{\partial #2 \partial #3}}
\newcommand{\fnDir}[2]{\frac{\delta #1}{\delta #2}}
\newcommand{\delRho}{\delta \hspace{-0.8mm} \rho}
\newcommand{\eqref}[1]{(\ref{#1})}
\newcommand{\chiArray}{ \left(\hspace{-2mm} \begin{array}{c} \chi \\ 1 \end{array} 
			\hspace{-2mm} \right)}
\newcommand{\matTT}[4]{\left(\hspace{-2mm}\begin{array}{cc} #1 & #2 \\ #3 & #4
			\end{array}\hspace{-2mm}\right)}

\makeatletter
\def\gsim{\compoundrel>\over\sim}
\def\lsim{\compoundrel<\over\sim}
\def\compoundrel#1\over#2{\mathpalette\compoundreL{{#1}\over{#2}}}
\def\compoundreL#1#2{\compoundREL#1#2}
\def\compoundREL#1#2\over#3{\mathrel
      {\vcenter{\hbox{$\m@th\buildrel{#1#2}\over{#1#3}$}}}}
\makeatother

\usepackage{prelim2e}\usepackage[none,bottom]{draftcopy}
\draftcopyName{preprint / }{1.2} 
\usepackage[usenames,dvipsnames]{color}
\usepackage{ulem}

\newcounter{countuwe}

\newcounter{countcoaut}

\begin{document}


\title[Modelling the evaporation of thin films of colloidal suspensions using DDFT]{Modelling the evaporation of thin films of colloidal suspensions using Dynamical 
         Density Functional Theory }
\author{M.J.~Robbins, A.J.~Archer and U.~Thiele}

\vspace{2mm}
\begin{center}
{\small Department of Mathematical Sciences,\\ Loughborough University,
	     Leicestershire LE11 3TU, UK}
\end{center}

\begin{abstract}
	Recent experiments have shown that various structures may be formed during the
	evaporative dewetting of thin films of colloidal suspensions. Nano-particle deposits
	of strongly branched `flower-like', labyrinthine and network structures are observed. 
	They are caused by the different transport processes and the rich phase behaviour 
	of the system. We develop a model for the system, based on a dynamical density functional theory, 
	which reproduces these structures. The model is employed to determine the influences
	of the solvent evaporation and of the diffusion of the colloidal particles and of the
	liquid over the surface.  Finally, we investigate the conditions needed for `liquid-particle' phase 
	separation to occur and discuss its effect on the self-organised nano-structures.
\end{abstract}

\maketitle

\section{Introduction}
\label{secIntro}

Surface patterns resulting from structure formation occur naturally in
many different systems and are extensively studied in various
scientific fields.  A classic example are branched patterns,
e.g.~found in river networks \cite{GMB95} or formed by bacterial
colonies \cite{MSM00}, that sometimes form fractals. Other examples
are the labyrinth patterns such as those formed via calcification and
mineralisation processes \cite{YTDG08}.  The spontaneous self-assembly
and self-organisation of atoms, molecules and nano-particles at interfaces is a
widely researched topic not only because the resulting structures are
interesting but also because they may be used in the manufacturing of
nanostructures, e.g.~assembling colloids to form photonic bandgap
crystals \cite{VBSN01}.  A large variety of intricate structures can
be formed even during the dewetting process of films of non-volatile
fluids on solid substrates.  The dewetting process starts with the
rupture of the initially homogeneous film caused either by a surface
instability (spinodal dewetting \cite{Mitl93}) or by the nucleation of
holes which often occurs at surface defects
\cite{Reiter92,Xie98,SHJ01,TVN01,Thie03,Beck03,BeTh10}. The 
holes then grow \cite{RBWR91} and the rims meet to form a global network, drop or
labyrinth pattern \cite{ShRe96,ShKh98,BeNe01}.

The evaporative dewetting of polymer/macromolecule solutions
\cite{TMP98,GRDK02,XSDA07} and colloidal (nanoparticle) suspensions
\cite{MMNP00,GeBr00,MTB02,PVSM08,Stan11} can produce a wide variety of
patterns and has been intensely studied in various experimental settings
over recent years. Although the motivation for the work presented here is
mainly drawn from the latter, we believe that our results
also explain the basic features in the case of evaporative dewetting
of solutions. The particular experiments that directly motivate our
theoretical work presented here are those described in
Refs.~\cite{MTB02,Stan11,MBM04,MBPA07,PVSM08} that use a suspension of
thiol-passivated gold nano-particles dispersed in an organic solvent.
A drop of the suspension is spin-coated onto a flat silicon substrate
to form a thin film over the surface. 
The solvent evaporates during the spin-coating and leaves a nano-particle
pattern on the surface.  In another experimental setup a drop of
suspension is placed on the surface within a teflon ring
\cite{PaMo07}. The evaporative dewetting is slower than in the case of
spin-coating and the structuring is observed using video-microscopy
\cite{PVSM08}.  What these experiments show is that branched
structures are formed by transverse instabilities of the receding
mesoscopic contact line.  However, more intriguing are the patterns
that are formed in an ultrathin layer that is left behind this contact
line.  Three different types of structures have been observed: a
labyrinth pattern formed during spinodal dewetting, a two-scaled
network structure formed via the nucleation and growth of holes and a
branched structure formed by a fingering instability that occurs at
the dewetting front of nucleated holes.  The diffusive mobility of the
nano-particles and the interaction energies between the particles can
be altered by changing the thiol chain length.  The fingering
instability is found for relatively low nano-particle mobilities.  
Because these structures
form in the ultra-thin layer, the height of film is of the same
order of magnitude as the diameter of the colloids.

To model these processes, simple two-dimensional (2d) lattice kinetic
Monte-Carlo (KMC) models were proposed
\cite{RRGB03,MBPA07,PVSM08,VTPV08}.  Vancea et.~al.~\cite{VTPV08} made
a detailed investigation of the characteristics of the branched
structures and their dependence on the model parameters.  They
observed that as in the experiments the fingering instability becomes
stronger the smaller the nano-particle mobility gets.  A pseudo
three-dimensional KMC model has also been considered \cite{MBPA07,
  SMPM08} which can reproduce dual-scale network patterns.

An alternative approach that may be used for modelling such systems
is based on thin-film hydrodynamical models, which are derived by
making a long-wave approximation \cite{ODB97}. Recently, for example,
line-pattern formation has been observed in a simple long-wave model
for a thin film of colloidal suspension evaporatively dewetting from a
surface \cite{FAT11}.  Such thin film equation based models provide a
good description of the system on mesoscopic length scales and
reproduce the experimental results \cite{YaSh05,Xu06,XXL07,BDG10}. However,
they are unable to describe the dynamics of the system at the
microscopic (single particle) level. In Ref.\ \cite{TVAR09} a more
detailed account of the different approaches that may be used for
modelling the evaporative dewetting of colloidal suspensions is given,
so we do not re-review the subject here.

In the present work, we give a detailed account of an alternative model
for this system that has been briefly discussed before
\cite{TVAR09,ART10}.  It is based on Dynamical
Density Functional Theory (DDFT) \cite{MaTa99, MaTa00, ArEv04, ArRa04}
and goes beyond the 2d KMC model by also allowing us to investigate
the influence of liquid diffusion over the surface.  This paper is
laid out as follows: First, in Sec.~\ref{secModel}, we present the
coarse-grained model for the Hamiltonian and the resulting
approximation for the free energy used in our model.  This is followed
in Sec.~\ref{secDynamics} by an introduction to the dynamical
equations of the DDFT.  In Sec.~\ref{secPhase} we discuss the phase
diagram and perform a linear stability analysis of homogeneous steady
films, whereas in Sec.~\ref{secNumerics} we present fully nonlinear
simulation results.  In Sec.~\ref{secCon} we summarise our findings
and draw some conclusions.

\section{Free Energy for the System}
\label{secModel}

\begin{figure}[tbh]
	\begin{center}
		\includegraphics[width=0.7\columnwidth]{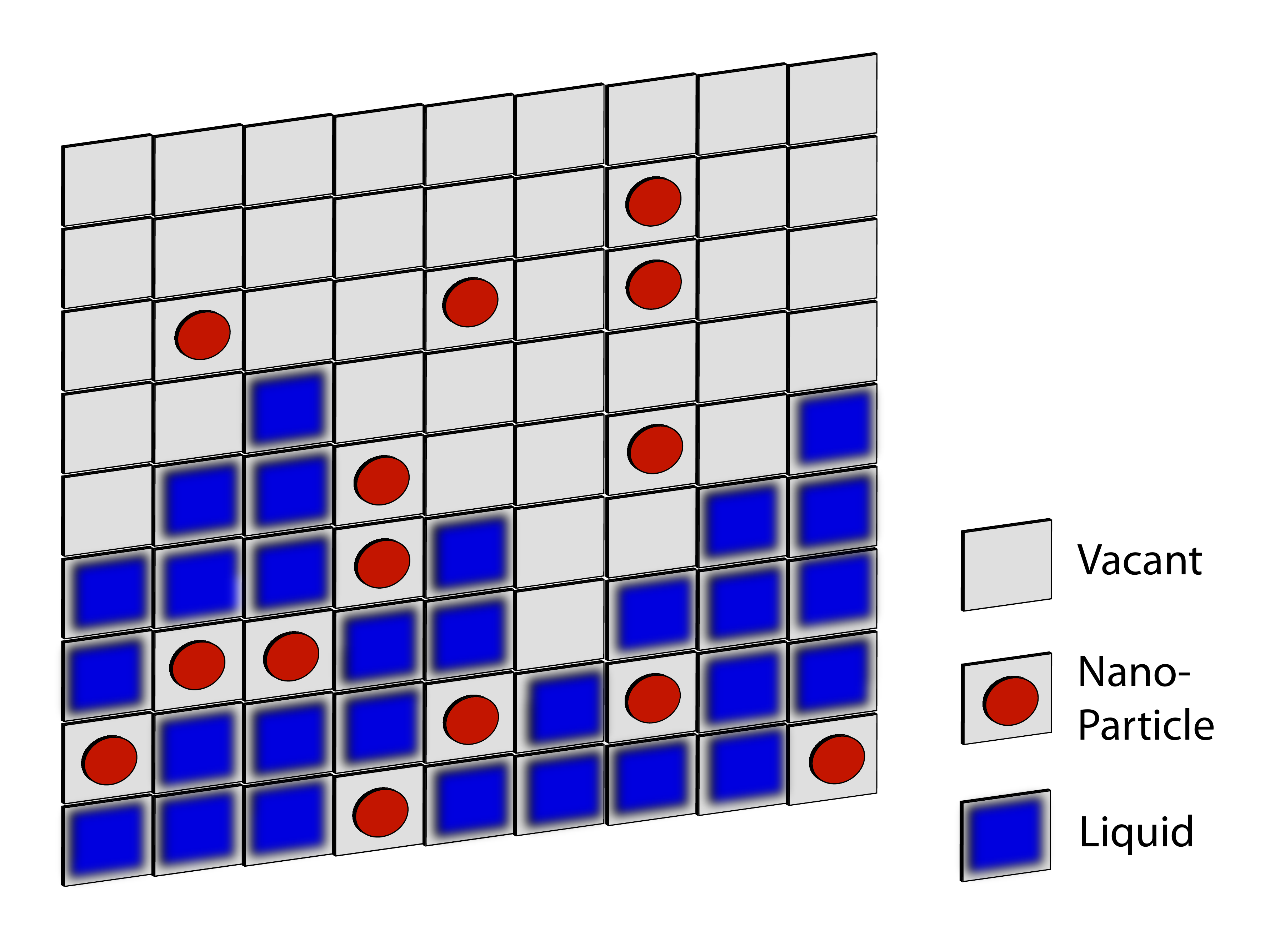}
		\caption{The sketch indicates how the substrate is divided
			      into lattice sites and shows the three possible states of each lattice site.
                              \label{coGrModel}}
	\end{center}
\end{figure}

We consider a simplified coarse-grained two-dimensional model for the system.  The surface 
of the substrate is divided into a square  array of discrete lattice sites.  We choose the cell
size so that each cell may be occupied by at most one nano-particle, i.e. the lattice spacing 
$\sigma$ is roughly equal to the diameter of the nano-particles.  As a consequence, each 
lattice site must be in one of three possible states: (i) occupied by a nano-particle, (ii) 
occupied by liquid or (iii) unoccupied (as displayed in Fig.~\ref{coGrModel}).  To 
characterise the state of the system, we introduce two occupation numbers for each 
lattice site $i$: $n_i$ for nano-particles and $l_i$ for liquid.  These occupation numbers 
are binary values which describe the state of each site.  The occupation numbers for (i) a 
lattice site containing a nano-particle would be $n_i = 1$, $l_i = 0$, for (ii) a site occupied 
by a film of liquid we have $n_i = 0$, $l_i = 1$ and for (iii) a vacant site we have 
$n_i = 0$, $l_i = 0$.  This type of lattice model has the following Hamiltonian 
\cite{RRGB03}: 
%
\begin{eqnarray}
	E &=& -\epsilon_l \sum_{<ij>} l_i l_j - \epsilon_n \sum_{<ij>} 
	n_i n_j - \epsilon_{nl} \sum_{<ij>} n_i l_j \nonumber \\ 
	&& -\mu \sum_i l_i + \sum_i \phi_i^l l_i + \sum_i \phi_i^n n_i
	\hspace{2mm}, 
	\label{eqHam}
\end{eqnarray}
where $\sum_{<ij>} $ denotes a sum over pairs of nearest neighbours.  $\phi^l_i$ and 
$\phi^n_i$ are external potentials acting on the liquid and nano-particles respectively, at 
lattice site $i$.  Three interaction terms are included which determine the strength of 
attraction between neighbouring cells.  $\epsilon_l$ is the interaction energy between 
two adjacent cells containing films of liquid, $\epsilon_n$ is for adjacent cells which 
both contain nano-particles and $\epsilon_{nl}$ is the energy between a cell containing a 
nano-particle and a cell containing liquid.  The amount of liquid on the surface is not a 
conserved quantity because it can evaporate to and condense from a reservoir of vapour 
above the surface.  $\mu$ is the chemical potential of this reservoir.	

From the Hamiltonian \eqref{eqHam} we can derive an expression for the free 
energy of the system.  The probability that a lattice site $i$ is
covered by a liquid film in 
an equilibrium configuration is given by the following integral over time $t$:
%
\begin{equation}
	\rho_i^l = \lim_{\tau \to \infty} \frac{1}{\tau} \int_0^\tau l_i(t) dt. 
	\label{eqRhoL}
\end{equation}
Similarly, the probability that a lattice site $i$ contains a nano-particle is given by the 
following expression:
%
\begin{equation}
	\rho_i^n = \lim_{\tau \to \infty} \frac{1}{\tau} \int_0^\tau n_i(t) dt. 
	\label{eqRhoN}
\end{equation}
By choosing the grid spacing $\sigma$ to be our unit of length, equal to one, these 
probabilities are also equal to the local number densities for the liquid and 
nano-particles.  Following the approach described in Refs.\ \cite{GPDM03, ChLu00}, 
we make a Bragg-Williams mean field approximation for the (semi-grand) free energy 
of the system:
%
\begin{eqnarray}
	F &=& k_B T \sum_i [ \rho_i^l \ln \rho_i^l + (1 - \rho_i^l) \ln (1 - \rho_i^l) 
	+  \rho_i^n \ln \rho_i^n \nonumber \\ && + (1 - \rho_i^n) \ln (1 - \rho_i^n)]  
	- \epsilon_l \sum_{<ij>} \rho_i^l \rho_j^l - \epsilon_n \sum_{<ij>}\rho_i^n 
	\rho_j^n \nonumber \\ &&  - \epsilon_{nl} \sum_{<ij>} \rho_i^n \rho_j^l  - \mu 
	\sum_i \rho_i^l + \sum_i \phi_i^l \rho_i^l + \sum_i  \phi_i^n \rho_i^n
	\hspace{2mm},
	\label{eqFreeEngLat} 
\end{eqnarray}
where $k_B$ is Boltzmann's constant and T is the temperature.  This is
a semi-grand free energy because the liquid is treated in the grand
canonical ensemble (the reservoir of vapour fixes the chemical
potential $\mu$), whereas the number of nano-particles in the system
is a conserved quantity (so these are treated canonically).
Eq.~\eqref{eqFreeEngLat} is derived using a `zeroth-order' mean-field
approximation and thus higher order terms are omitted (e.g. the terms
involving $\ln(1 - \rho_i^n - \rho_i^l)$ which describe the excluded
area correlations between the nano-particles and the liquid). If we
assume that the density values only vary on length scales $\gg
\sigma$, then we can define a gradient operator for this discrete
system:
%
\begin{eqnarray}
	\nabla \rho^l_{x, y} &\equiv& \left( \rho^l_{x + 1, y} - \rho^l_{x, y},
	\rho^l_{x, y+1} - \rho^l_{x, y} \right), \nonumber \\
	\nabla \rho^n_{x, y} &\equiv& \left( \rho^n_{x + 1, y} - \rho^n_{x, y},
	\rho^n_{x, y+1} - \rho^n_{x, y} \right),
	\label{eqGradOp} 
\end{eqnarray}
where each lattice site $i$ is now represented by the coordinates $(x, y)$.  Using the
operators (\ref{eqGradOp}) we can express the summation over pairs of nearest 
neighbours as
%
\begin{equation}
	\sum_{<ij>} \rho_i^\alpha \rho_j^\beta = 4 \sum_i \rho_i^\alpha \rho_i^\beta -
	\sum_i (\nabla \rho_i^\alpha) \cdot (\nabla \rho_i^\beta),
	\label{eqSNN}
\end{equation}
where $\alpha, \beta = n, l$.  Substituting \eqref{eqSNN} 
into our lattice free energy \eqref{eqFreeEngLat} we obtain
%
\begin{eqnarray}
	F &=& k_B T \sum_i [ \rho_i^l \ln \rho_i^l + (1 - \rho_i^l) \ln (1 - \rho_i^l) 
	 + \rho_i^n \ln \rho_i^n \nonumber \\ &&  + (1 - \rho_i^n) \ln (1 - \rho_i^n)] 
	- \sum_i \bigg[ \frac{4}{2} \epsilon_l (\rho_i^l)^2 + \frac{4}{2} 
	\epsilon_n (\rho_i^n)^2 + 4\epsilon_{nl} \rho_i^n \rho_i^l \bigg] 
	\nonumber \\ && + \sum_i \bigg[ \frac{\epsilon_l}{2} (\nabla \rho_i^l)^2 + 
	\frac{\epsilon_n}{2} (\nabla \rho_i^n)^2 + \epsilon_{nl} (\nabla \rho_i^n) 
	\cdot(\nabla \rho_i^l) \bigg] \nonumber \\ && - \mu \sum_i \rho_i^l + \sum_i 
	\phi_i^l \rho_i^l + \sum_i \phi_i^n \rho_i^n \hspace{2mm},
	\label{eqFreeEngLat2} 
\end{eqnarray}
where the factor of $\frac{1}{2}$ is included to avoid double counting.  Taking 
the continuum limit, so that $\sum_i \to \int d{\bf r}$, $\rho_i^n \to \rho_n \vcr$, 
$\rho_i^l \to \rho_l\vcr$, $\phi_i^n \to \phi_n \vcr$\ and $\phi_i^l \to \phi_l \vcr$ where 
the vector ${\bf r} = (x, y)$ is a continuous variable, we obtain for the free energy of the system:
%
\begin{eqnarray}
	\label{eqFreeEngCont1}
	F[\rho_l, \rho_n]  &=&  \int d{\bf r} \bigg[ f(\rho_l\vcr, \rho_n\vcr)
	+ \frac{\epsilon_l}{2} (\nabla \rho_l\vcr)^2
	\nonumber \\ && + \frac{\epsilon_n}{2}(\nabla \rho_n\vcr)^2  
	+ \epsilon_{nl}  (\nabla \rho_n\vcr) \cdot (\nabla \rho_l\vcr) \bigg] 
	\nonumber \\ && + \int d{\bf r} \rho_l\vcr(\phi_l\vcr - \mu)  + \int d{\bf r} 
	\rho_n\vcr \phi_n\vcr, \\ \mbox{where} \nonumber  \\ \nonumber \\
	f(\rho_l, \rho_n) &=& k_B T [\rho_l \ln \rho_l + (1-\rho_l) \ln(1-\rho_l) +
	\rho_n \ln \rho_n \nonumber \\ &&  + (1-\rho_n) \ln (1-\rho_n)] 
	-2 \epsilon_l \rho_l^2 -2 \epsilon_n \rho_n^2 - 4 \epsilon_{nl} \rho_n \rho_l.
	\label{eqFreeEngCont2}
\end{eqnarray} 
This free energy functional may be employed to determine the phase
diagram of the system -- i.e.\ the state of the system in the
thermodynamic limit (see Section~\ref{secPhase}). However, the
observed patterns are often non-equilibrium structures that are `dried
in', i.e., that evolve towards the equilibrium state on a time
scale that is much longer than the typical observation times. To model
the non-equilibrium processes that result in the observed self-organised structures,
one needs kinetic equations for the time evolution of the densities. They are developed in
the following section.

\section{Modelling the Dynamics of the System}
\label{secDynamics}
	
The chemical potential of the nano-particles may be calculated using the following 
functional derivative \cite{HaMc06, Evans79, Evans92Book}:
%
\begin{equation}
	\mu_n = \frac{\delta F[\rho_n,\rho_l]}{\delta \rho_n}.
	\label{eqNanoChem}
\end{equation}
In equilibrium systems the chemical potentials take a uniform value
throughout the system. However, this is not the case for
non-equilibrium configurations that the system takes during its time
evolution. There, the chemical potential varies temporally and
spatially over the surface. In particular, non-equilibrium density
profiles $\rho_l({\bf r},t)$ and $\rho_n({\bf r},t)$ give, via
Eq.~\eqref{eqNanoChem}, a non-equilibrium chemical potential for the
nano-particles $\mu_n({\bf r}, t)$.  Thus, the time-dependent
densities are coarse-grained `average' quantities.  Assuming that locally,
the system is in 
equilibrium, we may define these non-equilibrium density fields in a
similar way as the equilibrium densities [see Eqs.~\eqref{eqRhoL} and
\eqref{eqRhoN}]:
%
\begin{eqnarray}
	\rho_i^l &=& \frac{1}{\tau_M} \int_0^{\tau_M} l_i(t) dt, \\
	\rho_i^n &=& \frac{1}{\tau_M} \int_0^{\tau_M} n_i(t) dt,	
\end{eqnarray}
where $\tau_M$ is now a finite time that is large compared to the solvent molecular collision 
time, but is small compared to the time scale for a nano-particle to move from one lattice site 
to a neighbouring lattice site.  If we now assume that the driving force which causes a flux of 
the nano-particles over the surface of the substrate to be given by the gradient of the chemical 
potential $\mu_n$, then the nano-particle current is given by
%
\begin{equation}
	{\bf{j}}_n = -M_n({\bf r},t)\nabla\frac{\delta F[\rho_n,\rho_l]}
	{\delta \rho_n({\bf r},t)},
	\label{eqNanCurr}
\end{equation}
where $M_n({\bf r},t)$ is a mobility coefficient which we assume 
to depend on the local densities $\rho_n({\bf r},t)$ and $\rho_l({\bf r},t)$.  Since 
the number of nano-particles in the system is conserved we can combine   
Eq.~\eqref{eqNanCurr} with the continuity equation to get
%
\begin{equation}
	\paDir{\rho_n({\bf r},t)}{t} = \nabla \cdot \left[ M_n(\rho_n,\rho_l) \nabla 
	\frac{\delta F[\rho_n,\rho_l]}{\delta \rho_n({\bf r},t)} \right].
	\label{eqNanoDyn}
\end{equation}
We expect that when the liquid density is uniform throughout the system ($\rho_l\vcr=\rho$) 
and the density of the nano-particles is small everywhere ($\rho_n\vcr \ll 1$) then the 
nano-particle dynamics given by Eq.~\eqref{eqNanoDyn} must reduce to the diffusion 
equation (i.e.~Eq.~\eqref{eqNanCurr} becomes Fick's law).  When we apply this to 
Eq.~\eqref{eqNanoChem}, for small $\rho_n$, we get the leading order term:
%
\begin{equation}
	\mu_n\vcr \approx k_BT \ln \rho_n \vcr + C + O(\rho_n),
	\label{eqDiffRed}
\end{equation}
where $C = C(\rho_l)$ represents constant terms.  Substituting Eq.~(\ref{eqDiffRed}) into 
Eq.~(\ref{eqNanoDyn}) we see that in order for Eq.~\eqref{eqNanoDyn} to reduce to the diffusion 
equation the mobility $M_n$ has to be proportional to $\rho_n({\bf r}, t)$.  Therefore, we write the 
nano-particle mobility as:
%
\begin{equation}
	M_n = \rho_n({\bf r}, t)m(\rho_l({\bf r}, t)),
	\label{eqNMob}
\end{equation}
where $m(\rho_l({\bf r}, t))$ is a function of the local liquid
density.  We assume that the diffusion of the nano-particles over the
surface is caused by the Brownian `kicks' from the molecules in the
liquid.  Therefore when the liquid density is low (on the dry
substrate) the nano-particles are almost immobile.  However, when the
density of the liquid on the surface is high (where the substrate is
covered by the liquid film), the nano-particles are much more
mobile.  We model this behaviour by a function $m(\rho_l({\bf r}, t))$
which switches between a very small value, when $\rho_l$ is small
and $\alpha$, which is the mobility coefficient
for the nano-particles in a liquid film, when $\rho_l>0.5$.  The precise form of
$m(\rho_l)$ has a negligible effect on the qualitative behaviour of
the system.  Here we use
 %
 \begin{equation}
	m(\rho_l) = \frac{\alpha}{2} \{1+ \tanh[30(\rho_l - \frac{1}{2})]\}.
	\label{eqm}
\end{equation}
Next we consider the time evolution of the liquid density $\rho_l({\bf r}, t)$.  The dominant 
process governing the dynamics of the liquid is the evaporation and condensation of the 
liquid between the surface and the vapour reservoir above the substrate.  We define two 
different chemical potentials: $\mu$ is the chemical potential of the liquid in the reservoir 
 (cf.~Eq.~\eqref{eqFreeEngLat}) and $\mu_S({\bf r},t) = \fnDir{F}{\rho_l} + 
\mu$ denotes the local chemical potential of the liquid on the substrate.  We assume that the 
evaporative contribution to the time evolution of the liquid density is proportional to the 
difference between $\mu_S({\bf r},t)$ and $\mu$.  This gives us the following expression:
%
\begin{equation}
	\paDir{\rho_l({\bf r}, t)}{t} = - M_l^{nc} 
	\fnDir{F[\rho_n,\rho_l]}{\rho_l({\bf r},t)},
\end{equation}
where the dynamical coefficient $M_l^{nc}$ is assumed to be a constant.  The value of 
$M_l^{nc}$ determines the rate of the non-conserved (evaporation) dynamics of the liquid.  
We should also allow for (conserved) diffusive motion of the liquid over the surface.  We 
assume from DDFT that the diffusion of the liquid takes a similar form as the diffusion of the 
nano-particles given in Eq.~\eqref{eqNanoDyn}. We therefore
model the full liquid dynamics by combining the diffusive and the evaporative terms
%
\begin{equation}
	\paDir{\rho_l({\bf r}, t)}{t} =\nabla \cdot \left[ M_l^c \rho_l \nabla
	\fnDir{F[\rho_n,\rho_l]}{\rho_l({\bf r},t)} \right]  - M_l^{nc} 
	\fnDir{F[\rho_n,\rho_l]}{\rho_l({\bf r},t)}.
	\label{eqLiqDyn}
\end{equation}
The mobility coefficient $M_l^c$ for the conserved part of the dynamics is assumed to be constant.  
The ratio between the conserved and non-conserved mobility coefficients determines the 
influence that the diffusive/evaporative terms have on the overall dynamics of the liquid (i.e. 
$M_l^{nc}/M_l^c \gg 1$ corresponds to the case when the liquid dynamics are strongly dominated 
by evaporation and $M_l^{nc}/M_l^c \ll 1$ corresponds to the case when liquid diffusion plays an 
important role in the dynamics).  Thus, equations \eqref{eqFreeEngCont1}, \eqref{eqFreeEngCont2}, 
\eqref{eqNanoDyn} and \eqref{eqLiqDyn}, taken together, define our model equations, which govern 
the dynamics of the system. 
	
Note that when the liquid density $\rho_l({\bf r}, t)$ is a constant,
the theory reduces to the DDFT developed by Marconi and Tarazona
\cite{MaTa99}, that may be obtained by approximating the Fokker-Planck
equation for a system of Brownian particles with overdamped stochastic
equations of motion
\cite{MaTa99,MaTa00,ArEv04,ArRa04,GPDM03}. Equations similar to
  \eqref{eqLiqDyn} can also be derived in the context of
  hydrodynamics. The resulting mesoscopic hydrodynamic thin film
  equations contain different mobilities and local energies
  \cite{LGP02,Thiele10}. The combination of the diffusive and the
  evaporative terms can also be seen as a combination of a conserved
  Cahn-Hillard-type dynamics with a non-conserved Allen-Cahn-type
  dynamics \cite{CaHi58,Cahn65,Lang92}.

\section{Phase Behaviour}
\label{secPhase}
\subsection{One-component fluid}
\label{secPhaseOneComp}

It is important to understand the equilibrium behaviour of the fluid in our system as this 
gives us some insight into how the system behaves when it is out of equilibrium.  
Of particular importance is to determine what phases we may observe and the stability
of these phases.  Since we are modelling the evaporative dewetting of the liquid we 
initially seek parameter values which lead to a high density liquid phase (liquid film) 
coexisting with a low density phase (`dry' substrate).
Employing a linear stability analysis, we calculate the spinodal curve, i.e., the 
limit of linear stability for an infinitely extended system.  The 
spinodal curve is defined as the locus of points where the curvature of the free energy is 
zero, $\frac{d^2F}{d\rho^2} = 0$, which is equivalent to the isothermal compressibility 
being zero \cite{HaMc06}. Note that in this section we set $\rho_l=\rho$, for simplicity.
We also calculate the binodal curve, i.e., the coexisting density
values for a system in equilibrium, by equating the chemical
potentials, temperature and pressure in each of the coexisting phases. The area outside 
the binodal curve is a stable region where we see no phase separation.  Inside the 
binodal curve we have phase separation in the thermodynamic limit.
However, the linear stability 
of the fluid depends on whether the curvature of free energy is positive or negative.  
When we have positive curvature (outside the spinodal curve), the system at this state 
point rests within a local minimum of the free energy, i.e., it is
linearly but not absolutely stable.  There is a free energy barrier 
that must be traversed to cross into the (absolutely stable) equilibrium phase.  This is 
known as the metastable region, where local fluctuations in the density (if sufficient in size) 
create nucleation points for the phase transition to occur.  When the curvature of the 
free energy is negative (inside the spinodal curve) there is no free energy barrier.  This is 
the unstable region where we have spontaneous phase separation, i.e. where
fluctuations in the densities spontaneously grow.  One may also say that the homogenous 
fluid layer is linearly unstable to harmonic perturbations with certain wave numbers.

\begin{figure}[t]
	\begin{center}
		\includegraphics[width=0.49\columnwidth]{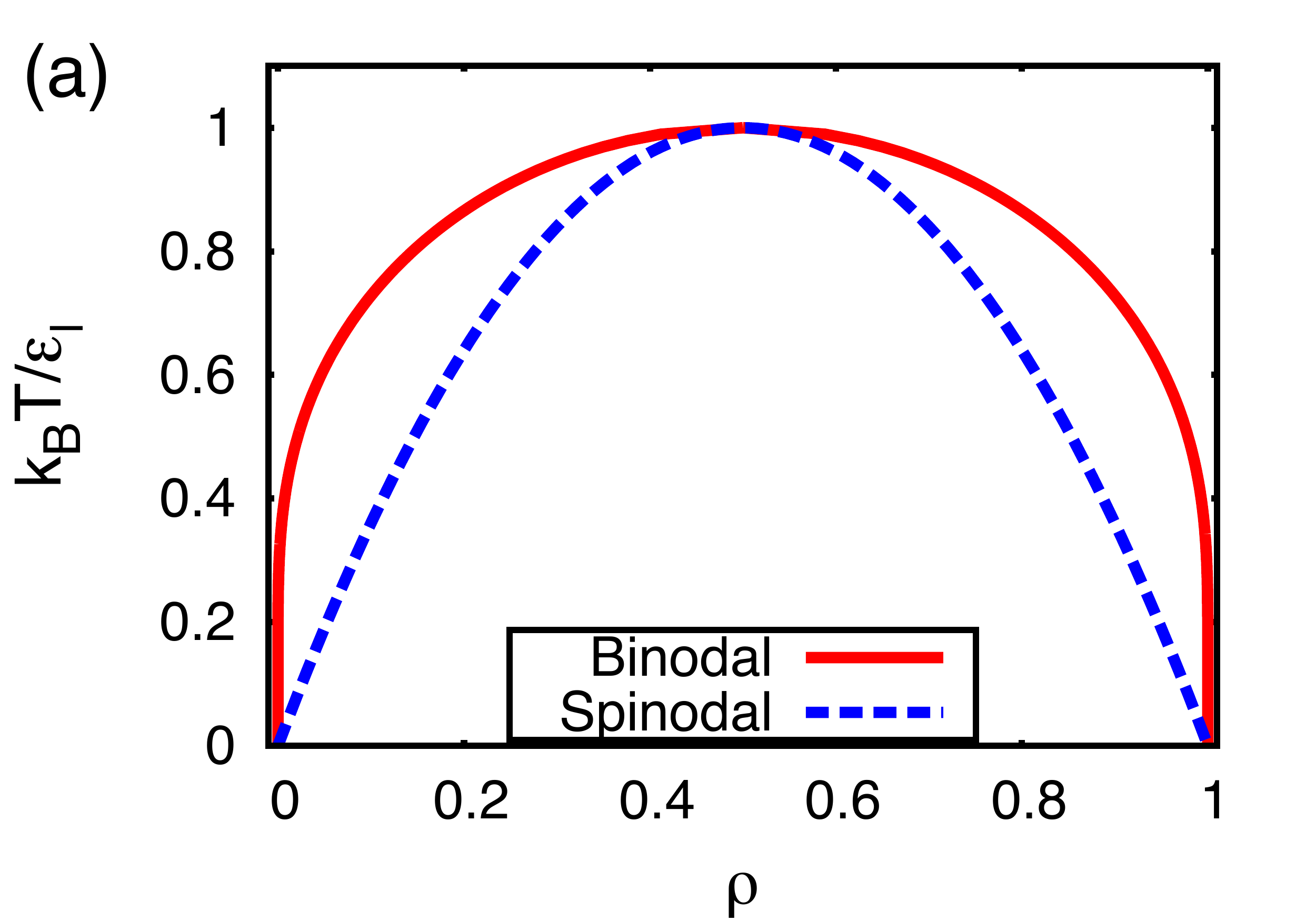}
		\includegraphics[width=0.49\columnwidth]{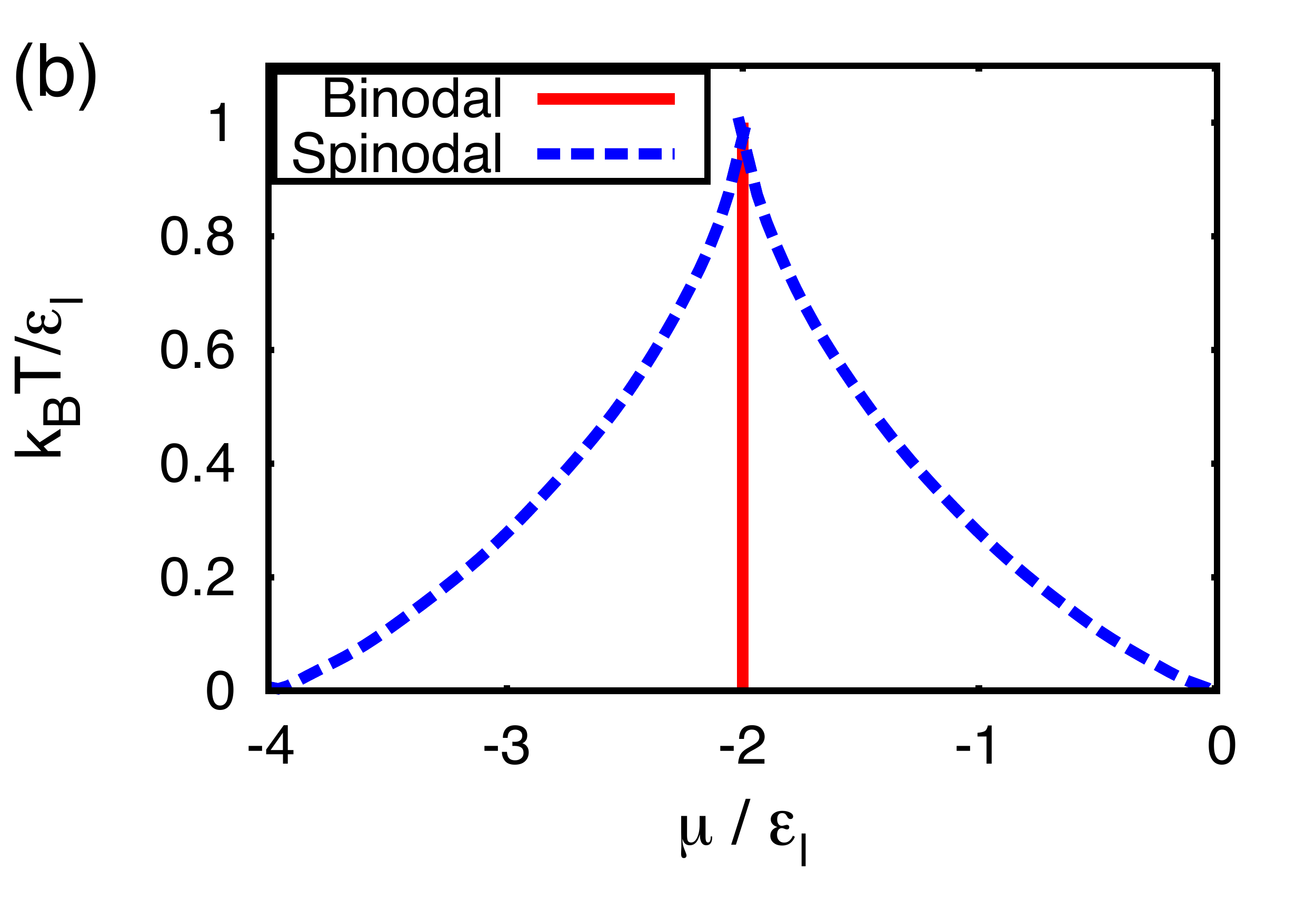}
		\caption{Phase diagrams showing the binodal (solid line) 
			      and spinodal (dotted line) for the one component pure fluid.
			      In (a) we plot the phase diagram in the temperature $T$ versus
			      density $\rho$ plane and in (b) we display the same phase 
		               diagram in the temperature versus chemical potential $\mu$
		               plane.
		\label{figOneCompPh}}
	\end{center}
\end{figure}

 We first consider the phase behaviour of the reduced case where we have a single 
 component fluid with no nano-particles (i.e.\ $\rho_n = 0$), as shown in 
 Fig.~\ref{figOneCompPh}.  We set the external potential $\phi_l\vcr = 0$ which results in 
 a uniform fluid density $\rho_l\vcr = \rho = \frac{N}{A}$, where N is the number of 
 `particles' of liquid (i.e.~filled lattice sites) and A is the area of the system.  From 
Eq.~(\ref{eqFreeEngCont2}) we find that the Helmholtz free energy per unit area of the 
uniform system is
%
\begin{equation}
	f = \frac{F}{A} = k_B T \left[ \rho \ln \rho + (1 - \rho) \ln 
	(1 - \rho) \right] - 2\epsilon_l \rho^2. 
	\label{eqOneCompFEng}
\end{equation}
We define the Helmholtz free energy per `particle' as
%
\begin{equation}
	a \equiv \frac{F}{N} = k_B T \left[ \ln \rho + \frac{(1 - \rho)}{\rho} \ln 
	(1 - \rho) \right] - 2\epsilon_l \rho.
	\label{eqOneCompFree}
\end{equation}
From this, we may calculate other thermodynamic quantities: the pressure $P$ and 
chemical potential $\mu$, which are given by the following relations \cite{HaMc06}:
%
\begin{eqnarray}
	P &=& \rho^2 \left( \frac{\partial a}{\partial \rho} \right), \label{eqPress} \\
	\mu &=& a + \rho \left( \frac{\partial a}{\partial \rho} \right).
\end{eqnarray}
We can calculate the spinodal for the one component fluid from the free energy 
Eq.~(\ref{eqOneCompFEng}) and the definition of the spinodal 
$\frac{d^2f}{d\rho^2} = 0$, giving us the following equation:
%
\begin{equation}
	\frac{k_BT}{\epsilon_l} = 4 \rho (1 - \rho).
\end{equation}
Equations (\ref{eqOneCompFree}) and (\ref{eqPress}) give the following expression for 
the pressure in the system:
%
\begin{equation}
	P = -k_BT \ln(1-\rho) - 2\epsilon_l \rho^2.
	\label{eqOneCompPress}
\end{equation}
In order to simplify the task of determining the phase diagram, we may
use the symmetry of the Hamiltonian \eqref{eqHam}.  The
  Hamiltonian remains unchanged under the exchange $l_i \to (1 -
  l_i)$, i.e., it has a symmetry between `holes' (nearly dry
  substrate) and `drops' (liquid layer). This means that for the one
component fluid, the phase diagram is symmetric around the line $\rho=
\frac{1}{2}$, i.e.~for a phase on the binodal with a density of $\rho
= \rho_1$ the coexisting phase has a density of $\rho_2 = (1 -
\rho_1)$.  Using Eq.~(\ref{eqOneCompPress}) and this symmetry of the
Hamiltonian, we obtain the following expression for the density along the
binodal, by equating the pressure in the two phases ($P_1 = P_2)$:
%
\begin{equation}
	\frac{k_BT}{\epsilon_l} = \frac{2 (2 \rho - 1)}{\ln [ \rho/(1-\rho)]}
	\label{eqCoExDen}
\end{equation}
The maximum on this curve is at $\rho = \frac{1}{2}$ and corresponds to the critical temperature 
$k_BT/\epsilon_l = 1$.  Below the critical temperature 
there are two solutions; these are the coexisting densities.  The binodal and 
spinodal curves for the pure liquid are plotted in Fig.~\ref{figOneCompPh}(a).  
Fig.~\ref{figOneCompPh}(b) shows the value of the chemical potential along these curves.

The spinodal region can also be calculated from the dynamical equations \eqref{eqNanoDyn} 
and \eqref{eqLiqDyn} employing a linear stability analysis.  This allows us to determine the 
typical length scales of the density fluctuations in the liquid film which might exist during the 
evaporation process.  To perform the linear stability analysis we consider a liquid density which 
varies in space and time $\rho_l = \rho({\bf r},t)$.  The free energy for the single component fluid 
($\rho_n=0$) is given by:
%
\begin{equation}
	F = \int d {\bf r}  \bigg[ f(\rho) + \frac{\epsilon}{2} (\nabla \rho)^2 - 
	       \mu \rho \bigg],
	\label{eqOneCompFreeEng}
\end{equation}
where the subscript on the liquid interaction variable $\epsilon_l$ is dropped for simplicity.
The steady state solutions of the liquid dynamical equation Eq.~\eqref{eqLiqDyn} represent 
the equilibrium density configurations.  There are several steady states for this system, 
(e.g. a density profile containing a free interface between two co-existing densities with
$\mu = \mu_\mathrm{coex}$) but here we consider the simplest steady state: a flat 
homogeneous film with a density $\rho = \rho_0$ which is defined by:
%
\begin{equation}
	\fnDir{F}{\rho} \bigg|_{\rho_0} = 0.
	\label{eqRho0Def}
\end{equation} 
We consider small amplitude perturbations $\delRho$ from $\rho_0$ of the form 
$\rho = \rho_0 + \delRho = \rho_0 + \phi e^{i{\bf k\cdot r}}e^{\beta
  t}$, where the amplitude $\phi \ll 1$ 
is a small positive constant, $k = |{\bf k}|$ is the wave number and $\beta$ is the rate of
growth/decay  with time (for positive/negative values) of the perturbation.
We substitute this expression for 
$\rho$ into the dynamical equation \eqref{eqLiqDyn} and expand in powers of $\delRho$.  
Then taking just the leading order terms allows us to derive a simple expression for $\beta$ 
which can be solved analytically. 

A Taylor series expansion of the functional derivative of the free energy 
\eqref{eqOneCompFreeEng} yields:
%
\begin{eqnarray}
	\fnDir{F}{\rho} &=& \paDir{f}{\rho} - \epsilon \nabla^2 \rho - \mu
	\nonumber \\ &=& \paDir{f}{\rho} \bigg|_{\rho_0} + \paDirH{f}{\rho}{2}
	\bigg|_{\rho_0} \delRho + \epsilon k^2 \delRho - \mu +
	 O(\delRho^2).
	 \label{eqFuncDerOne}
\end{eqnarray}
Substituting this approximation for the functional derivative \eqref{eqFuncDerOne} into 
the dynamical equation \eqref{eqLiqDyn} we obtain
%
\begin{eqnarray}
	\beta \delRho &=& M_c \nabla \cdot \bigg[ \rho_0(ik \paDirH{f}{\rho}{2} 
	\bigg|_{\rho_0}\delRho + i\epsilon k^3 \delRho)\bigg] - \nonumber \\ && 
	M_{nc} \bigg( \paDir{f}{\rho}\bigg|_{\rho_0} + \paDirH{f}{\rho}{2} 
	\bigg|_{\rho_0} \delRho + \epsilon k^2 \delRho -\mu \bigg) 
	+ O(\delRho^2). 
\end{eqnarray}
Using the definition of $\rho_0$ [Eq.~\eqref{eqRho0Def}] gives $\paDir{f}{\rho}\big|_{\rho_0} - 
\mu = 0$ and neglecting second order terms $O(\delta\rho^2)$ we arrive at the
expression for the growth rate
%
\begin{equation}
	\beta = -\bigg( M_c \rho_0 k^2 + M_{nc}\bigg) 
	\bigg( \paDirH{f}{\rho}{2} \bigg|_{\rho_0} + \epsilon k^2 \bigg). 
\end{equation}
When $\beta$ is positive, small perturbations from the steady state $\rho_0$ with the wave 
number $k$ grow in amplitude over time.  Conversely, if $\beta$ is negative then small 
perturbations decay.  Since $M_c$, $\rho_0$, $k^2$, $M_{nc}$ and $\epsilon$ 
are all positive quantities, $\beta$ is always negative (i.e.~the fluid is stable for all wave 
numbers $k$) when the second derivative $\paDirH{f}{\rho}{2} \big|_{\rho_0}$ is positive.  
However when $\paDirH{f}{\rho}{2} \big|_{\rho_0}$ is negative,  we find that $\beta$ is positive 
for small values of $k$ and negative when $k$ is large - i.e.\ the fluid is unstable against long 
wavelength (small wave number) fluctuations in density.  This corresponds to the 
thermodynamic definition of the spinodal as previously discussed.  We may define a critical 
wave number $k_c$, as the wave number at which $\beta(k_c) = 0$.  When $M_{nc} \ne 0$ 
and $\paDirH{f}{\rho}{2} \big|_{\rho_0} < 0$, the critical wave number $k_c$ is given by:
%
\begin{equation}
	 k_c = \pm \sqrt{ -\frac{1}{\epsilon} \paDirH{f}{\rho}{2} \bigg|_{\rho_0} }.
 \end{equation}
 The real system we are modelling is very large ($L \gg 2\pi/k_c$),
 which means fluctuations can occur on the full spectrum of wave
 numbers.  The mode with wave number $k=k_m$, which has the largest
 positive value for $\beta$ grow the fastest.  The wave number
 $k_m$ corresponds to a typical length scale $2 \pi/k_m$ that are
 visible during the early stages of spinodal decomposition.  However,
 at later stages (beyond the linear stage) the length scale of the
 modulations is likely to deviate from this value as the pattern
 coarsens. For the purely evaporative case when $M_{c} = 0$,
   the maximum value of $\beta(k)$ occurs at $k = k_m = 0$, which
   means there is no typical length scale in the early stages of the
   evaporation process.  However, in the purely diffusive case $M_{nc}
   = 0$ (and $\paDirH{f}{\rho}{2} \big|_{\rho_0} < 0$), $\beta(k=0)=0$
   due to mass conservation, and the maximum value of $\beta(k)$ occurs
   at a non-zero value of $k_m$, so the typical length scale $2
   \pi/k_m$ is visible during the spinodal decomposition--evaporation. In all
 other cases the following expression is both a necessary and
 sufficient condition for the existence of the typical length scale
 (i.e. $k_m \ne 0$):
%
\begin{equation}
	\frac{M_{nc}}{M_c} < - \frac{\rho_0}{\epsilon} \paDirH{f}{\rho}{2}
	\bigg|_{\rho_0}.
	\label{eqMaxCond}
\end{equation}
If a typical length scale does exist, then the corresponding wave number $k_{m}$ is given 
by:
%
\begin{equation}
	k_m = \sqrt{-\frac{1}{2}\bigg( \frac{1}{\epsilon} \paDirH{f}{\rho}{2}
	\bigg|_{\rho_0} + \frac{M_{nc}}{\rho_0 M_c}\bigg)} . 
 \end{equation}
 %
\begin{figure}[t]
\begin{center}
	\includegraphics[width=0.49\columnwidth]{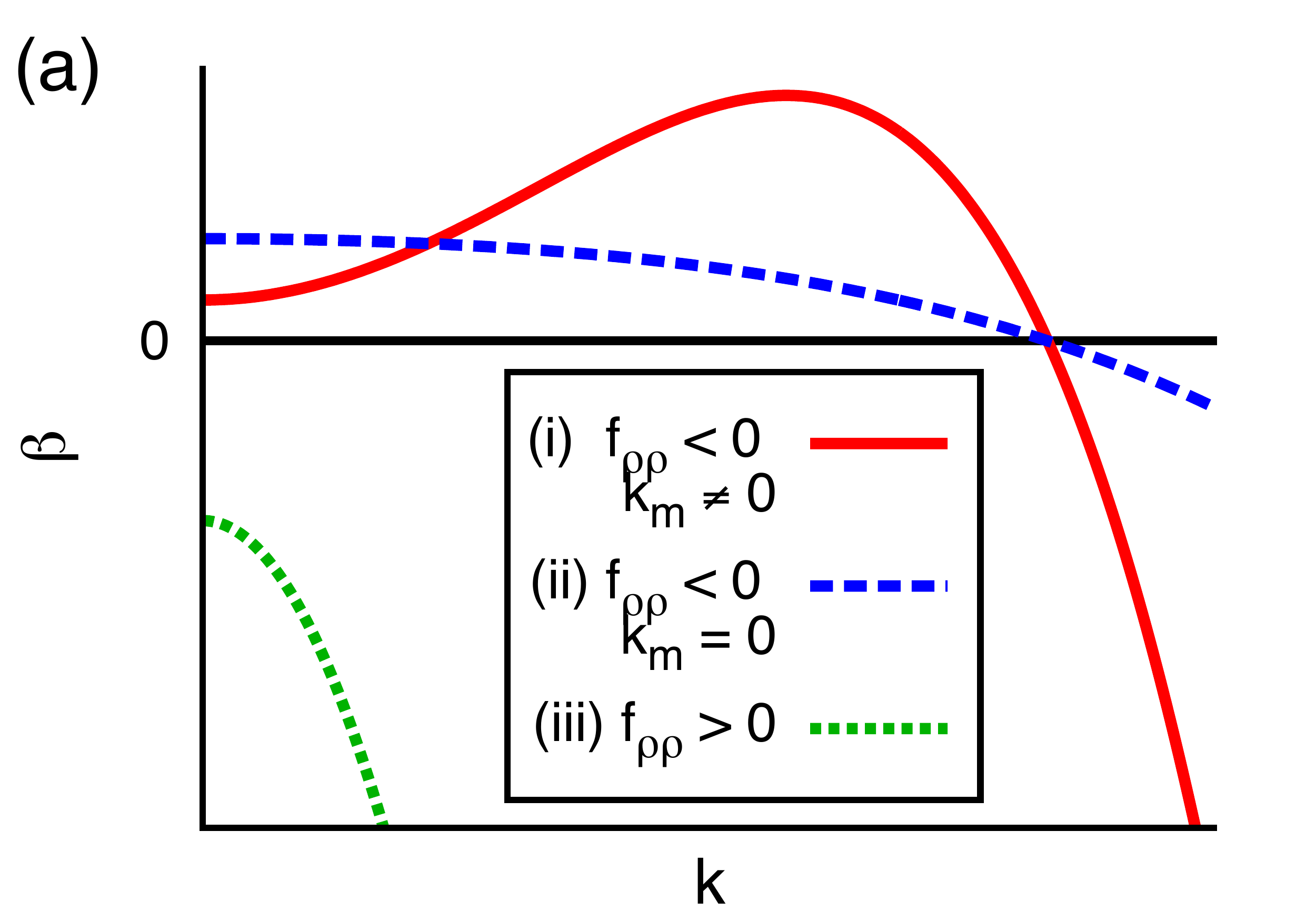}
	\includegraphics[width=0.49\columnwidth]{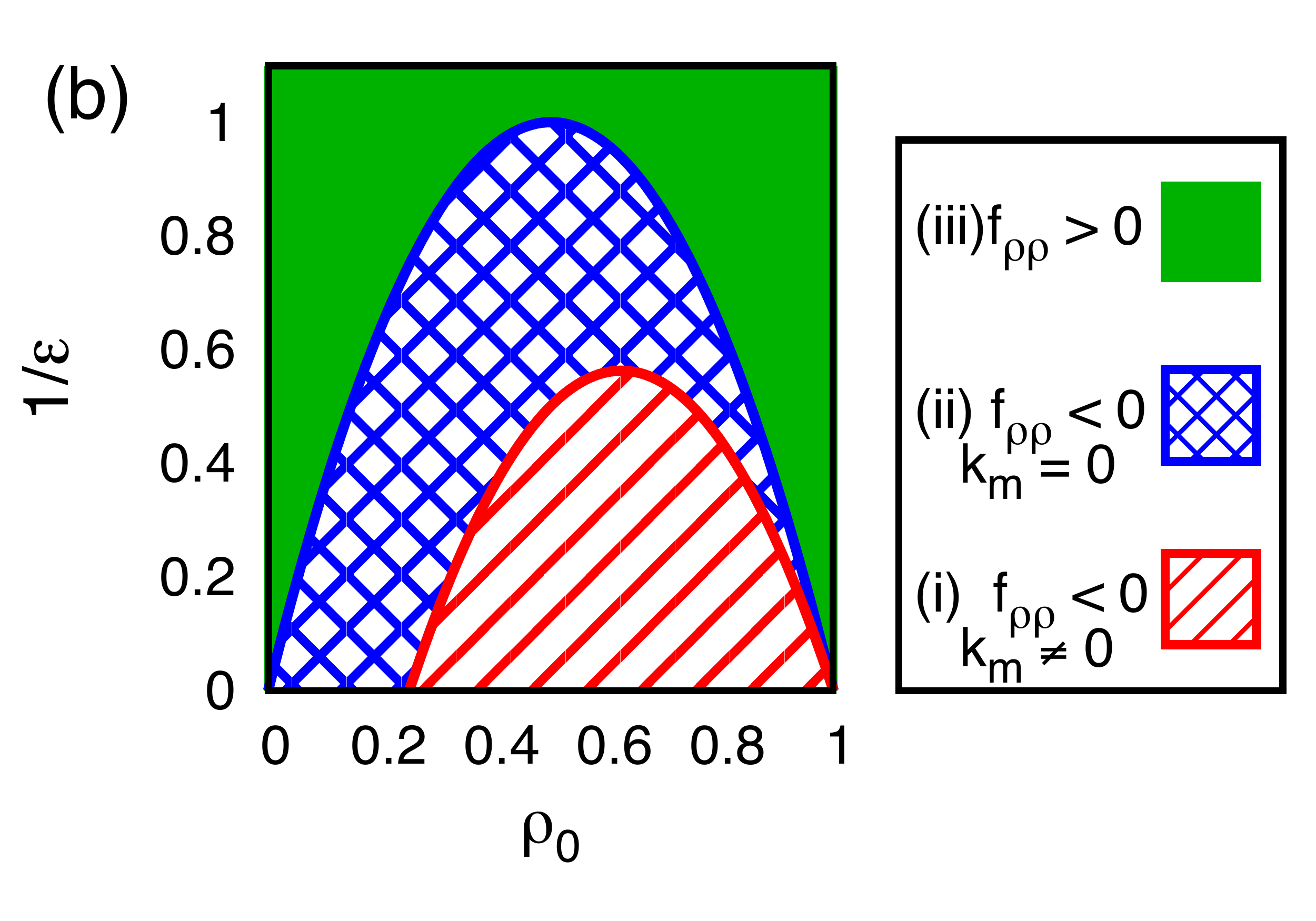}
	\caption{There are three possible forms for $\beta (k)$: (a) shows an example of
		      each form and (b) shows the regions in parameter space where these 
		      different types of $\beta(k)$ curves occur, for the case when 
		      $M_{nc}/M_c = 1$ and $k_B T = 1$.}
	\label{figOneComp}
\end{center}
\end{figure}

Figure \ref{figOneComp} displays (a) typical $\beta(k)$ curves for a one-component fluid 
and (b) the values of $1/\epsilon$ and $\rho_0$ where each type of curve is found (for the 
case when $M_{nc}/M_c = 1$ and $k_B T = 1$).  Three distinct cases are displayed: Case 
(i) the solid line (red online) in (a) and the striped area (red online) in (b) display the case when one observes 
the growth of density fluctuations with a typical length scale. Case (ii), the dashed line (blue online) in 
(a) and the hashed area (blue online) in (b) correspond to the situation when the fluid is unstable for 
density fluctuations corresponding to small wave numbers $k$ but no typical length scale 
is observed, because the fastest growing mode is the $k = 0$ fluctuation.  Case (iii), 
the dotted line (green online) in (a), corresponding to the (green) solid area in (b), displays the situation 
when the liquid is linearly stable against density fluctuations with all wave numbers $k$.  

\subsection{Binary mixture}
\label{secPhaseBinary}

To determine the binodal and spinodal curves for the binary mixture is
more challenging than for the one component system.  There are many
more parameters defining the model. In particular, the behaviour of
the system strongly depends on the ratios between the different
interaction strengths $\epsilon_l$, $\epsilon_n$ and $\epsilon_{nl}$.
Ref.~\cite{WoSc06} gives a good overview of the equilibrium bulk fluid
phase behaviour for binary fluid mixtures and the different classes of
phase diagrams that may be observed.  Here, we only describe the
influence of the chemical potential of the nano-particles $\mu_n$ on
the densities in the coexisting phases and on the critical point.

\begin{figure}[t]
\begin{center}
	\includegraphics[width=0.49\columnwidth]{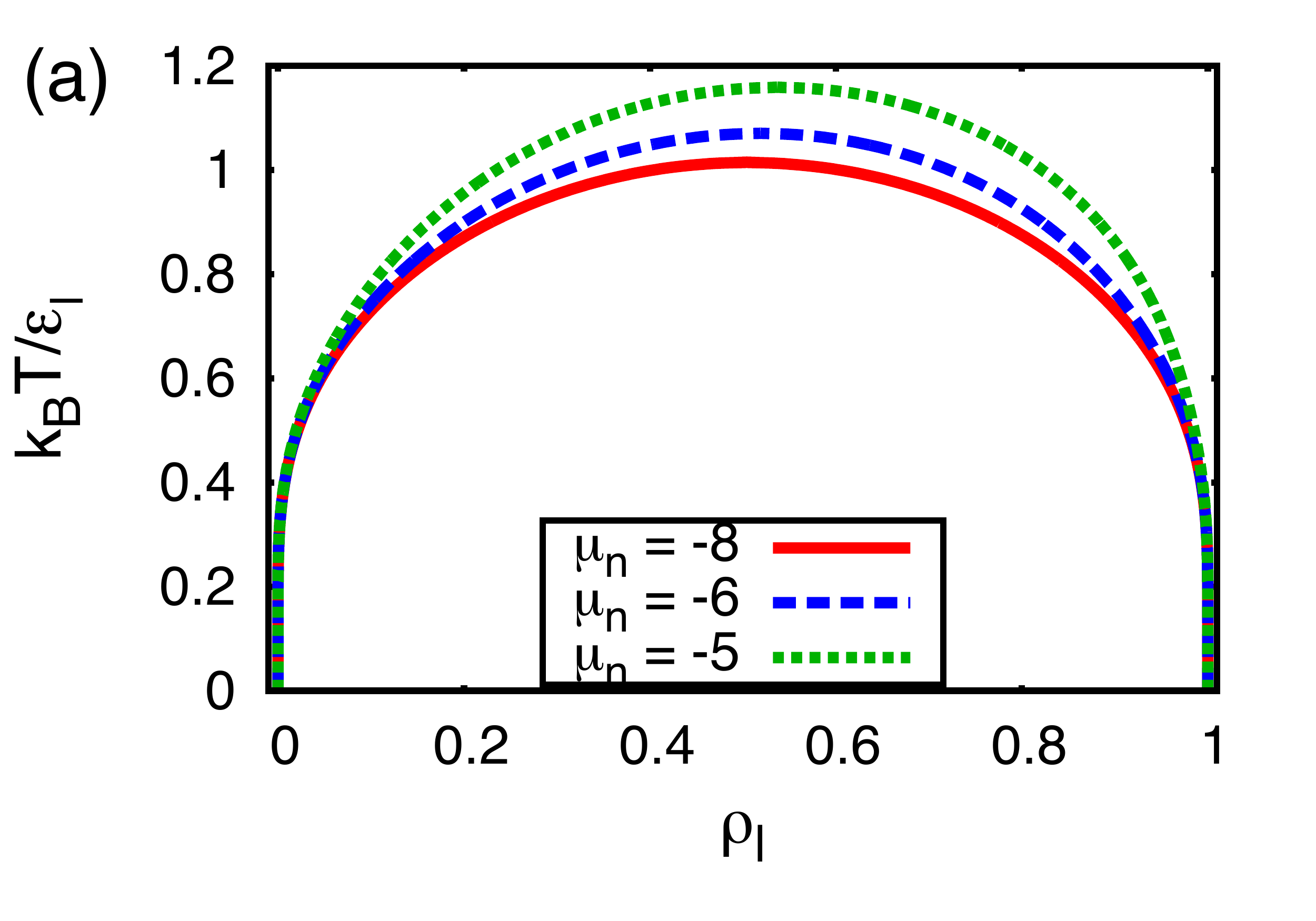}
	\includegraphics[width=0.49\columnwidth]{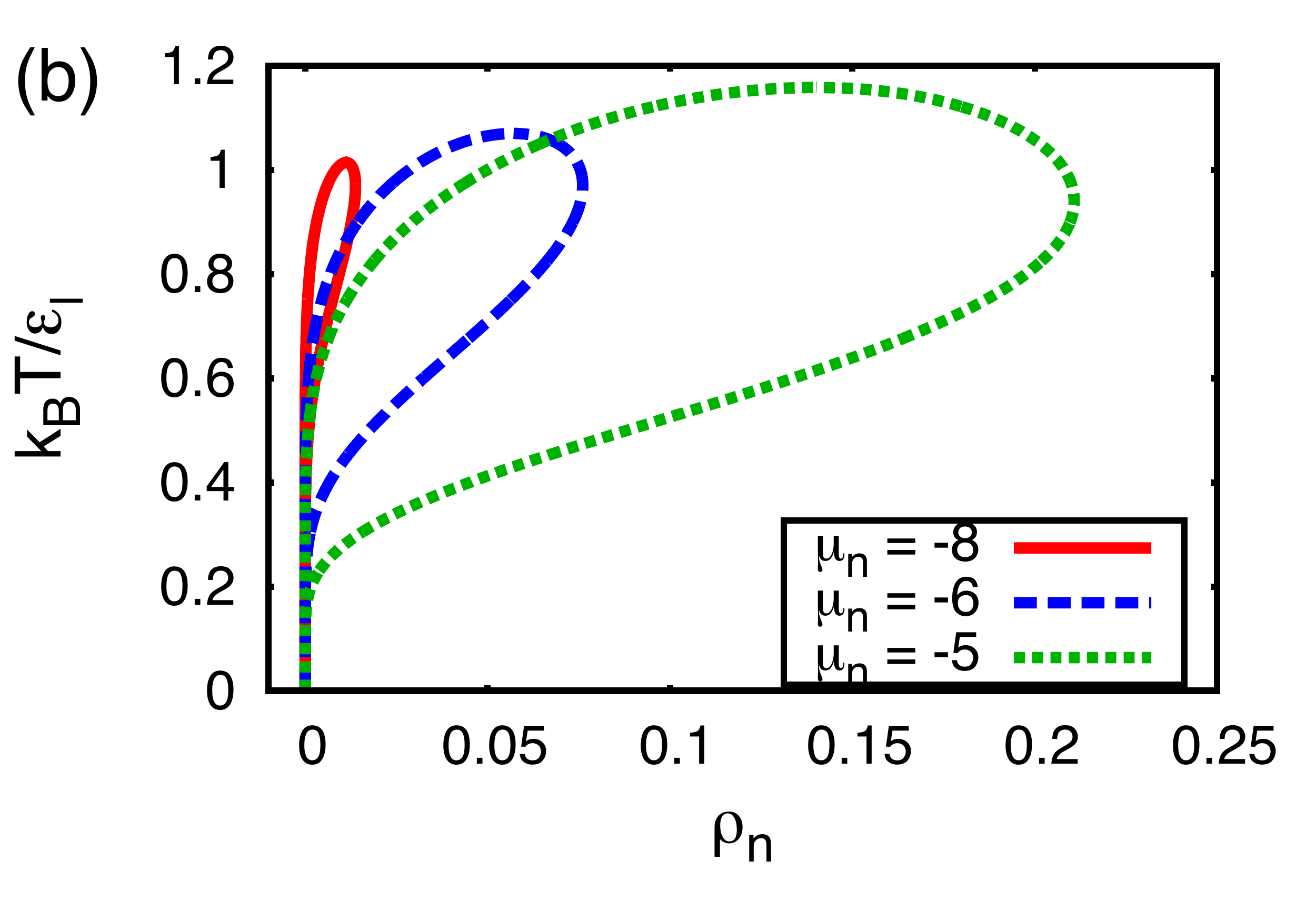}
	\caption{Binodal curves for the binary mixture for varying fixed values
		      of the chemical potential of the nano-particles $\mu_n$, for 
		      the case when $\epsilon_n/\epsilon_l = 0.43$ and
		      $\epsilon_{nl}/\epsilon_l = 0.57$.}
	\label{figBinBin}
\end{center}
\end{figure}

For two phases of a binary mixture to coexist in thermodynamic
equilibrium, there are four conditions that must be satisfied.
We denote these two phases as (i) the low density phase (LDP)
or `dry' substrate and (ii) the high density phase (HDP) or
substrate covered by a colloidal film:
%
	\begin{equation}
		T^{LDP} = T^{HDP}, \label{eqCoExPh1}
	\end{equation}
	\begin{equation}
		\mu_l^{LDP} = \mu_l^{HDP}, \label{eqCoExPh2}
	\end{equation}
	\begin{equation}
		\mu_n^{LDP} = \mu_n^{HDP}, \label{eqCoExPh3}
	\end{equation}
	\begin{equation}
		P^{LDP} = P^{HDP}, \label{eqCoExPh4}
	\end{equation}
%
where $T$ is the temperature, $\mu_l$ and $\mu_n$ are the
chemical potentials of the liquid and nano-particles respectively, $P$
is the pressure and the superscript denotes the phase.  The first of
these equations is trivial to satisfy in our model.  We may then fix
the chemical potential of the nano-particles, to some value $\eta$,
and then solve equations \eqref{eqCoExPh2}, \eqref{eqCoExPh4},
$\mu_n^{LDP} = \eta$ and $\mu_n^{HDP} = \eta$ for the four density
values: $\rho_l^{LDP}$, $\rho_n^{LDP}$, $\rho_l^{HDP}$ and
$\rho_n^{HDP}$.  The curves of the coexisting density values
(binodals) for the parameters $\epsilon_{nl}/\epsilon_l = 0.57$ and
$\epsilon_n = 0.43$ are displayed in Fig.~\ref{figBinBin}.  The density
values calculated for the two phases meet at a critical temperature to
form binodal curves similar to the one found for the one component
system Fig.~\ref{figOneCompPh}a. However, we find that the
reflection symmetry w.r.t.\ the liquid density $\rho_l = 1/2$ seen for
the one component fluid is broken. In particular, the critical point
of the liquid is no longer at $\rho_l = 1/2$.  The liquid binodal
reduces to the one for the pure liquid as $\mu_n \to \pm \infty$.  The
curves for $\mu_n \to -\infty$ and $\mu_n \to \infty$ are identical
within our model
due to the symmetry of the Hamiltonian (\ref{eqHam}).  We observe
that the density of the nano-particles $\rho_n \to 0$ becomes very
small as the chemical potential decreases $\mu_n \to -\infty$.
Conversely, the density becomes very large $\rho_n \to 1$ as the
chemical potential increases $\mu_n \to \infty$.  The critical point
on the nano-particle binodal curve shifts in a similar manner to that
of the liquid binodal.
%

We now consider the linear stability of the fluid for the two component case.  There are 
many different possible steady states for the binary system including some with interfaces 
between coexisting phases.  However, here we limit ourselves to investigating the linear 
stability of the simplest steady state: where both components have a uniform constant 
density over the surface.  The uniform density of the nano-particle film is denoted by 
$\rho_n = \rho_n^0$ and the density of the liquid is given by $\rho_l=\rho_l^0$ where 
$\rho_l^0$ is determined from the condition:
%
\begin{equation}
	\fnDir{F}{\rho_l} \bigg|_{\rho_l^0} = 0.
\end{equation}
We consider small amplitude perturbations in the density of both components from this 
steady state.  We assume these perturbations to take the form:
%
 	\begin{equation}
	\rho_l = \rho_l^0 + \delRho = \rho_l^0 + \phi e^{i {\bf k\cdot r}}e^{\beta t},
	\end{equation}
	\begin{equation}
	\rho_n = \rho_n^0 + \chi \delRho = \rho_n^0 + \chi \phi e^{i{\bf k\cdot r}}e^{\beta t}, 
	\end{equation}
%
where the amplitude $|\phi| \ll 1$ and the parameter $\chi$ is the ratio between the 
perturbation in the densities of the two components.  The sign of $\chi$ determines 
whether any instabilities are in-phase (positive) or anti-phase (negative) between 
the two coupled density fields.  From the magnitude of $\chi$ we can determine whether 
an instability is driven by the liquid component ($|\chi| \ll 1$), the nano-particles 
($|\chi| \gg 1$) or stems from the interaction between the two components 
($|\chi| =O(1)$).  Making a Taylor series expansion of the functional derivative of the 
free energy with respect to the liquid density $\fnDir{F}{\rho_l}$, keeping only terms linear 
in $\delRho$ and substituting this expression into the liquid dynamical equation 
\eqref{eqLiqDyn} we obtain
%
\begin{equation}	
	\beta = - (M^l_c \rho_l^0 k^2 + M^l_{nc}) 
	\bigg(\paDirH{f}{\rho_l}{2} \bigg|_{\rho_l^0, \rho_n^0} 
	\hspace{-4mm} + \chi \frac{\partial^2 f}{\partial \rho_l \partial 
	\rho_n} \bigg|_{\rho_l^0, \rho_n^0} \hspace{-4mm} + k^2 
	\epsilon_l + k^2 \chi \epsilon_{nl} \bigg).
\label{eqRhoBeta}
\end{equation}
Turning our attention now to the nano-particles: in order to linearise 
Eq.~\eqref{eqNanoDyn}, we must first examine the nano-particle mobility function 
$M^n(\rho_l,\rho_n) = \rho_nm(\rho_l)$, given by Eqs.~\eqref{eqNMob} and \eqref{eqm} 
in our model.  Making a Taylor series expansion of the function $m(\rho_l)$ we obtain
%
\begin{equation}
	\frac{m}{\alpha} = \hspace{0.8mm} \gamma_0 + \gamma_1 \delRho + 
	O(\delRho^2),
	\label{eqNanMobTaylor}
\end{equation}
where $\gamma_0 \approx 0$ for small values of $\rho_l$ ($\rho_l < 0.45$) and 
$\gamma_0 \approx 1$ for large values of $\rho_l$ ($\rho_l > 0.55$).  Substituting a Taylor 
series expansion of the functional derivative, together with 
Eq.~\eqref{eqNanMobTaylor}, into the time evolution equation for the nano-particles 
Eq.~\eqref{eqNanoDyn} we find
%
\begin{equation}
	\chi \beta = -\alpha \rho_n^0 \gamma_0 k^2 \bigg(\chi \paDirH{f}{\rho_n}{2} 
	\bigg|_{\rho_l^0, \rho_n^0} \hspace{-4mm} + \frac{\partial^2 f}{\partial \rho_n 
	\partial \rho_l} \bigg|_{\rho_l^0, \rho_n^0} \hspace{-4mm} + k^2 \chi \epsilon_n 
	+ k^2 \epsilon_l \bigg).
	\label{eqNuBeta}
\end{equation}
When $\rho_l$ is small, $\gamma_0 \approx 0$. In consequence,
  $\chi=0$ and the Eqs.~\eqref{eqRhoBeta} and \eqref{eqNuBeta} reduce
  to the one of the one-component fluid (with a local free energy that
  depends also on $\rho_n^0$).  We now address the case when $\rho_l >
  0.55$, and we therefore assume $\gamma_0 = 1$.  The expressions for
  the time dependency of the amplitudes of the two density fields
  (Eq.~\eqref{eqRhoBeta} and Eq.~\eqref{eqNuBeta}) can be solved
  simultaneously to determine $\beta$ and $\chi$ as a function of the wave number
  $k$.  This allows us to determine the stability of the fluid for
  different values of the system parameters.  A fact that simplifies
  the analysis is that these two equations may be written in matrix
  form (similar to the case of two coupled mesoscopic hydrodynamic equations
  for dewetting two-layer films \cite{PBMT05}):
%
\begin{equation}		
	\beta \chiArray = \mathbf{M} \cdot \mathbf{G} \chiArray ,
	\label{eqMatrixForm}
\end{equation}
where, 
\begin{eqnarray}
	\mathbf{M} &=& \matTT{-\alpha \rho_n^0 k^2}{\hspace{5mm}0}{0}
	{-(M^l_c\rho_l^0 k^2 + M^l_{nc})}, \nonumber \\ 
	\mathbf{G} &=& \matTT{k^2 \epsilon_n + f_{nn}}{k^2 \epsilon_{nl} + f_{nl}}
	{k^2 \epsilon_{nl} + f_{nl}}{k^2 \epsilon_l +f_{ll}}, \nonumber
\end{eqnarray}
and we have used the shorthand notation $f_{ij} \equiv \paDirM{f} {\rho_i}{\rho_j} 
\big|_{\rho_l^0 \rho_n^0}$, where $i,j = n,l$ and where $n$ denotes the nano-particles 
and $l$ denotes the liquid.  We can determine $\beta(k)$ using the following 
expression for the eigenvalues of a $2\times2$ matrix:
%
\begin{equation}
	\beta(k) = \frac{\mbox{Tr}(\mathbf{M \cdot G})}{2} \pm 
	\sqrt{\frac{\mbox{Tr}(\mathbf{M \cdot G})^2}{4} - |\mathbf{M \cdot G}|}.
	\label{eqBetaTwoComp}
\end{equation}
We define critical wave numbers $k = k_c$ for the density fluctuations in the 
two-component fluid as the wave numbers at which one of the two solutions of 
Eq.~\eqref{eqBetaTwoComp} is equal to zero. Below 
[Eqs.~\eqref{eqCritWaveNSimple} and \eqref{eqCritWaveN}] we derive explicit 
expressions for $k_c$ which can be used to determine the conditions for the linear stability 
of the two component fluid (i.e.\ the fluid is stable when there is no solution for $k_c$ and 
the function $\beta(k) < 0$ for all wave numbers $k$).  Since the matrix $\mathbf{M}$ is 
diagonal and all diagonal elements are non-zero the inverse $\mathbf{M^{-1}}$ exists, 
allowing us to rewrite Eq.~\eqref{eqMatrixForm}  as a generalised
eigenvalue problem
%
\begin{equation}
	(\mathbf{G} - \mathbf{M^{-1}} \beta) \chiArray = 0.
\end{equation}
Setting $\beta = 0$ in this equation in order to find the critical wave numbers $k_c$, we find 
that the determinant $|\mathbf{G}| = 0 $, implying that
\begin{equation}
	k_c^4(\epsilon_n \epsilon_l  -\epsilon_{nl}^2) + k_c^2 (\epsilon_n f_{ll} + 
	\epsilon_l  f_{nn} - 2 \epsilon_{nl} f_{nl})
	+ f_{nn} f_{ll} - f_{nl}^2 = 0.
\end{equation}
In the special case when $\epsilon_n \epsilon_l = \epsilon_{nl}^2$, the critical wave 
number is given by:
\begin{equation}
	k_c = \sqrt{\frac{f_{nl}^2 - f_{nn} f_{ll}}{\epsilon_n f_{ll} + \epsilon_l 
	f_{nn} - 2 \epsilon_{nl} f_{nl}}}.
	\label{eqCritWaveNSimple}
\end{equation}
However, more generally, when $\epsilon_n \epsilon_l \neq
\epsilon_{nl}^2$, we have
%
\begin{equation}
	k_c = \sqrt{\frac{-b \pm \sqrt{b^2-4ac}}{2a}},
	\label{eqCritWaveN}
\end{equation}
where
\begin{eqnarray}
	a &=& \epsilon_n \epsilon_l - \epsilon_{nl}^2, \nonumber \\		
	b &=&  \epsilon_n f_{ll} + \epsilon_l f_{nn} - 
	2 \epsilon_{nl} f_{nl}, \nonumber \\
	c &=&  f_{nn} f_{ll} - f_{nl}^2.
	\label{eqBetaRootsABC}
\end{eqnarray} 

\begin{figure}[t]
\begin{center}
	\includegraphics[width=0.275\columnwidth]{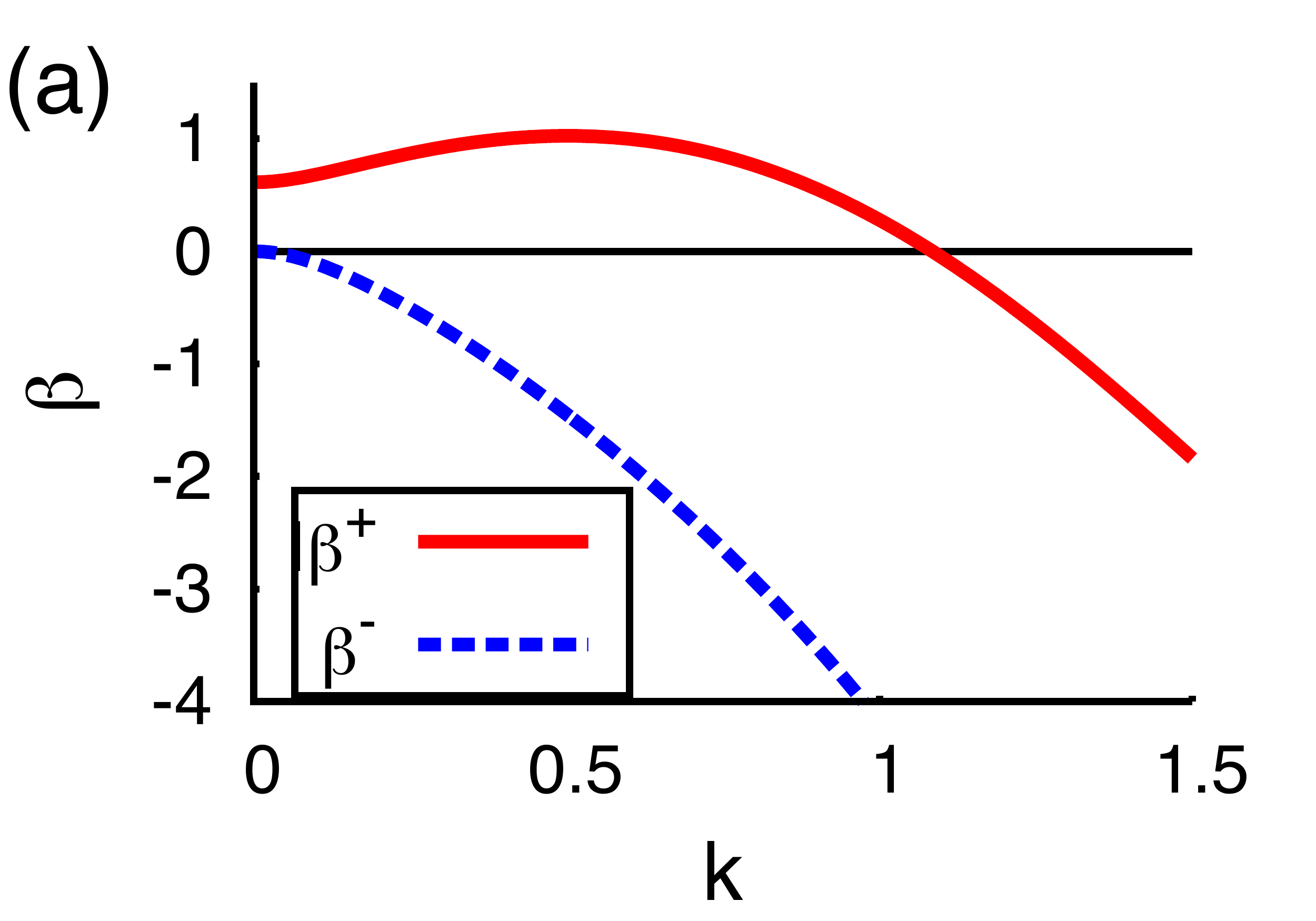} \hspace{2mm}
	\includegraphics[width=0.275\columnwidth]{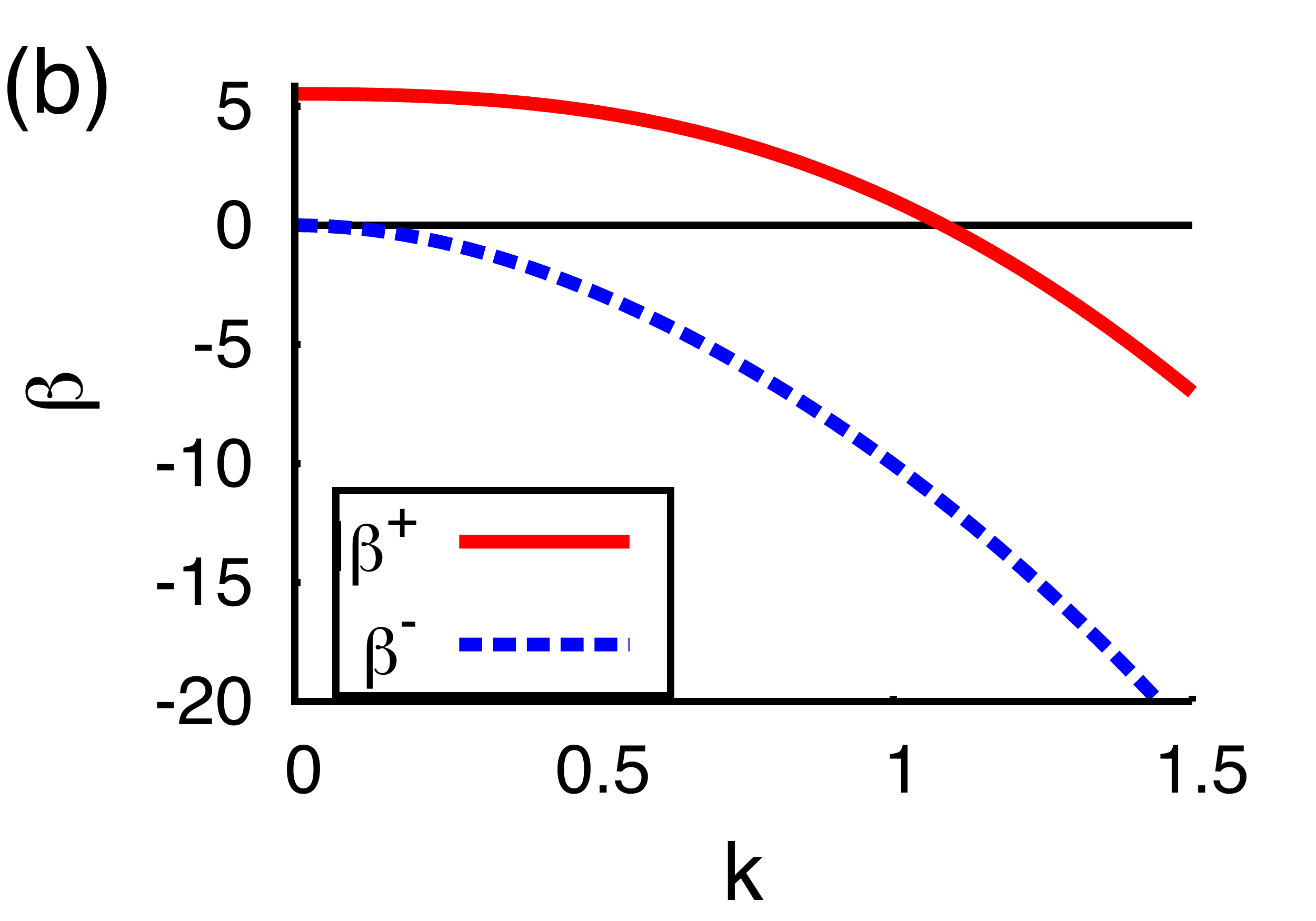} \hspace{2mm} 
	\includegraphics[width=0.275\columnwidth]{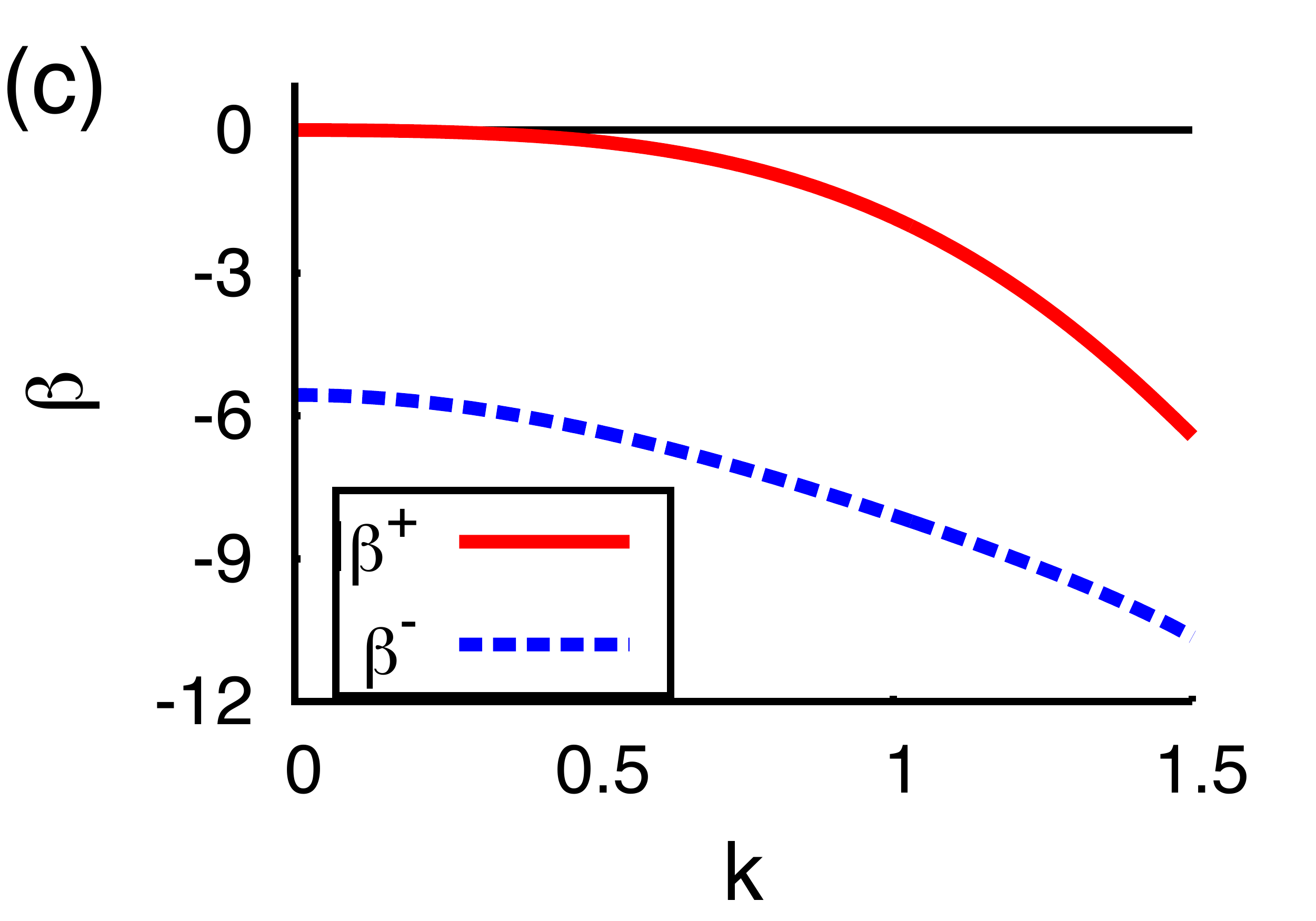}
	\includegraphics[width=0.275\columnwidth]{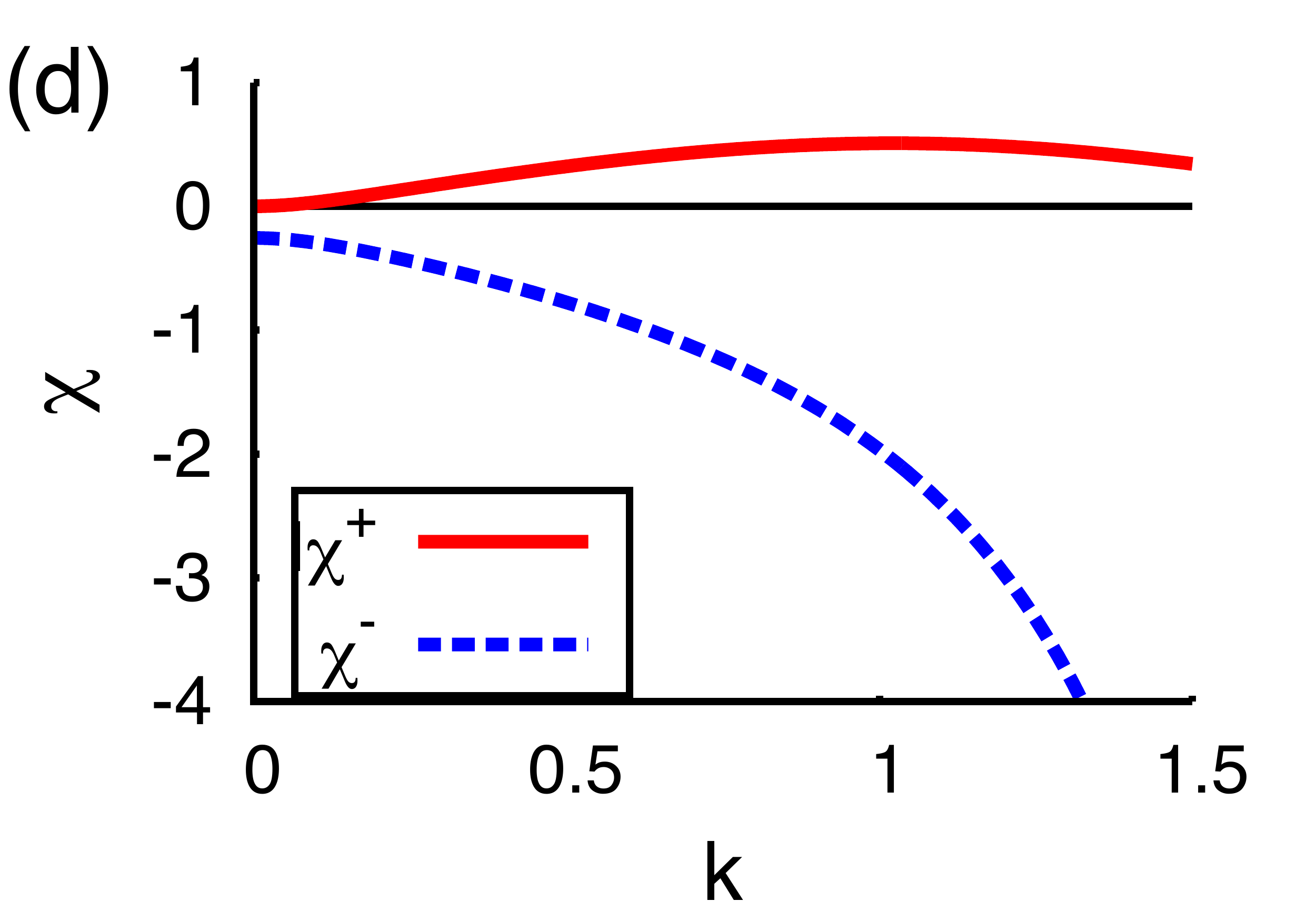} \hspace{2mm}
	\includegraphics[width=0.275\columnwidth]{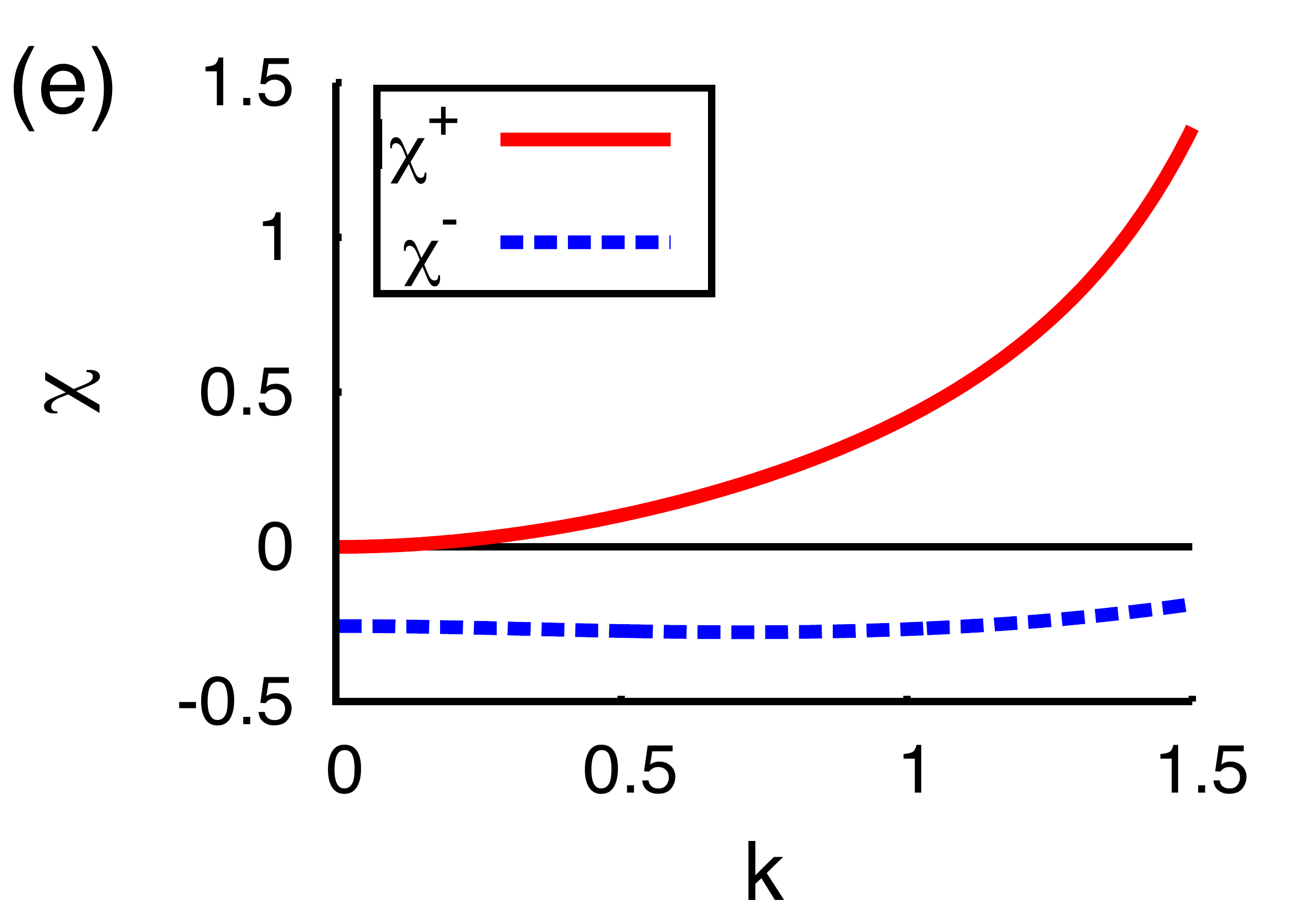} \hspace{2mm}
	\includegraphics[width=0.275\columnwidth]{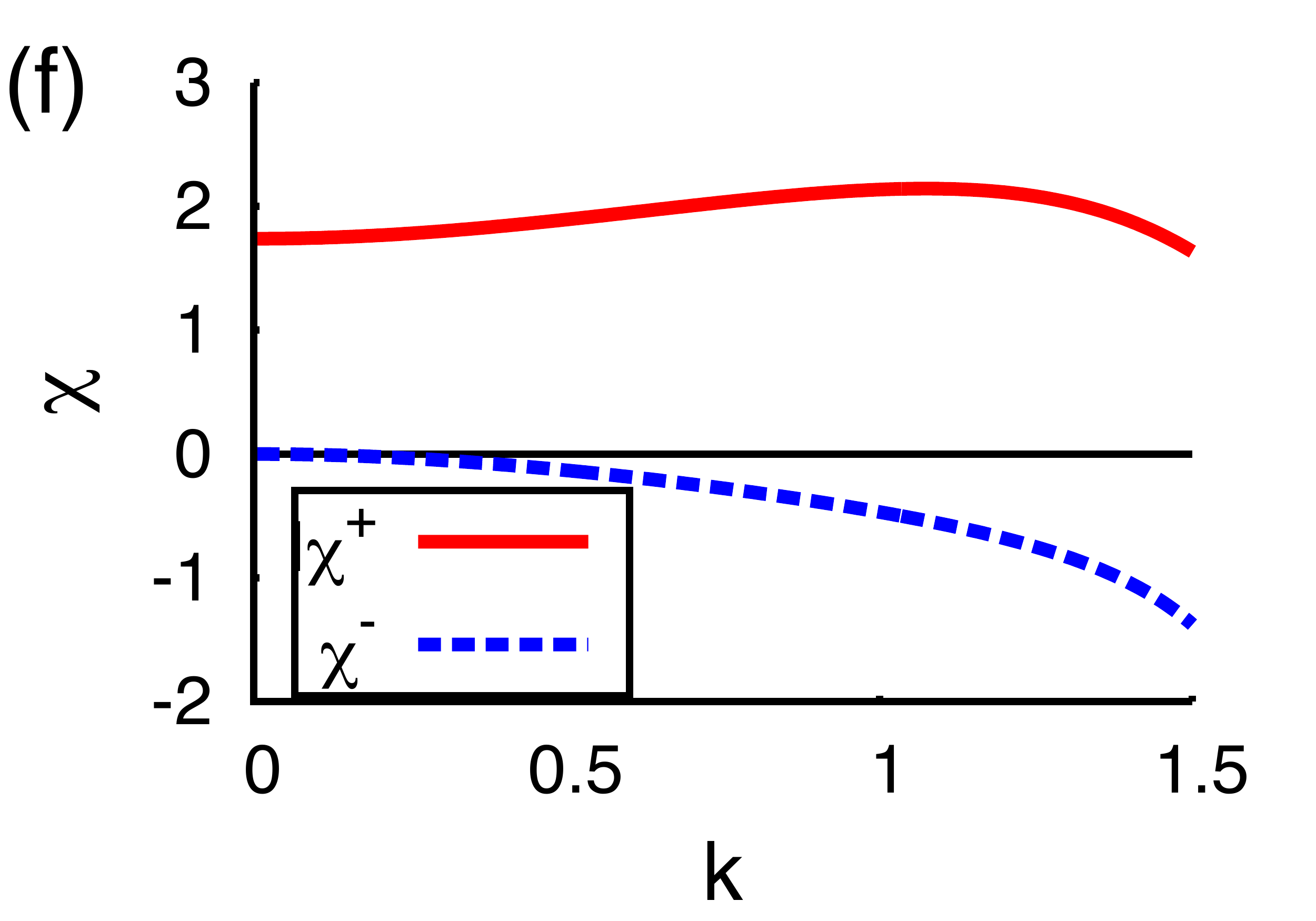} 
	\includegraphics[width=0.35\columnwidth]{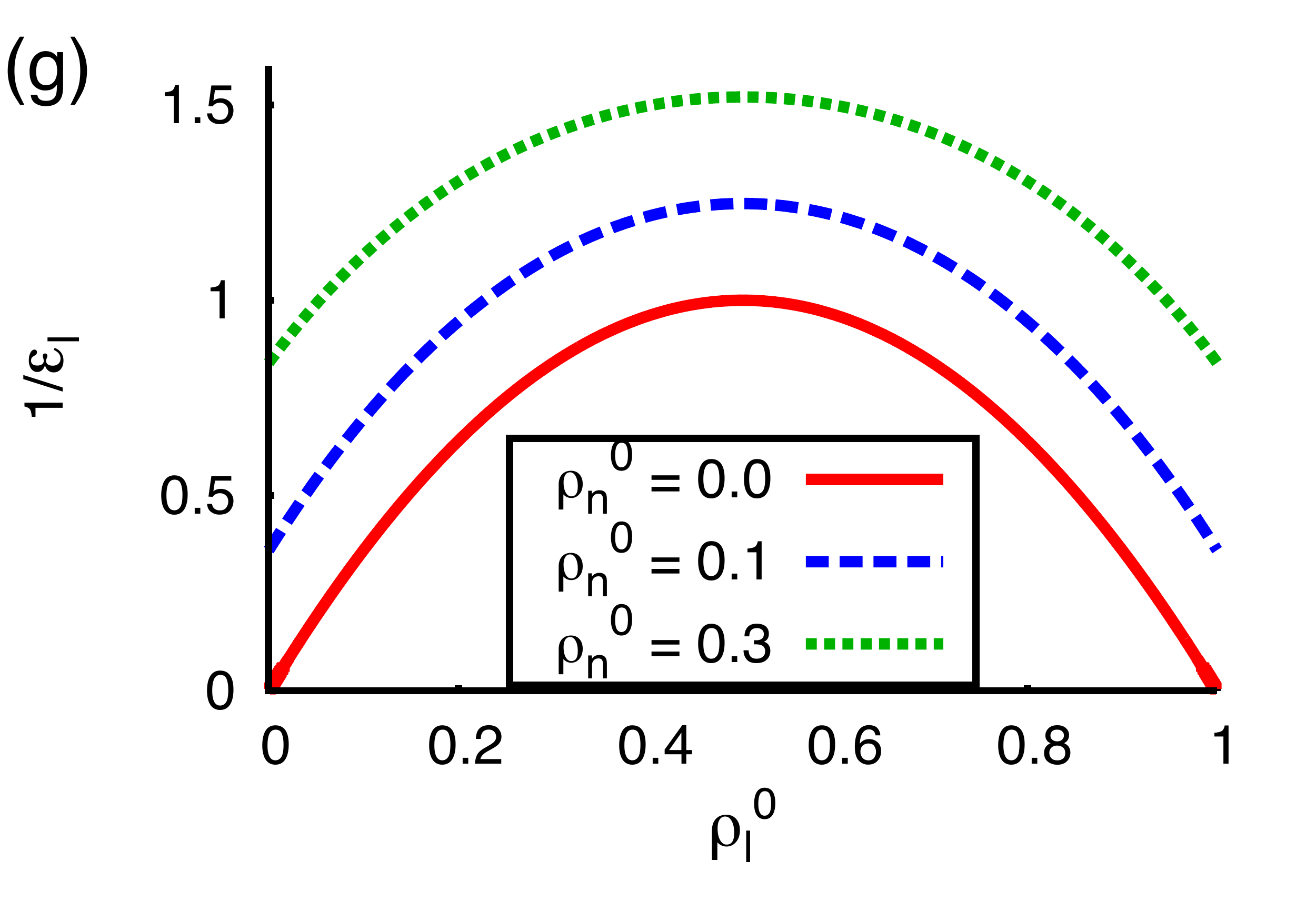}
	\includegraphics[width=0.35\columnwidth]{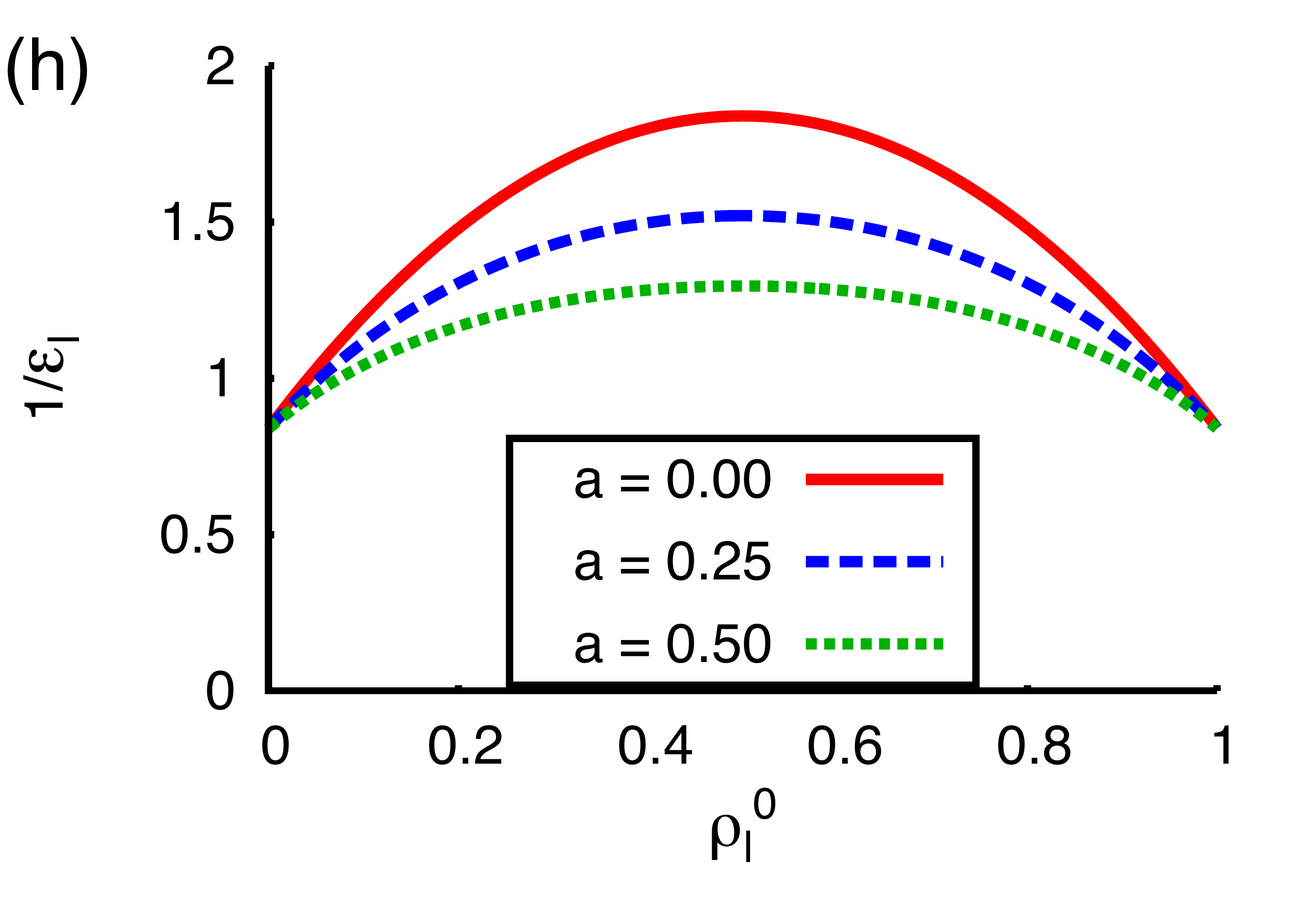}	
	\caption{(a) - (f) Dispersion relations for the case when 
		      $\epsilon_n \epsilon_l > \epsilon_{nl}^2$, \hspace{1mm} 
		      $\epsilon_n f_{ll} + \epsilon_l f_{nn} - 2 \epsilon_{nl} f_{nl} > 0$ 
		      (i.e. in Eq.~\eqref{eqBetaRootsABC} $a > 0, b > 0$) and $k_B
		      T = 1$.  The graphs in the top row show $\beta(k)$ and the graphs 
		      in the middle row show $\chi(k)$.  In (a) and (d) the parameters are: 
		      $\rho_l^0 = 0.648, \rho_n^0 = 0.3, \epsilon_l = 1.25, 
		      \epsilon_n = 0.5, \epsilon_{nl} = 0.6, \mu = -3.35, M^n = 1, 
		      M^l_c = 0 \mbox{ and } M^l_{nc} = 1.$  In (b) and (e) the 
		      parameters are: $\rho_l^0 = 0.648, \rho_n^0 = 0.3, \epsilon_l 
		      = 1.25, \epsilon_n = 0.5, \epsilon_{nl} = 0.6, \mu = -3.35, 
		      M^n = 1, M^l_c = 0 \mbox{ and } M^l_{nc} = 9$.  In (c) and (f) the 
		      parameters are: $\rho_l^0 = 0.901, \rho_n^0 = 0.3, \epsilon_l 
		      = 1.4, \epsilon_n = 0.6, \epsilon_{nl} = 0.8, \mu = -3.8, M^n =
		      1, M^l_c = 0 \mbox{ and } M^l_{nc} = 1.$  (g) shows how the 
		      spinodal line shifts for increasing $\rho_n^0$, where $\epsilon_n 
		      \epsilon_l - \epsilon_{nl}^2 = 0.25$, $\epsilon_n = \epsilon_l$ 
		      and $k_BT = 1$.  (h) shows how the spinodal line shifts as 
		      the value of $a = \epsilon_n \epsilon_l - \epsilon_{nl}^2$ increases, 
		      when $\rho_n^0 = 0.3$, $\epsilon_n = \epsilon_l$ and 
		      $k_BT = 1$.}
	\label{figDispRelAPBP}
\end{center}
\end{figure}

Note that the locus $c = 0$ is the spinodal curve \cite{ArEv01,
  RoSw82}.  We categorise the linear behaviour of the system by
  the signs of $a$ and $b$ in Eq.~\eqref{eqBetaRootsABC} as they have
  a profound impact on the shape of the dispersion curves.  In
Fig.~\ref{figDispRelAPBP} we display for the case when $a > 0$ and $b
> 0$ all the different possible $\beta(k)$ curves, together with the
corresponding $\chi(k)$.  From Eq.~\eqref{eqBetaTwoComp} we see that
there are two branches for $\beta(k)$, which we denote $\beta^+(k)$
and $\beta^-(k)$. The corresponding $\chi(k)$ curves are denoted
$\chi^+$ and $\chi^-$, respectively.  The $\beta^+$ branch (red solid
lines in Fig.~\ref{figDispRelAPBP}(a) - (c)) corresponds to the second
term in Eq.~\eqref{eqBetaTwoComp} being positive and the $\beta^-$
branch (dashed lines) corresponds to the second term being
negative. Due to mass conservation of the nano-particles,
  there is always one $\beta(k)$ branch that is zero at $k=0$, and
  $\chi=0$ at $k = 0$ for the curve that corresponds to the $\beta(k)$
  that is not zero at $k=0$. If $f_{ll} \ge 0$ then $\beta^+(k=0) =
0$ and $\chi^-(k=0) = 0$.  Alternatively, if $f_{ll} \le 0$ then
$\beta^-(k=0) = 0$ and $\chi^+(k=0) = 0$.

Inspecting Fig.~\ref{figDispRelAPBP}, we find that there are three possible forms for 
$\beta(k)$, similar to the one component fluid case (Fig.~\ref{figOneComp}).  In  
Fig~\ref{figDispRelAPBP}(a) there is a maximum in $\beta^+$ at $k \ne 0$.  This indicates 
that the fluid is unstable and that one will observe the growth of density fluctuations with a 
typical length scale $2\pi/k_m^+$ during the early stages of the spinodal process.  Here 
$k_m^+$ refers to the wave number at the maximum of $\beta^+$.  In the second 
case [Fig.~\ref{figDispRelAPBP}(b)]  the fluid is linearly unstable for density fluctuations with 
small wave numbers $k$ as in Fig.~\ref{figDispRelAPBP}(a), but the maximum in $\beta^+(k)$ 
occurs at $k_m^+=0$, i.e. there is no typical length scale visible in the density profiles.  
In the last case, shown in Fig.~\ref{figDispRelAPBP}(c), the fluid is stable for all wave numbers 
$k$.  We can use Eq.~\eqref{eqCritWaveN} for the critical wave number $k_c$ to determine the 
stability of the system.  Since $a > 0$ and $b > 0$ we observe that
there is at most one positive 
root of $\beta(k)$, which only exists if $c < 0$ (i.e.~inside the spinodal).  The spinodal curve 
when $a > 0$ and $b > 0$ is plotted in Fig.~\ref{figDispRelAPBP}(g) and (h).  In (g) we show 
how the spinodal line moves upwards in $1/\epsilon_l$ as the nano-particle density 
$\rho_n^0$ is increased.  In (h) we show how the shape of the spinodal changes as the 
value of the parameter $a$ is increased.  We find that larger values of $a$ make the spinodal 
curve flatter.

In the reduced case when $a = 0$, one can determine from Eqs.~\eqref{eqBetaRootsABC} 
that $b$ is always be positive.  We find the same three types of dispersion relations for this 
reduced case as we do for the $a > 0$, $b > 0$ case. For the 
case when $a > 0$ and $b < 0$ we observe that the fluid is unstable for all possible 
combinations of parameter values.  When we have $a < 0$ (where at least one of the 
components of the mixture is more attracted to the other component than to itself), we 
find that $b$ must be positive.  In this regime, we observe that as $k \to \infty, \beta^+(k) \to 
\infty$ and $\beta^-(k) \to - \infty$.  This indicates that the density fluctuations with an 
infinitesimally small typical length scale will grow fastest.  This behaviour corresponds to a 
mixture that exhibits micro-phase separation.  This behaviour is common in block copolymer 
systems, where chemical bonding prevents macroscopic demixing and instead demixing on 
the nano-scale is witnessed \cite{Leibler80, NiOh95}.  Micro-phase separation is also 
observed in certain colloidal suspensions \cite{ArWi07, SCMG99, CADB05, SSCP04, SEP04}.

An important consideration which must be made with the binary system
is whether `liquid-particle' phase separation can occur as
well as the `liquid-gas' (low density - high density) phase separation
that we have already discussed.  The former corresponds to the
coexistence of two phases, both having a high density.  In terms of
our system this would mean that we have coexisting phases with a high
liquid density (i.e. $\rho_l \gsim 0.6$) in each phase.  The
coexisting values depend upon the temperature $T$, chemical potential
$\mu$ and the interaction energies $\epsilon_l$, $\epsilon_n$ and
$\epsilon_{nl}$.  It is known that the following condition must be
satisfied for `liquid-particle' demixing to occur \cite{WoSc06}:
%
\begin{equation}
	\epsilon_{nl} < \frac{\epsilon_n + \epsilon_l}{2}.
	\label{eqLiqLiqCond}
\end{equation} 
The existence of liquid-particle phase separation in addition to the
gas-liquid phase separation implies that for certain parameter values
we may have three co-existing phases.  This situation has the
potential to lead to dramatic consequences for the pattern formation
in our dynamical system.  We return to this issue in
Sec.~\ref{secNumEnn} below.


\section{Nonlinear Dynamics}
\label{secNumerics}

We now go beyond the linear analysis presented above and discuss numerical results for the
fully non-linear time evolution.  

\subsection{Numerical setup}

Recall that the dynamics of our model is governed by the coupled partial differential 
equations \eqref{eqNanoDyn} and \eqref{eqLiqDyn}, together with the free energy functional 
Eq.~\eqref{eqFreeEngCont1}.  We numerically solve these non-linear partial 
differential equations using a finite difference scheme on a square lattice with grid spacing 
$\Delta x = 1$.  The time step size $\Delta t$ varies between simulations, as the stability of 
our numerical scheme depends strongly on the values of the parameters in the model.  
Central difference approximations are made for the partial derivatives with respect to space 
and forward difference approximations are made for the partial derivatives with respect to time.  
The Laplacian terms ($\nabla^2$) are approximated using the eight-neighbour discretisation 
\cite{BRSG97}:
%
\begin{equation}
	\nabla^2 \rho = \frac{1}{2(\Delta x)^2} \left( \sum \rho^{NN} + \frac{1}{2} 
	 \sum \rho^{NNN} - 6 \rho \right),
	\label{eqLap}
\end{equation}
where, $\sum \rho^{NN}$ denotes a sum over the nearest neighbour lattice sites and 
$\sum \rho^{NNN}$ denotes a sum over the next nearest neighbour lattice sites.  Alternative
approximations may be used \cite{FoCl97}.  However, the choice of Laplacian approximation has 
little effect on the qualitative behaviour of the system.

Our numerical results show that as the various parameters in the model are varied, several 
different patterns are formed.  We begin in Sec.~\ref{secNumChem} by considering the effect 
of changing the chemical potential $\mu$ of the vapour 
reservoir.  We then focus on the fingering instability.  In Sec.~\ref{secNumFing} we discuss the 
effect of varying the parameter $\alpha$ in the nano-particle mobility and also the role which the 
conserved part of the liquid dynamics plays in the overall dynamics of the system, by varying 
the mobility coefficient $M_l^{c}$.  Finally, in Sec.~\ref{secNumEnn} we discuss the influence of 
`liquid-particle' demixing at the receding front and how this affects the fingering mechanism.

For the simulation results shown in Sec.~\ref{secNumChem} and Sec.~\ref{secNumFing} we set 
the interaction energies to  $\epsilon_l = 1.4$, $\epsilon_n = 0.6$ and $\epsilon_{nl} = 0.8$.  
Using the linear stability analysis we have shown that it is possible to obtain stable phases with 
these parameter values [Fig.~\ref{figDispRelAPBP}(c), $a > 0$ and $b > 0$ in 
Eq.~\eqref{eqBetaRootsABC}].

\subsection{Influence of the vapour chemical potential $\mu$}
\label{secNumChem}

In section~\ref{secPhaseOneComp} we have discussed how changing the value of $\mu$, the chemical 
potential of the vapour reservoir above the surface, affects the structures displayed by the system 
as the pure liquid evaporatively dewets from the surface.  Recall that as $\mu$ is decreased below 
its coexistence value $\mu_\mathrm{coex}$, the dewetting mechanism is at first via the nucleation 
of holes and then when $\mu$ is further decreased, via spinodal dewetting.  This sequence as 
$\mu$ is decreased is also observed when there are nano-particles dispersed in the liquid, 
although as previously discussed, the phase boundaries are shifted and there is the possibility of 
other (liquid-particle) phase transitions.  Choosing a binary mixture with the parameter values 
$k_BT = 1.0$, $\epsilon_n = 0.6$, $\epsilon_l = 1.4$ and $\epsilon_{nl} = 0.8$ gives a phase with 
a high liquid density coexisting with a low density of the liquid.  We also set $M_l^c=0$ and 
$M_l^{nc}=1$ to allow us to initially focus on just the evaporative non-conserved dynamics of the 
liquid.  We set the initial density profiles corresponding to a (high density) uniform film of 
liquid with density $\rho_l(x,t=0) = 1 - 10^{-6}$ mixed with nano-particles having an average 
density of $\rho_n^{av} = 0.3$.  In order to allow for the growth of density fluctuations when the 
system is (linearly) unstable, we add a small amplitude random noise to the density profile of the 
nano-particles.  Thus the initial nano-particle density profile is $\rho_n(x,t=0) = \rho_n^{av} 
+ 2\lambda(Y - 0.5)$, where $Y$ is a random real number uniformly distributed between 0 and 1 
and $\lambda$ is the magnitude of the noise.  Without these random density fluctuations the 
density profiles would remain uniform under the evolution of the DDFT with a film of liquid 
remaining.  Our boundary conditions are periodic in all directions.

\begin{figure}[b]
\begin{center}
	\includegraphics[width=0.3\columnwidth]{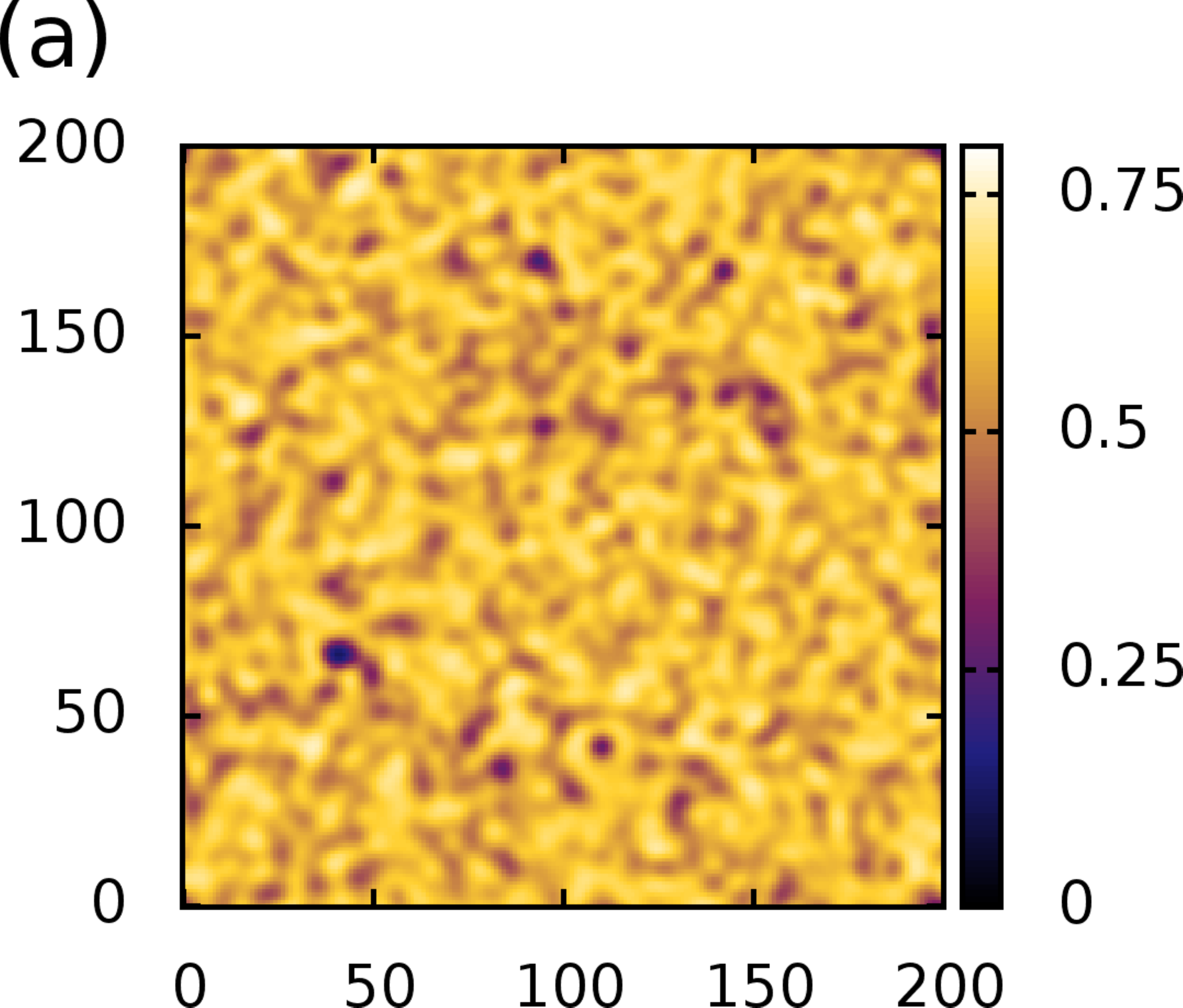} \hspace{1mm}
	\includegraphics[width=0.3\columnwidth]{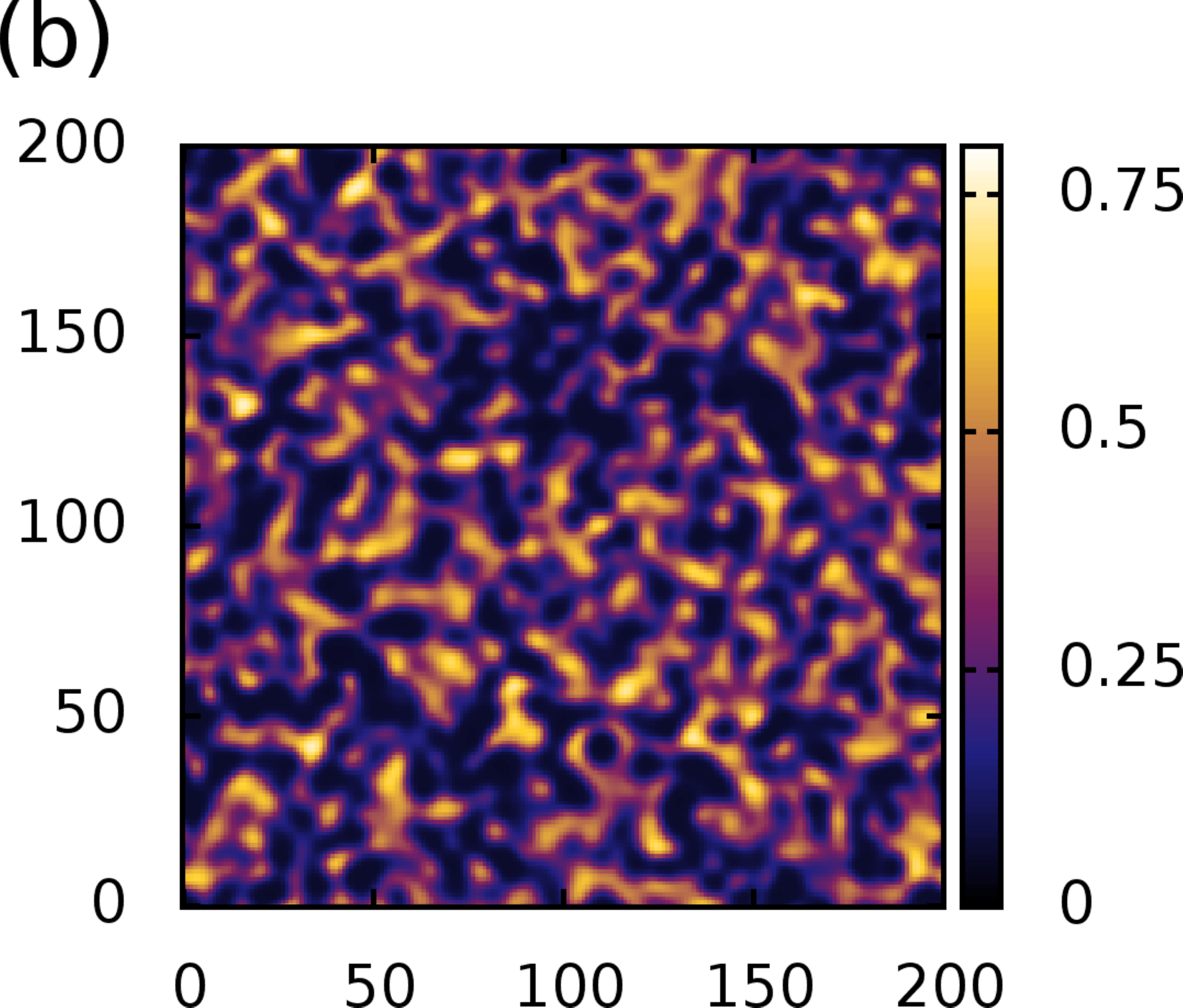} \hspace{1mm}
	\includegraphics[width=0.3\columnwidth]{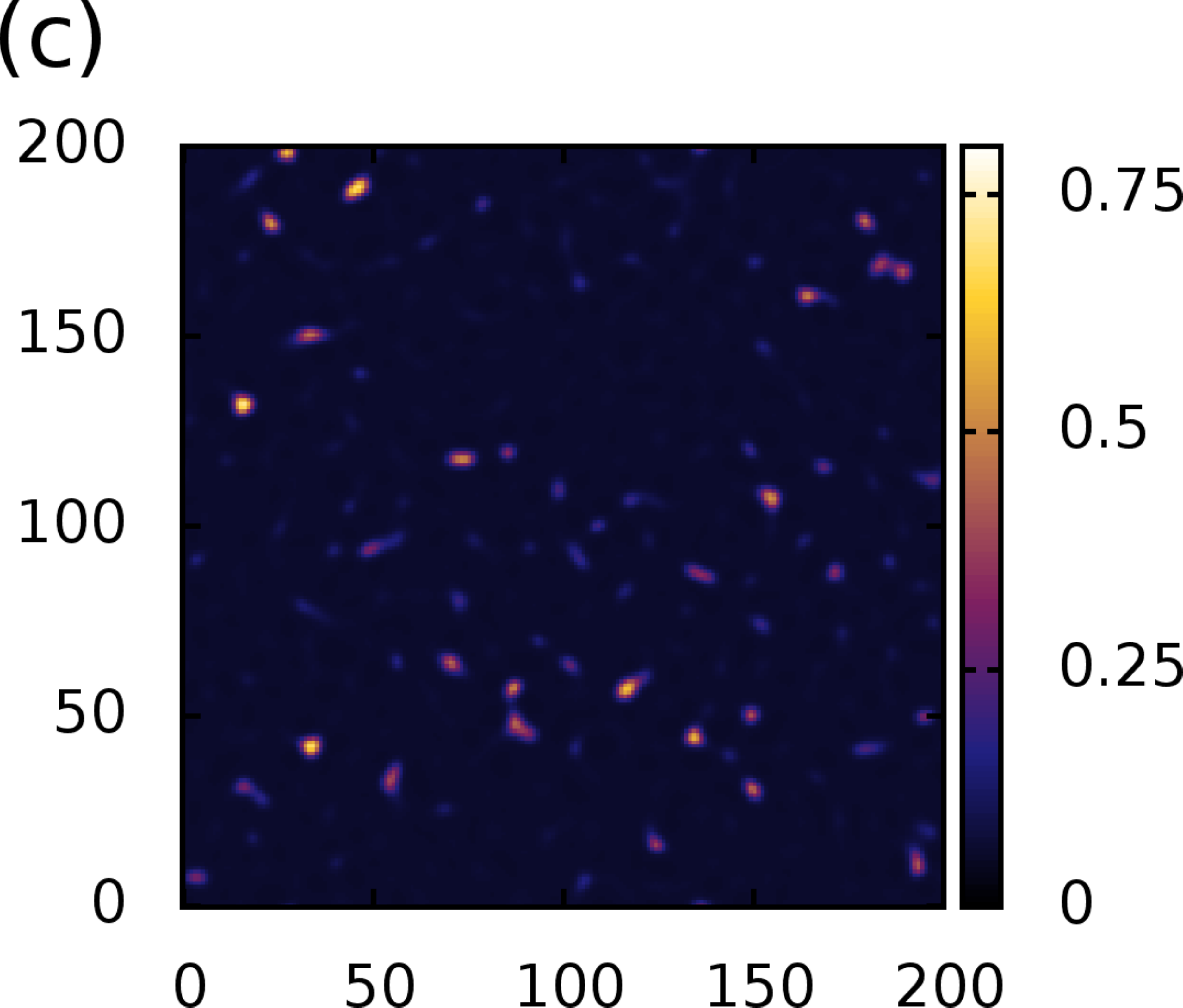} \\
	\includegraphics[width=0.3\columnwidth]{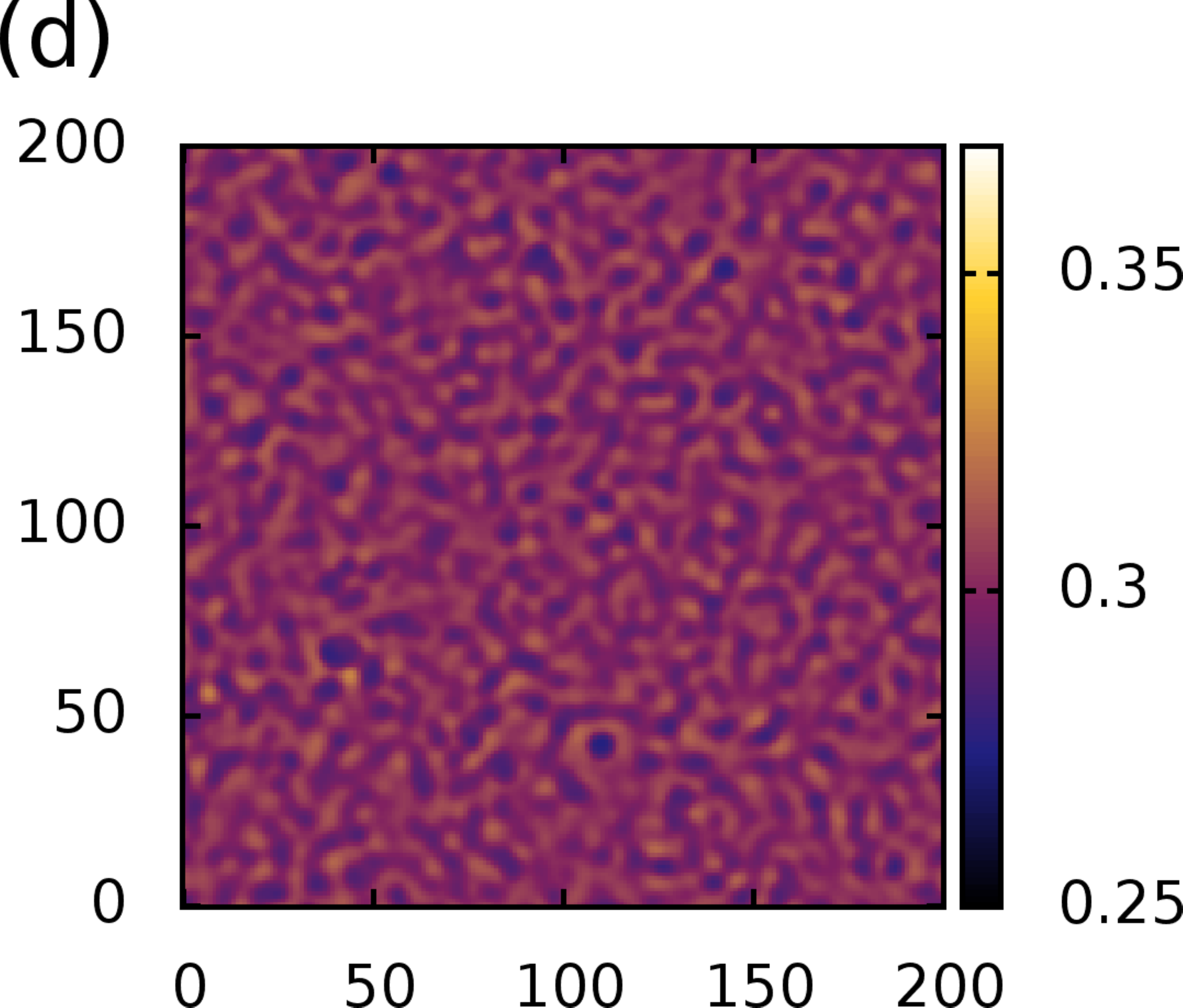} \hspace{1mm}
	\includegraphics[width=0.3\columnwidth]{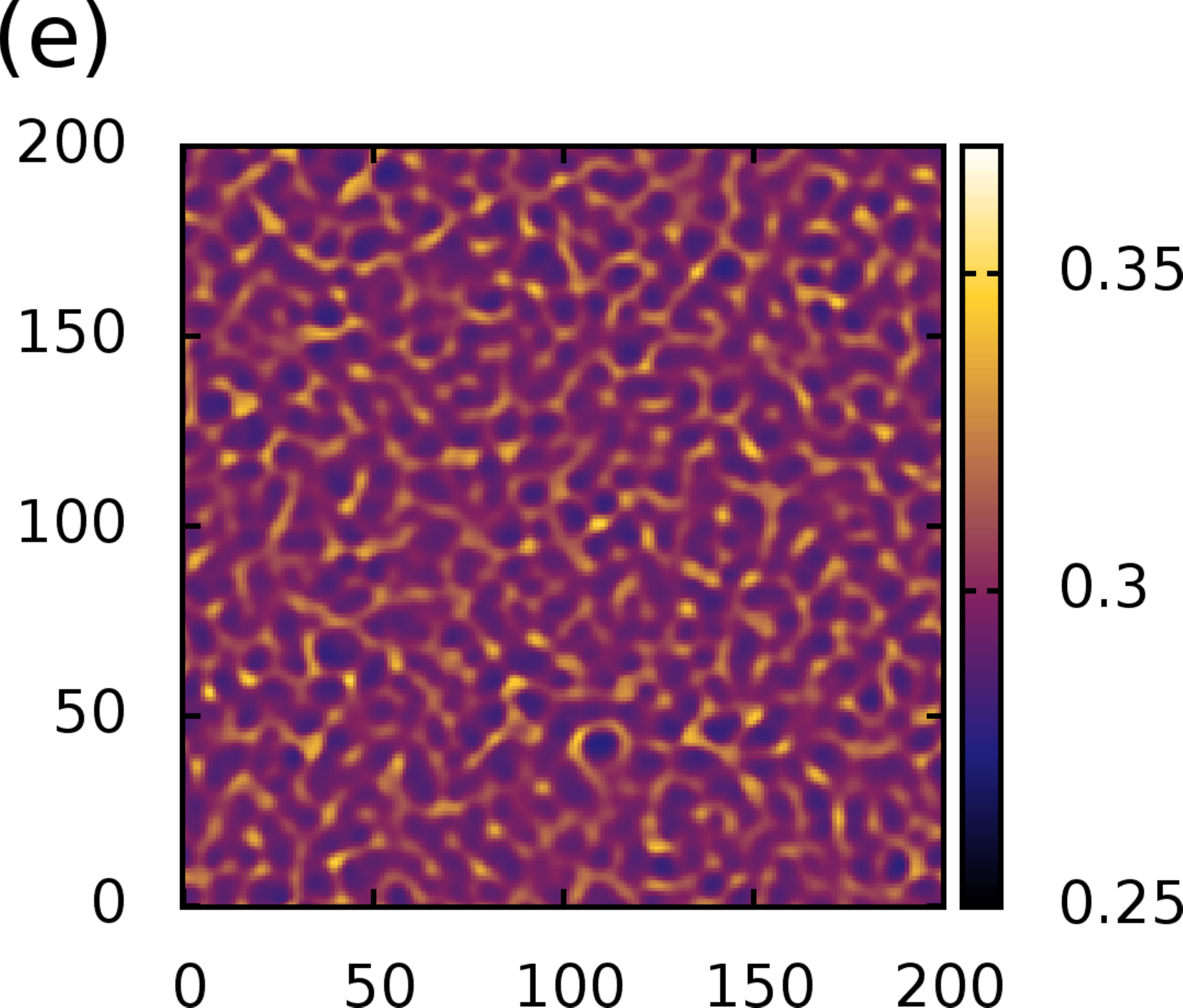} \hspace{1mm}
	\includegraphics[width=0.3\columnwidth]{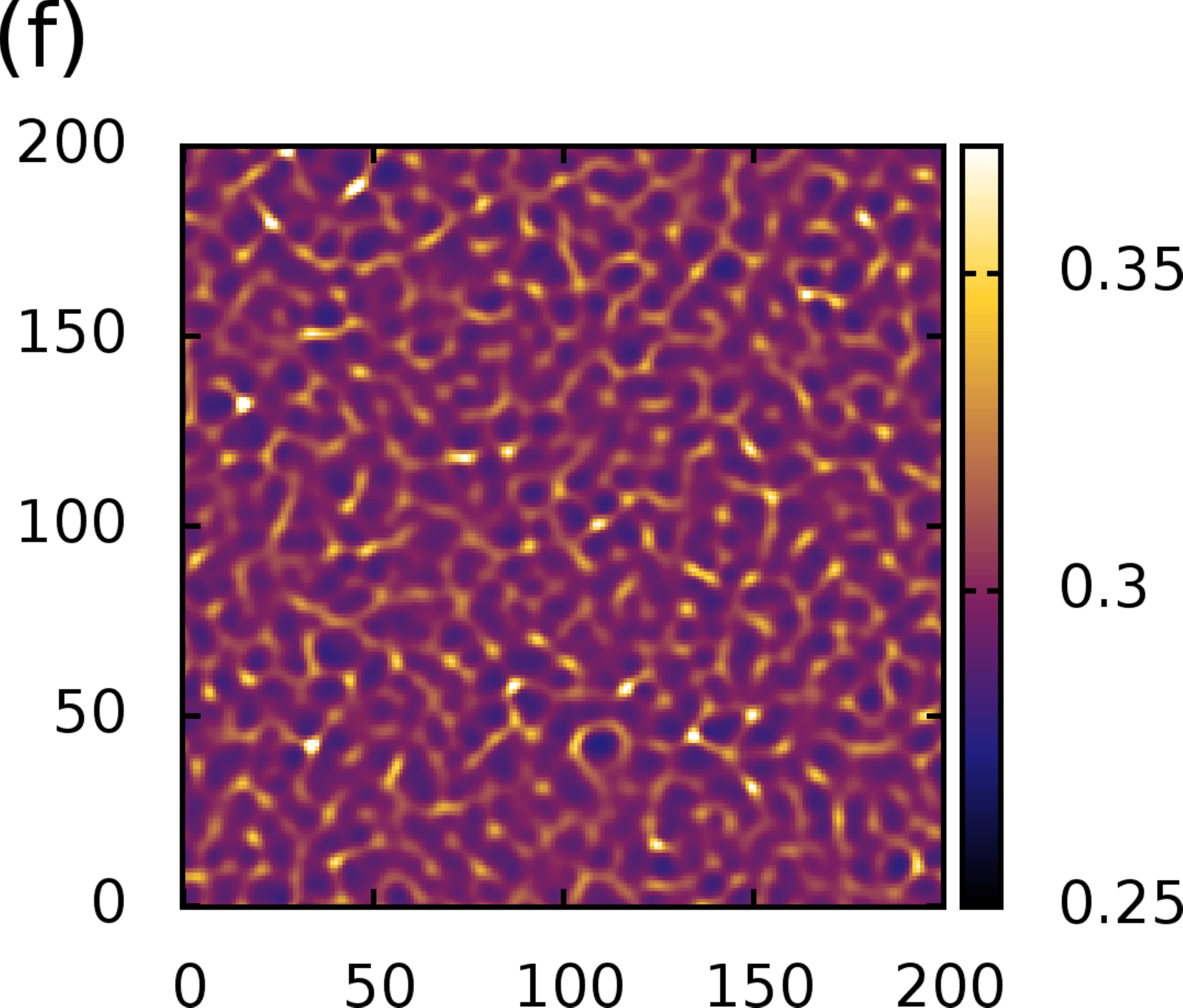}
	\caption{Density profiles displaying evaporation via spinodal 
	                decomposition.  The top row shows the liquid density 
	                profiles and the bottom row shows the nano-particle 
	                profiles at times $t/t_l=7$ (left), $t/t_l=8$ (centre) 
	                and $t/t_l=9 $ (right), where $t_l \equiv \frac{\beta}
	                {M_l^{nc}}$.  The system parameter values are:
		      $k_BT=1$, $\epsilon_l = 1.4$, $\epsilon_n =0.6$, 
		      $\epsilon_{nl} = 0.8$, $M_l^c=0$, $M_l^{nc}=1$, 
		      $\alpha=0.5$, $\beta \mu=-4.08$ and 
		      $\lambda = 0.005$.}
	\label{figSpinDe}
	\vspace{-0.5cm}
\end{center}
\end{figure}

The $c = 0$ spinodal curve (as calculated in the previous section, Fig.~\ref{figDispRelAPBP}) 
defines the limit of stability, i.e. inside this line the fluid becomes linearly unstable.  Thus, the fluid 
is unstable when $\beta \mu < -3.869$, where $\beta = 1/k_BT$.  The speed of the process increases 
with decreasing values of $\beta \mu$.  For very low values of $\beta \mu$ the evaporation process is 
so fast that we do not see any pattern formation in the nano-particles - the liquid evaporates too 
quickly for the nano-particles to diffuse.  For the parameters $\epsilon_l = 1.4$, $\epsilon_n = 0.6$, 
$\epsilon_{nl} = 0.8$, $\alpha = 0.5$, $M_l^c=0$ and $M_l^{nc}=1$ this occurs when 
$\beta \mu \lsim -4.2$.  Fig.~\ref{figSpinDe} shows the particular case when $\beta \mu = -4.08$.  
We see that the liquid behaves in a similar manner to that of a single component fluid by 
spontaneously dewetting everywhere.  Now the evaporation is slow enough for the nano-particles 
to move into areas with a high density of liquid during this evaporative process which creates a 
fine network structure [as shown in Fig.~\ref{figSpinDe}(e) and (f)].  However, this diffusion of the 
nano-particles is limited as it is still a much slower process than the evaporation of the 
liquid.  We observe that towards the end of the process 
small heaps of nano-particles are formed, where the density is significantly larger.  This effect is 
enhanced by the attraction between the nano particles ($\epsilon_n > 0$).  Increasing the value of $\mu$, 
we move from the linearly unstable (spinodal) region into the metastable region of the phase diagram 
(c.f. Fig.~\ref{figOneCompPh}).  Note however, that the actual values of $\beta \mu$ on the binodal 
and spinodal now differ from those in Fig.~\ref{figOneCompPh} because of the inclusion of 
nano-particles, c.f.~Fig.~\ref{figBinBin}.

\begin{figure}[b]
\begin{center}
	\includegraphics[width=0.3\columnwidth]{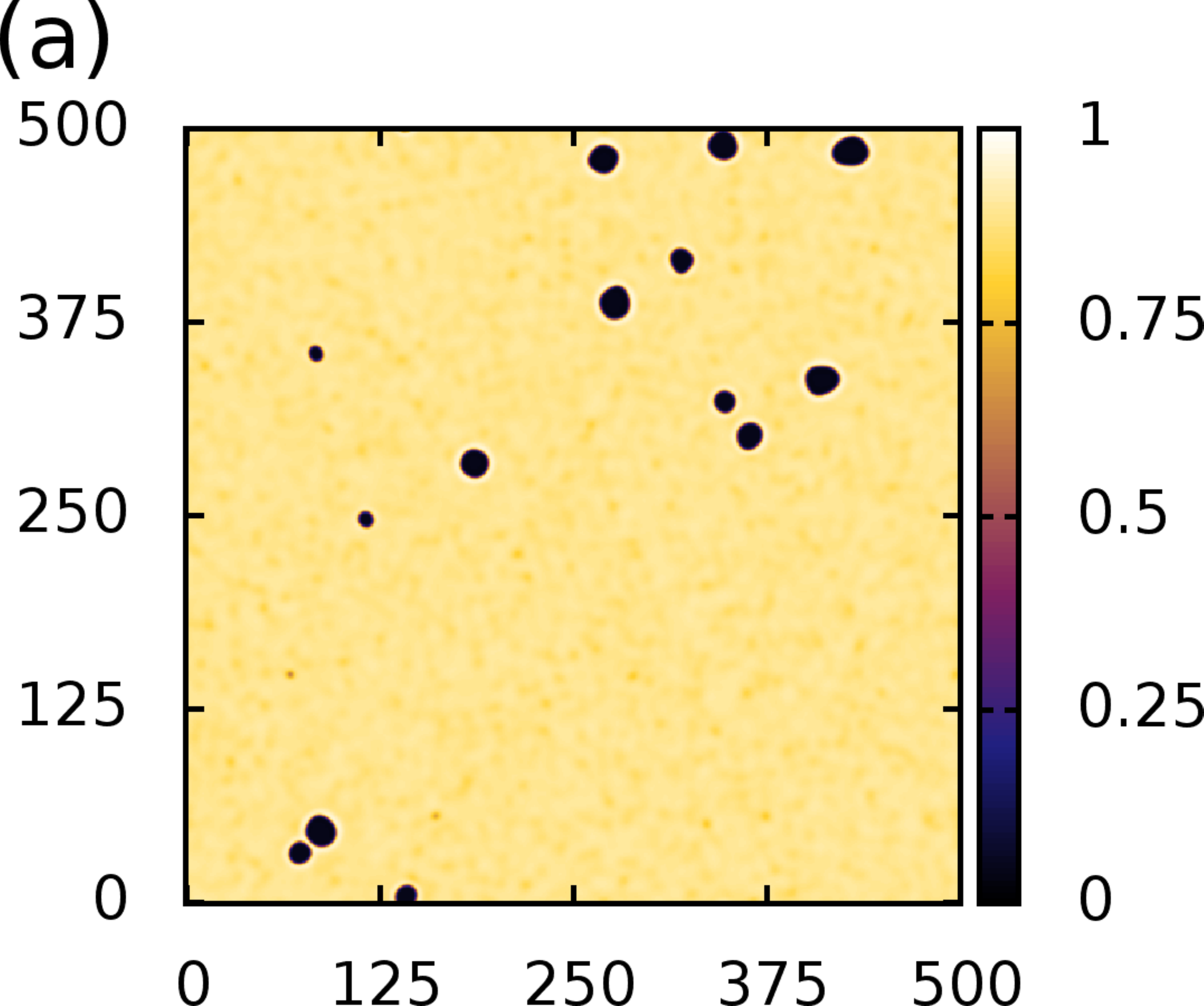} \hspace{1mm}
	\includegraphics[width=0.3\columnwidth]{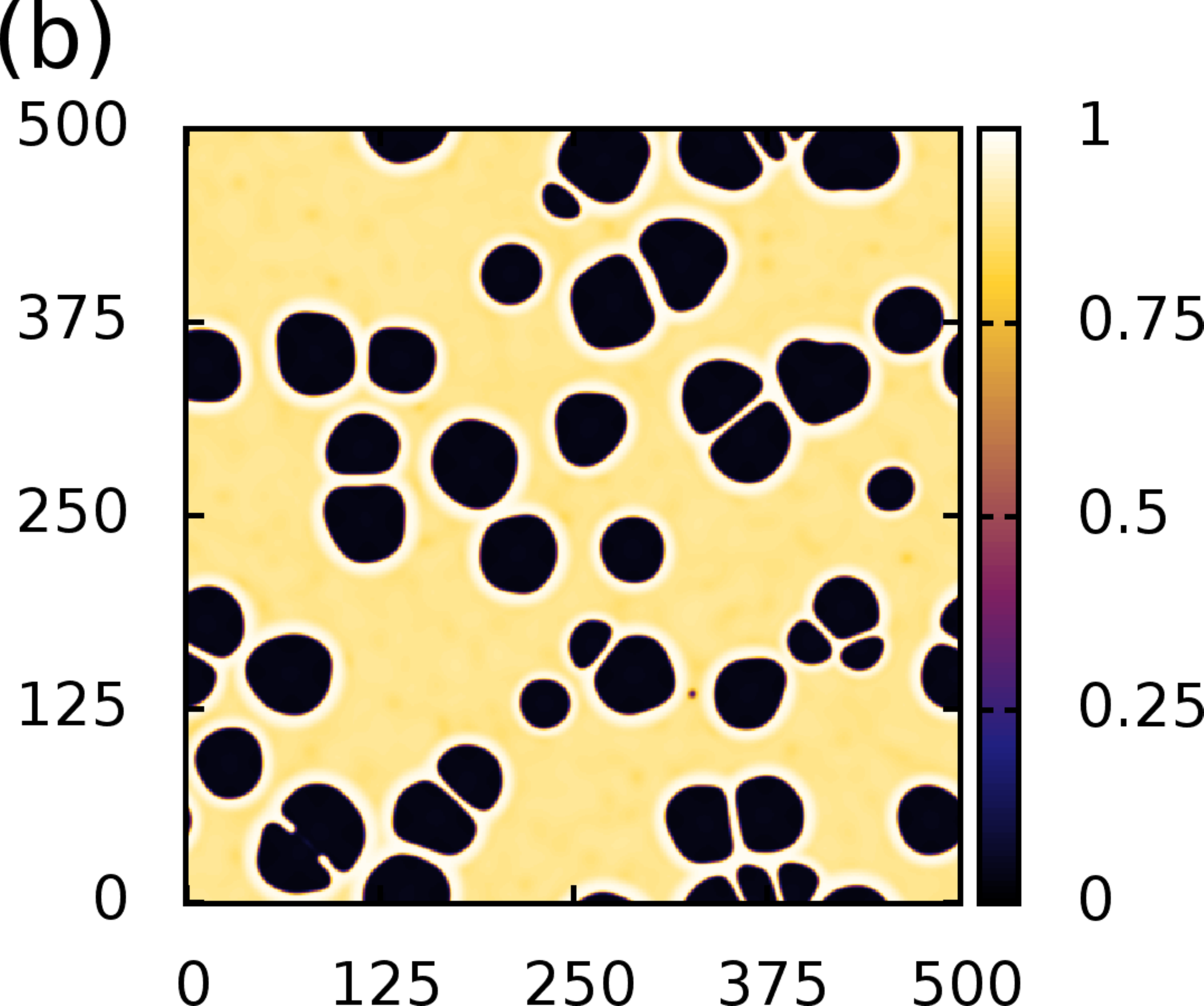} \hspace{1mm}
	\includegraphics[width=0.3\columnwidth]{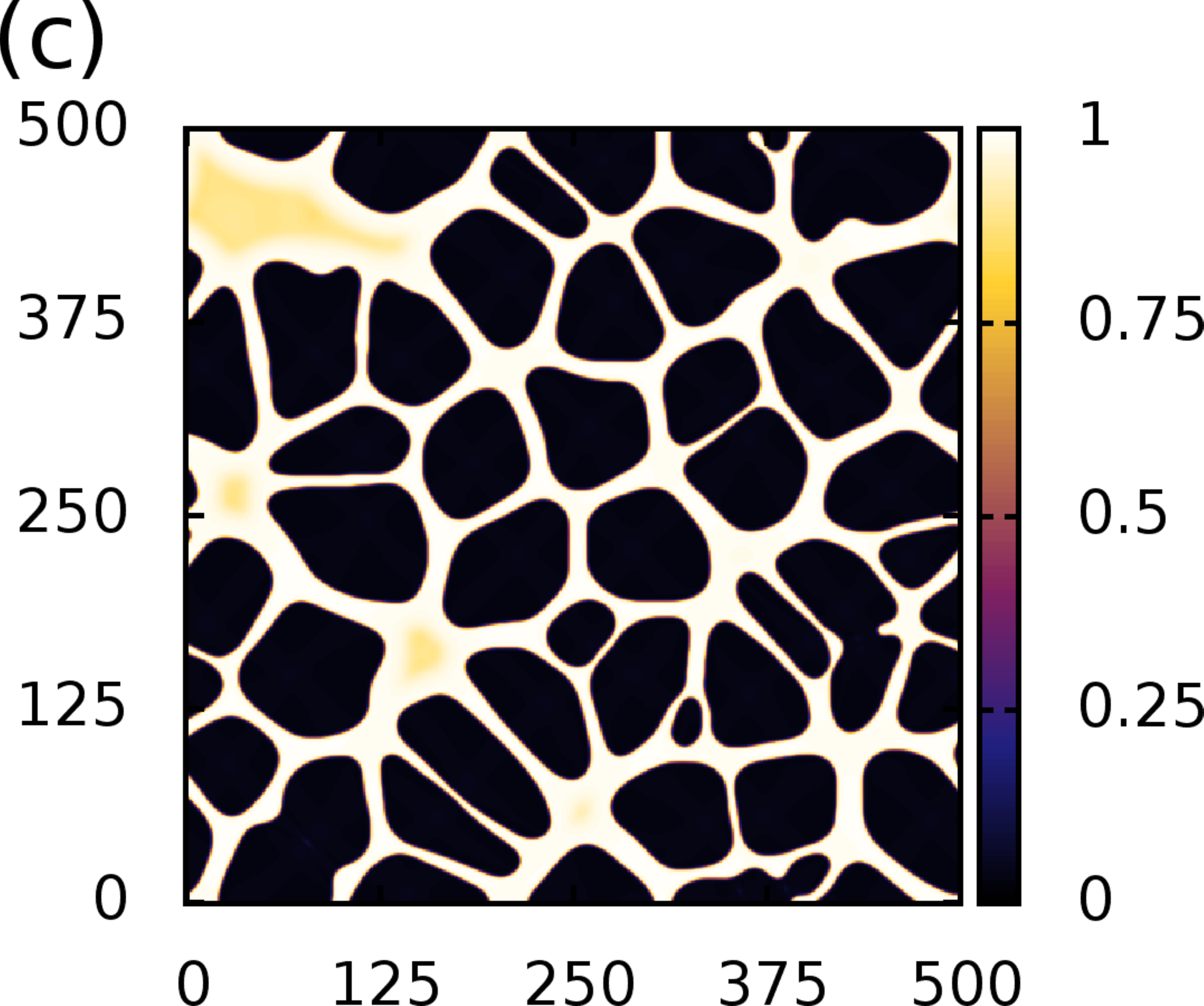} \\
	\includegraphics[width=0.3\columnwidth]{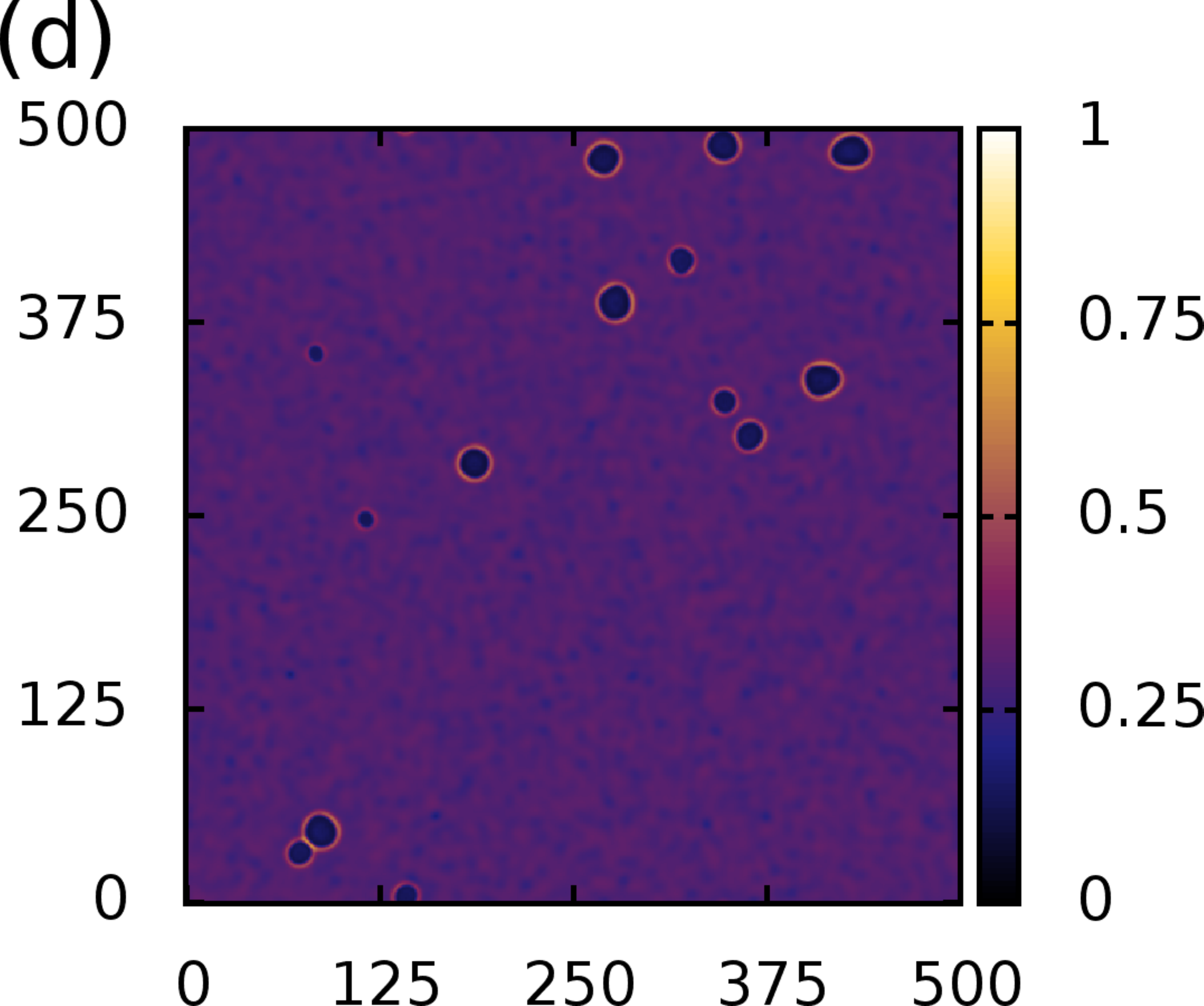} \hspace{1mm}
	\includegraphics[width=0.3\columnwidth]{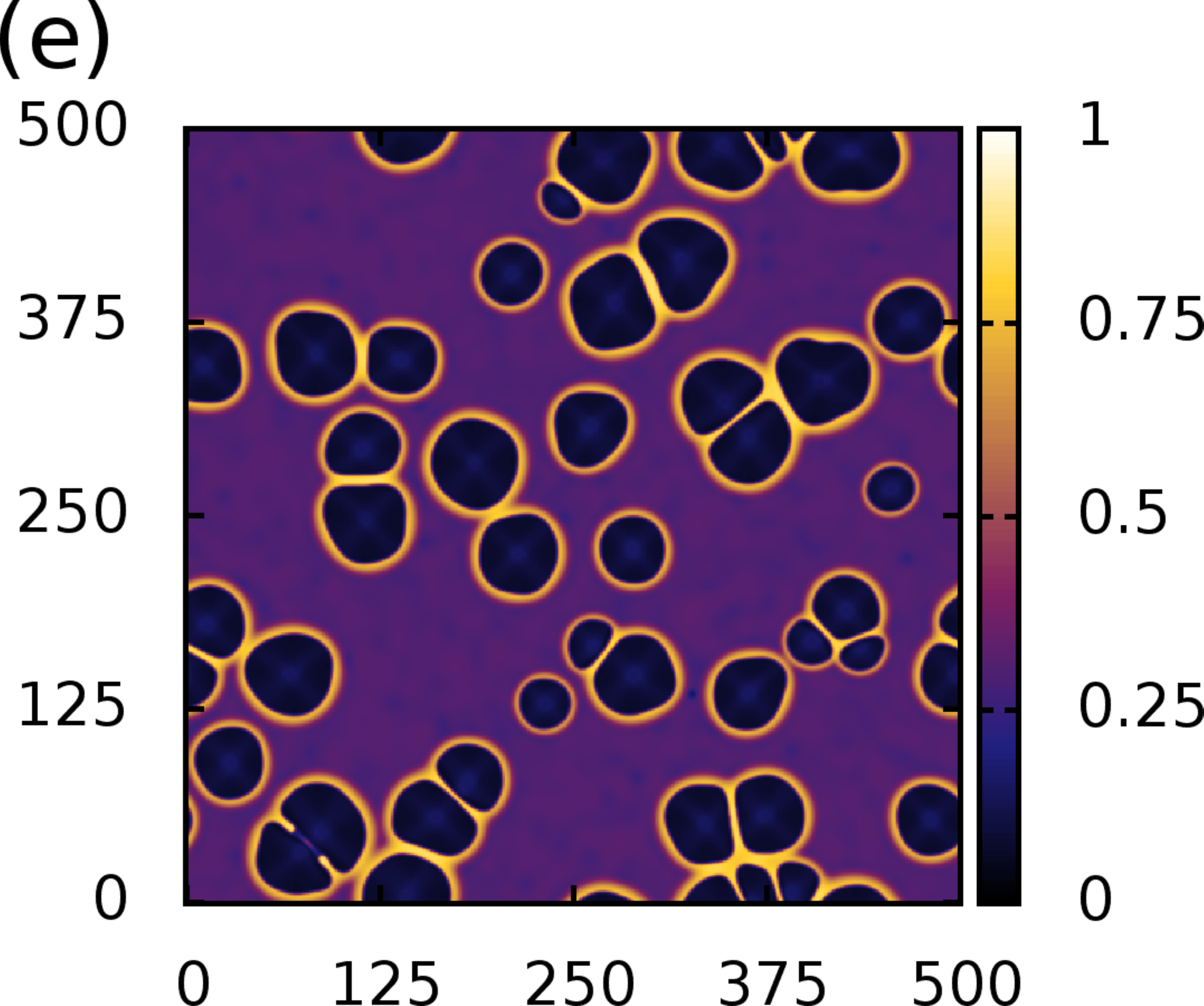} \hspace{1mm}
	\includegraphics[width=0.3\columnwidth]{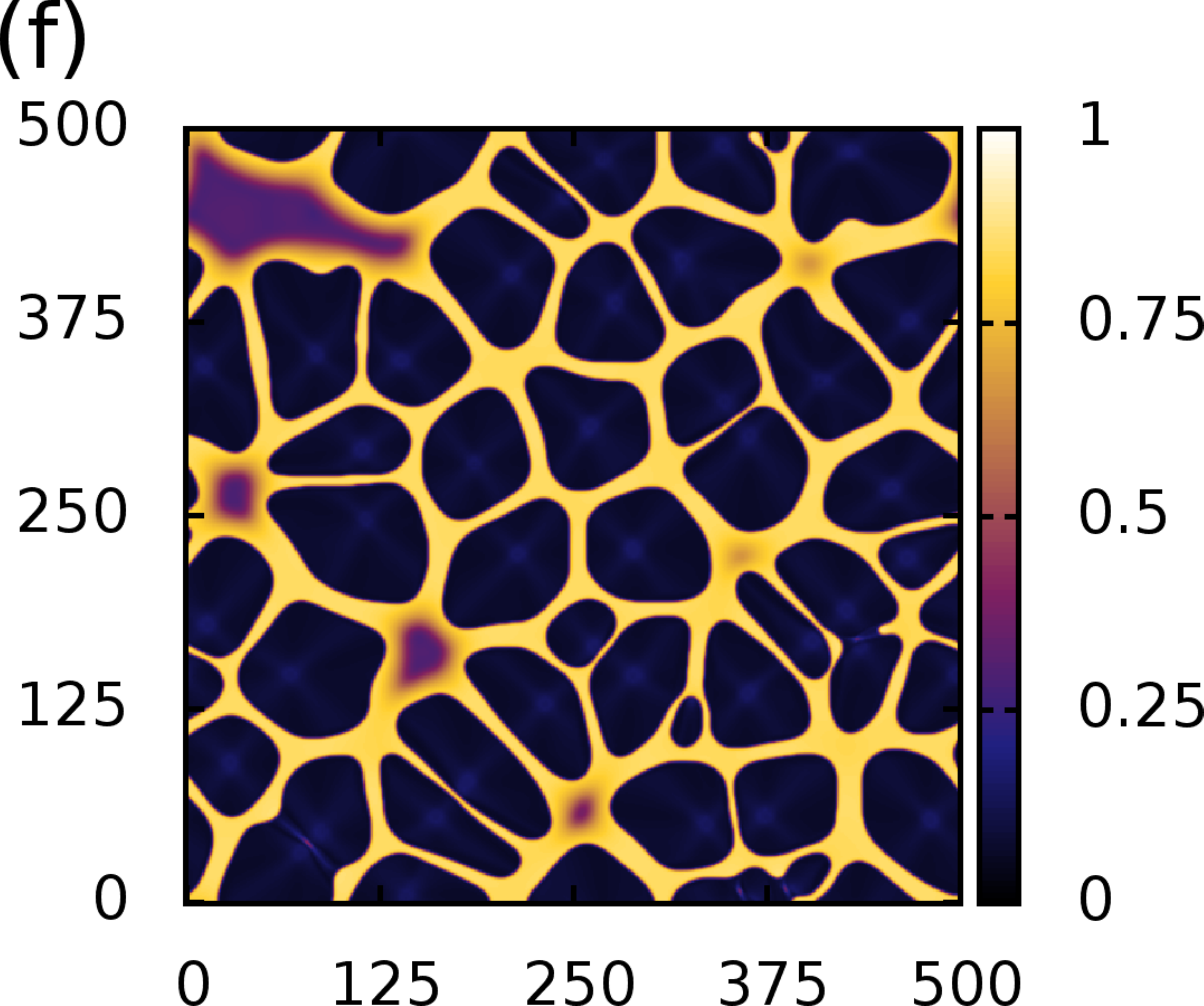}
	\caption{Density profiles displaying nucleation and growth of
		      holes which leads to the development of a network
		      pattern.  The top row shows the liquid density profiles 
		      and the bottom row shows the nano-particle profiles at 
		      times $t/t_l=30$ (left), $t/t_l=200$ (centre) and 
		      $t/t_l=800$ (right).  The system parameter values are: 
		      $k_BT=1$, $\epsilon_l = 1.4$, $\epsilon_n =0.6$, 
		      $\epsilon_{nl} = 0.8$, $M_l^c=0$, $M_l^{nc}=1$, 
		      $\alpha=0.5$, $\beta \mu=-3.86$ and $\lambda = 0.2$.}
	\label{figNuc}
	\vspace{-0.5cm}
\end{center}
\end{figure}

\begin{figure}[t]
\begin{center}
	\includegraphics[width=0.5\columnwidth]{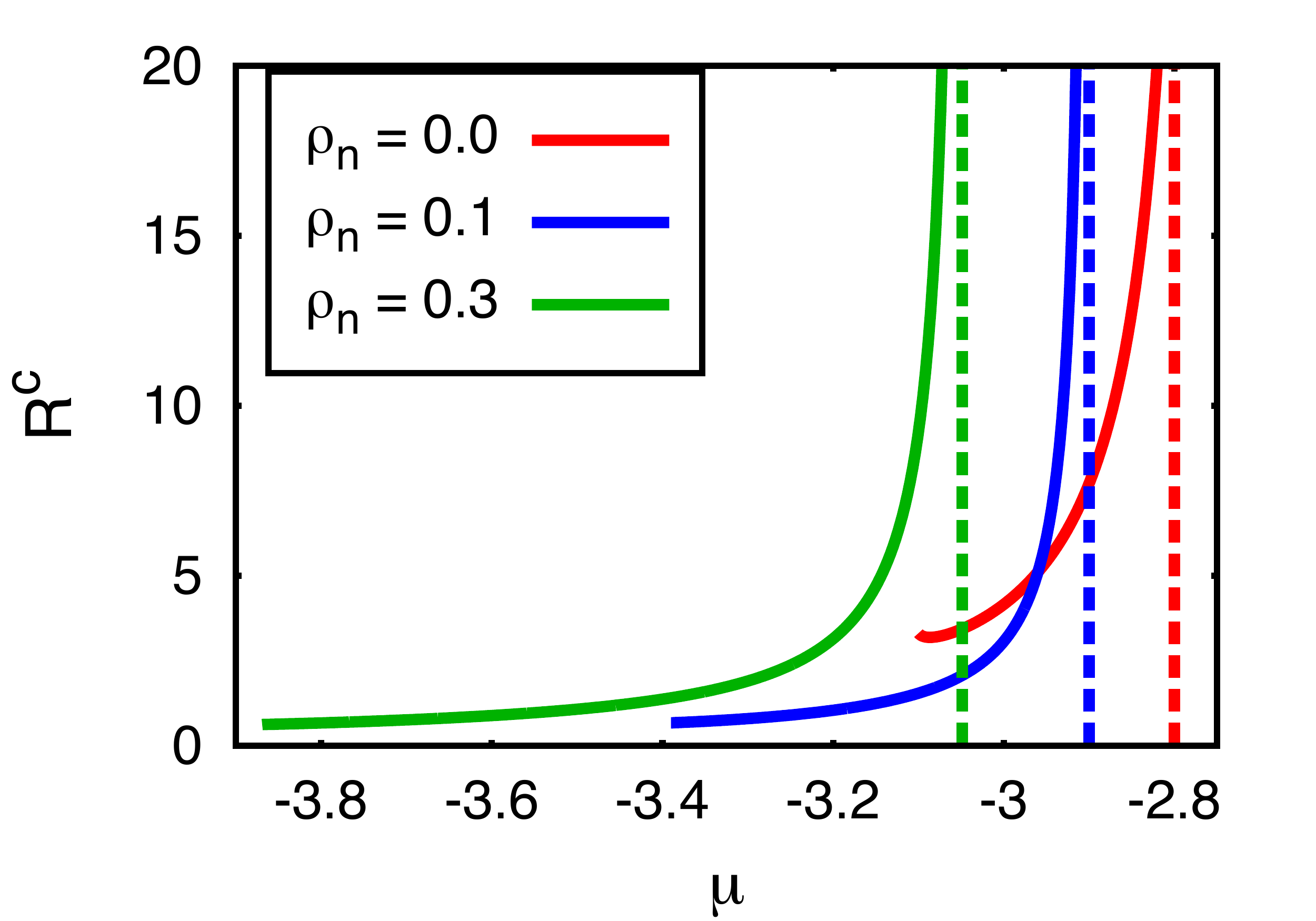}
	\caption{The solid lines display the critical hole radius $R^c$ in units of the 
		       lattice grid spacing $\sigma$ versus the chemical potential $\mu$ 
	                for different nano-particle densities $\rho_n$.  The lines start at the
	                lowest value of $\mu$ for which the system is linearly stable.  The 
	                dotted lines show the value of the chemical potential at coexistence
	                $\mu_\mathrm{coex}$,
	                which the curves approach asymptotically.  Here $k_BT = 1$, 
	                $\epsilon_l = 1.4$, $\epsilon_n = 0.6$ and $\epsilon_{nl} = 0.8.$}
		\label{figCritRad}
	 \end{center}
\end{figure} 

On increasing the chemical potential into the range $-3.869 < \beta \mu \lsim -3.8$, the liquid film 
becomes metastable but may still evaporate through the nucleation and growth of holes.  
Fig.~\ref{figNuc} shows the case when $\beta \mu = -3.86$.  The nucleation is caused by the 
random fluctuations in the density distributions (the initial density profiles are defined in a similar 
manner to the previous case: $\rho_l(x,t=0) = 1 - 10^{-6}$, $\rho_n(x,t=0) = \rho_n^{av} + 0.4(Y - 0.5)$.  
The amount of noise used and the free energy `barrier' for forming a hole determines the probability 
of a nucleation event occurring.  There is a critical hole radius $R^c$ which can be determined from 
the free-energy of the system.  If a hole is smaller than this critical radius then it will shrink and the 
liquid density will return to its bulk high density value in this region.  However, if the size of the hole 
is larger than this critical value then this hole will begin to grow.  We can apply classical nucleation 
theory to calculate an estimate for the critical hole radius $R^c$ by determining the change in free 
energy $\Delta F$ when a low density (thin film) circular `hole' with a radius of $R$ is inserted into the 
metastable liquid film.  The radius $R$ which corresponds to the maximum in $\Delta F$ is the 
critical hole radius $R^c$.  We approximate the change in free energy using the formula:
%
\begin{equation}
	\Delta F = \pi R^2 \Delta P + 2\pi R \gamma,
	\label{eqFreeEngGain}
\end{equation}
where $\Delta P$ is the pressure difference between the two phases (the hole and the
fluid film).  $\gamma$ is the interfacial tension (excess free energy) for creating
a straight interface between the two phases
at coexistence $\mu = \mu_\mathrm{coex}$.  It is important to note that the density
values at coexistence are different to the density values out of
coexistence.  The density of the thin film of fluid inside the hole is
such that its chemical potential is equal to that of the bulk film of
fluid surrounding the hole.  The critical hole radius $R^c$ is given
by the maximum of Eq.~\eqref{eqFreeEngGain} - i.e. when $\paDir{\Delta
  F}{R} = 0$.  Thus, the critical hole radius is $R^c =
-\frac{\gamma}{\Delta P}$.  Fig.~\ref{figCritRad} shows how $R^c$
depends on the chemical potential $\mu$ for different nano-particle
densities $\rho_n = 0$, $0.1$ and $0.3$.  This analysis only applies
to the metastable region of the phase diagram
(c.f.~Fig.~\ref{figOneCompPh}) and therefore the critical hole radius
$R^c$ curves are bounded on the left by the spinodal curve and on the
right by the binodal curve.  Note that classical nucleation theory
incorrectly predicts a finite value for $R^c$ at the spinodal, due to
the fact that the theory assumes a sharp interface as one approaches
the spinodal.  For further discussion on this see e.g.\  Ref.~\cite{OxEv88}.  The
probability for a hole to be nucleated by random thermal fluctuations
is proportional to $e^{-\beta \Delta F}$.  The curves in
Fig.~\ref{figCritRad} show that as we approach $\mu_\mathrm{coex}$ the
size of the critical hole increases.  Hence,
the probability of nucleation is greater nearer the limit of stability
(spinodal curve) and decreases greatly as we approach coexistence
(binodal curve).  We observe that increasing the density of the
nano-particles $\rho_n$ shifts the metastable region to lower chemical
potential values and increases the range of the metastable region.
This is due to the increase in the critical temperature associated
with the increase in the nano-particle density $\rho_n$, as previously
discussed.

In the case shown in Fig.~\ref{figNuc}, we are near the limit of stability where the critical radius of a 
hole is very small.  This results in many nucleation points where holes are formed and begin to grow.  
The nano-particles are picked up by these growing holes which creates a rim around each hole
with a high density of nano-particles in the rim.  The holes in the liquid film continue to grow until their 
rims meet, creating a random polygonal network pattern of nano-particles.  The liquid wets the surface of the 
nano-particles, which means the liquid remains on the surface in areas with a high density of 
nano-particles.  This is due to the positive interaction energy between the liquid and the 
nano-particles ($\epsilon_{nl}>0$).

\begin{figure}[b]
\begin{center}
	\includegraphics[width=0.3\columnwidth]{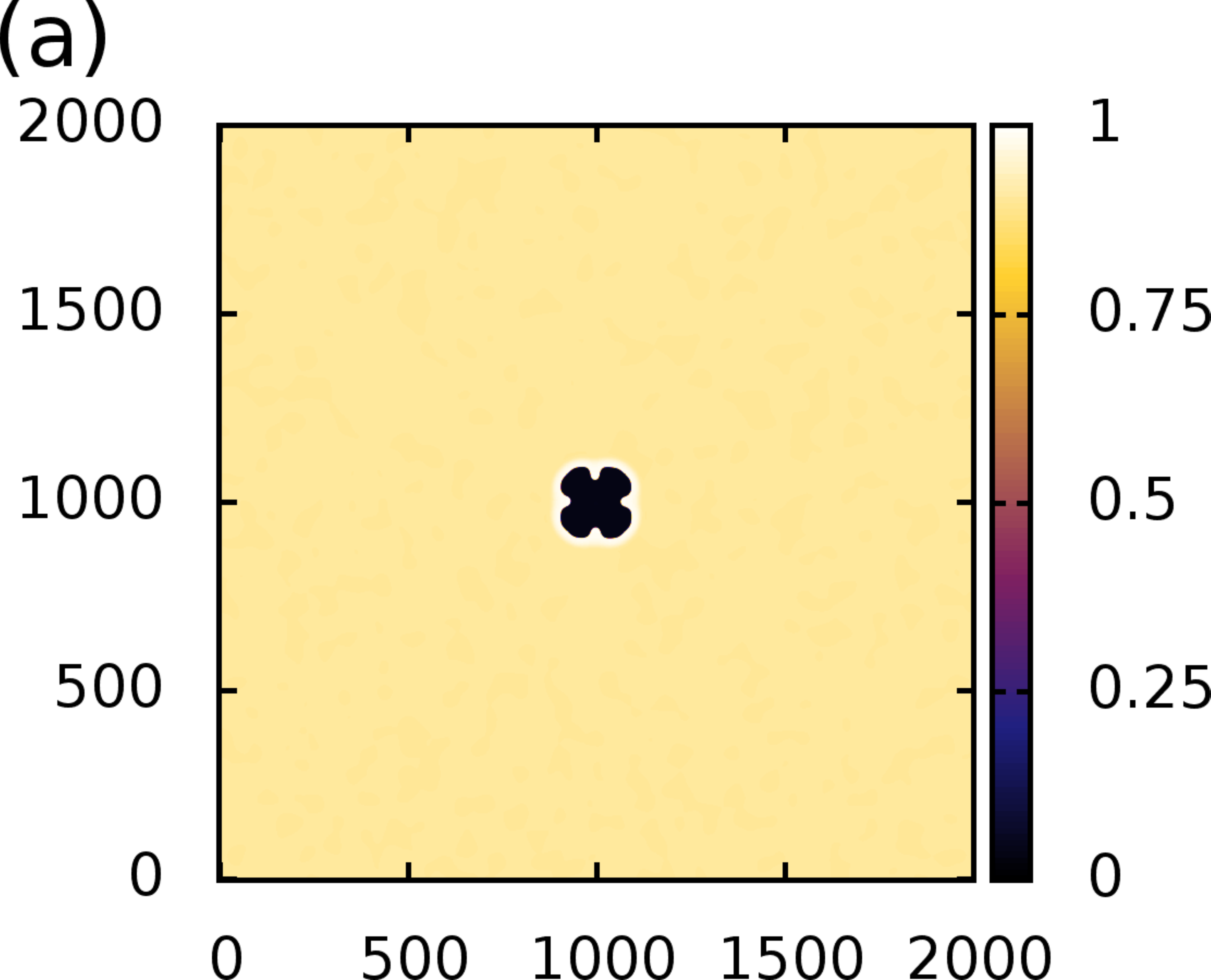} \hspace{1mm}
	\includegraphics[width=0.3\columnwidth]{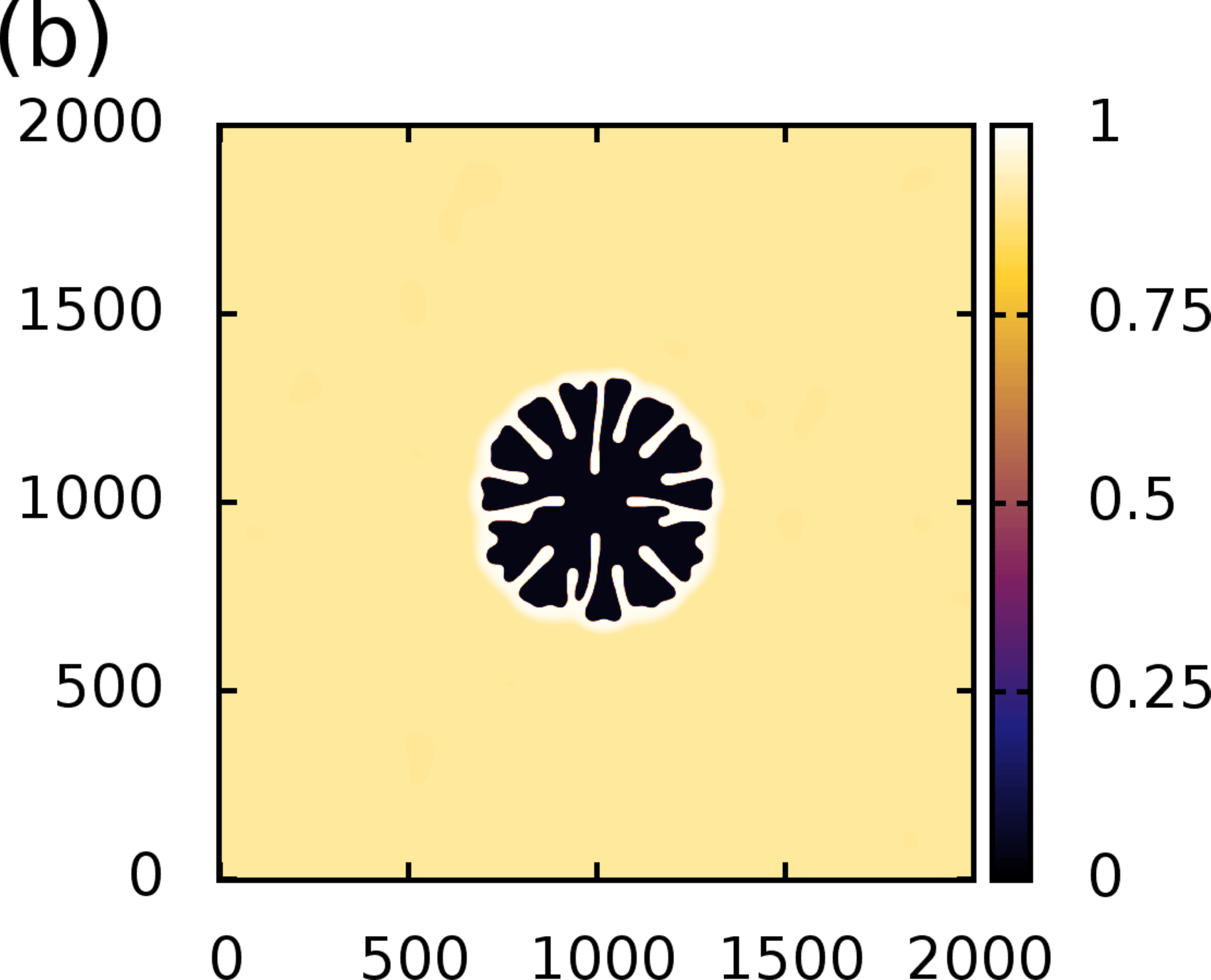} \hspace{1mm}
	\includegraphics[width=0.3\columnwidth]{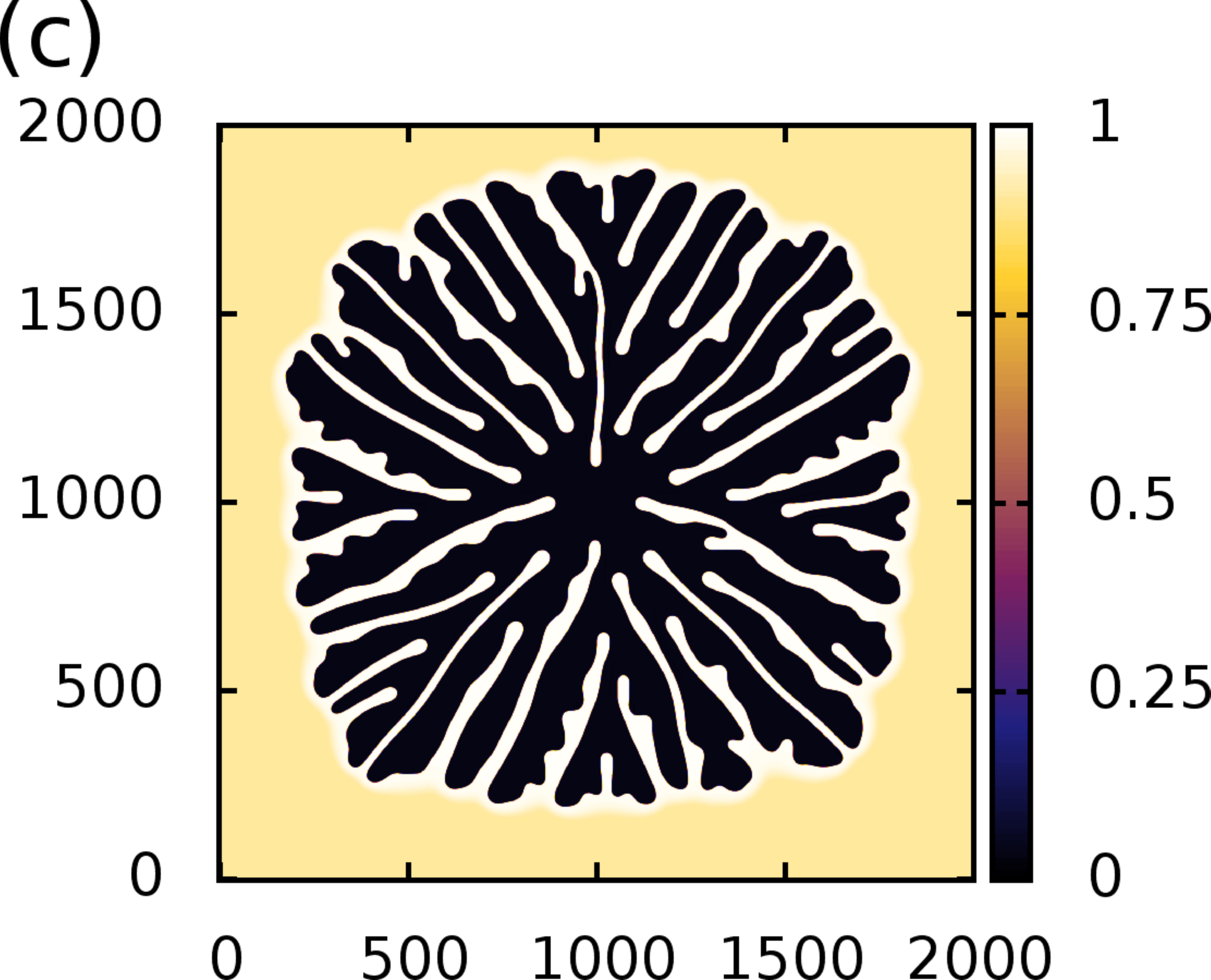} \\
	\includegraphics[width=0.3\columnwidth]{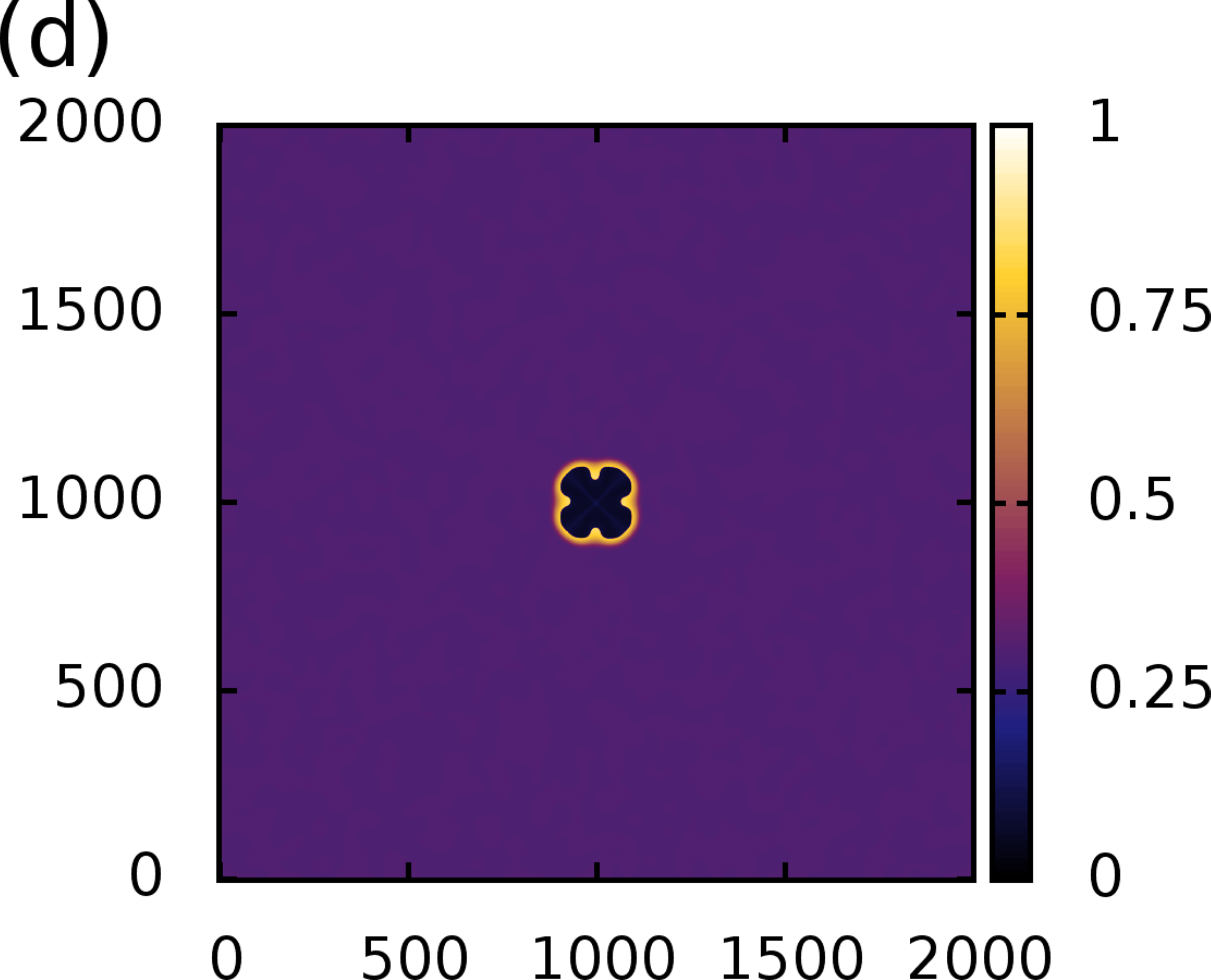} \hspace{1mm}
	\includegraphics[width=0.3\columnwidth]{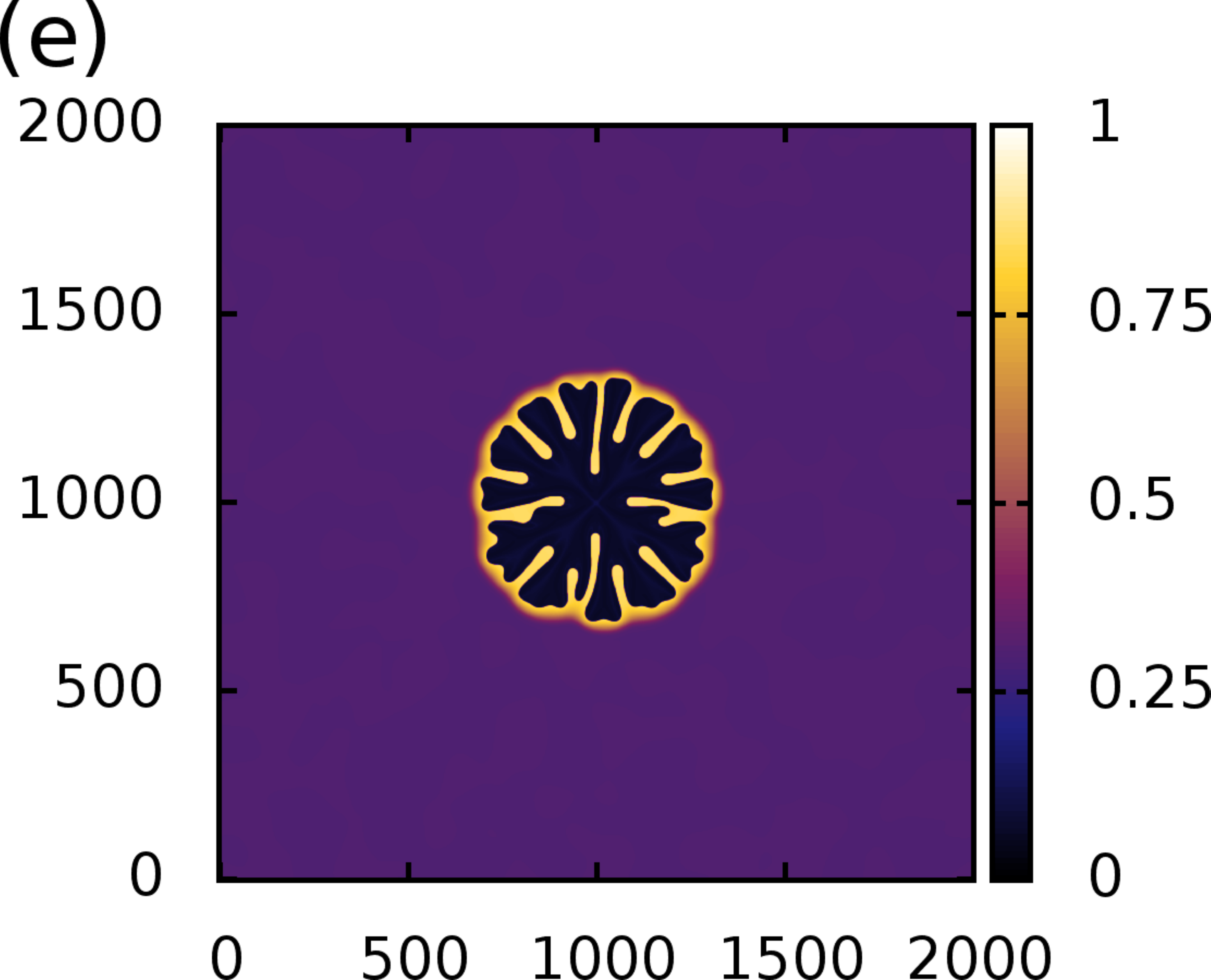} \hspace{1mm}
	\includegraphics[width=0.3\columnwidth]{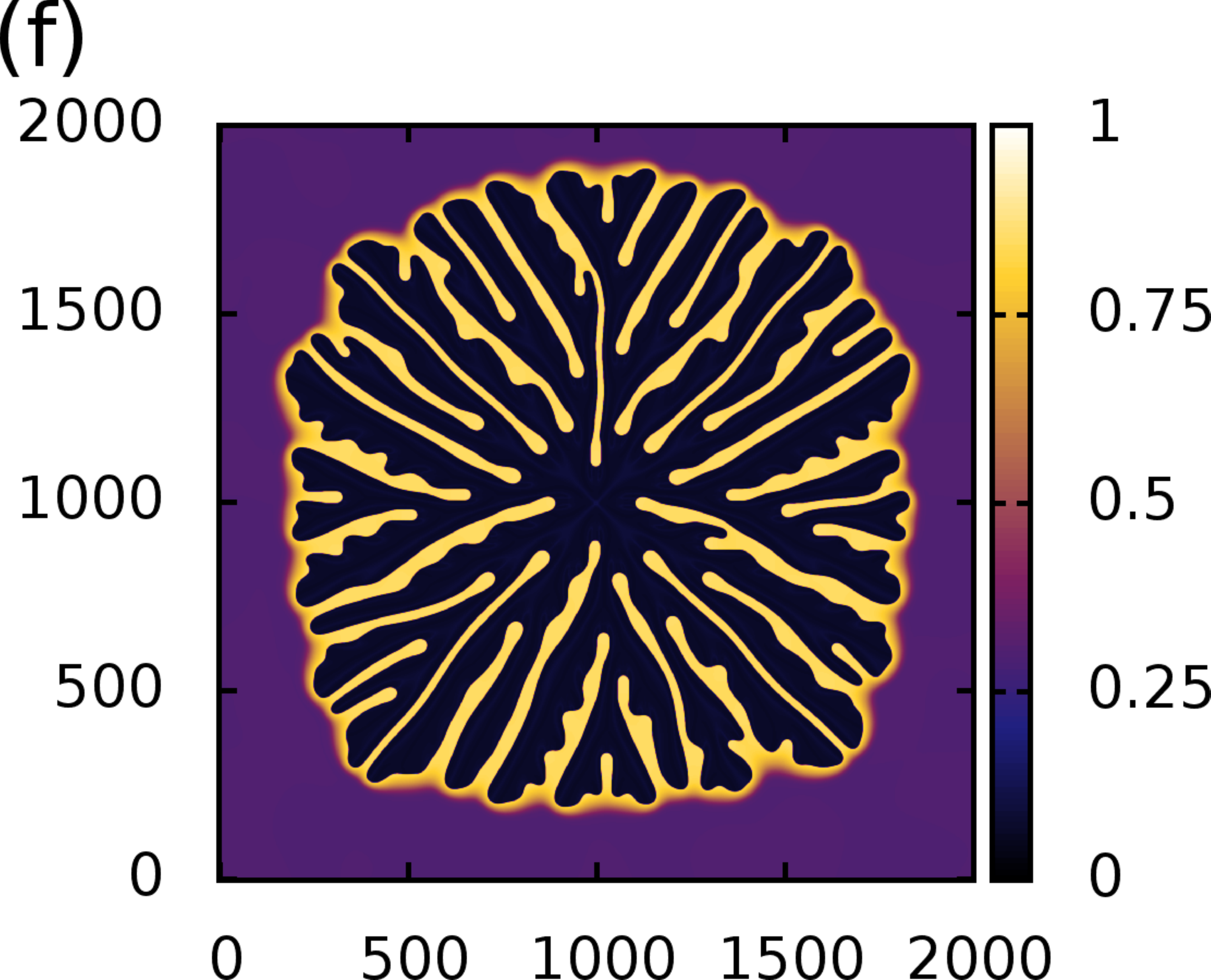}	
	\caption{Density profiles displaying the growth of an 
		      artificially nucleated point which develops branched 
		      structures.  The top row shows the liquid density profiles  
		      and the bottom row shows the nano-particle profiles at  		
		      times $t/t_l=2000$ (left), $t/t_l=10000$ (centre) and 
		      $t/t_l=30000$ (right).  The system parameter values are: 
		      $k_BT=1$, $\epsilon_l = 1.4$, $\epsilon_n =0.6$, 
		      $\epsilon_{nl} = 0.8$, $M_l^c=0$, $M_l^{nc}=1$, 
		      $\alpha=0.5$, $\beta \mu=-3.8$ and $\lambda = 0.1$.  Note
		      that instead of using Eq.\ \eqref{eqLap}, here the
		      Laplacian term is approximated using: $\nabla^2 \rho = 
		      \frac{1}{6(\Delta x)^2} \left( \sum 4 \rho^{NN} + \sum \rho^{NNN} 
		      - 20 \rho \right)$.}
 	\label{figFingering}
	\vspace{-0.5cm}
\end{center}
\end{figure}

If we increase the chemical potential further into the metastable
range $-3.8 \lsim \beta \mu < \mu_\mathrm{coex}$, 
then the probability of a hole being nucleated becomes much smaller.
In an experiment, in this parameter range, all
holes that are formed are normally nucleated at defects or impurities of
the surface (heterogeneous nucleation).  Any interfaces between a high
density liquid phase and a low liquid density phase will recede as the
liquid evaporates.  The velocity of the receding front depends on the
value of $\beta \mu$.  For small values of $\beta \mu$ we have a fast
front, but the speed of the front reduces as $\mu$ approaches
$\mu_\mathrm{coex}$. If we choose $\mu = \mu_\mathrm{coex}$ then any
straight front remains stationary.  On a completely structureless
substrate, one would usually expect such an interface to recede
homogeneously; this is certainly the case for the pure liquid, when
$\rho_n=0$.  However, in our system when $\rho_n > 0$ we see the
formation of fingers as the front recedes, due to the presence of the
nano-particles.  Fig.~\ref{figFingering} shows a case when $\beta \mu=
-3.8$.  The initial density profiles in this situation differ slightly
from the previous cases.  Here we create an artificial nucleation
point by setting the density of the liquid and the nano-particles to
$\rho_l = \rho_n = 10^{-6}$ in a central $2 \sigma \times 2 \sigma$
region.  Without this seed nucleus the initial noise on the density
profiles slowly decays and the densities of the two species return to
their (metastable) equilibrium values.  The liquid surrounding this
nucleation point slowly recedes creating a circular dewetting front.
As the front recedes, it begins to collect the nano-particles, as was
also observed for the case in Fig.~\ref{figNuc}.  However, here the
growing hole does not meet any other holes and there is time for an
instability to develop at the front which causes the liquid to
evaporate faster in some regions and slower in others creating a
`wavy' front, as seen in Fig.~\ref{figFingering}(a) and
Fig.~\ref{figFingering}(d).  The `bumps' at the front then appear to
stop moving while the rest of the front continues to recede. As
the front recedes and the hole circumference increases, more fingers
develop, leaving a branched `fingered' nano-particle structure
behind -- Figs.~\ref{figFingering}(e) and (f). The time scale for this dewetting process is
rather long and so we also observe some long-time coarsening effects
on the finger structures.

Recall that one of the goals of our work is to develop an
understanding of how the different self-organised structures of
nano-particles observed in the experiments
\cite{MTB02,MBM04,MBPA07,PVSM08} are formed.  Distinct observed
structures are a) network structures and b) branched structures.
Results from our model have shown how two different types of network
structures can develop: i) a fine network structure created by
a spinodal evaporation process (Fig.~\ref{figSpinDe}) in which
the nano-particle density varies over a fairly small range $0.27 \lsim
\rho_n \lsim 0.45$, ii) a large well defined network structure created
by the nucleation and growth of holes in the liquid
(Fig.~\ref{figNuc}), in which the nano-particle density varies over a
large range $0.05 \lsim \rho_n \lsim 0.9$.  Our model also shows how
instabilities at the evaporative dewetting front can create branched
structures for certain parameter values (Fig.~\ref{figFingering}).
Note that there is also evidence of early stages of the fingering
instability in the nucleation case shown in Fig.~\ref{figNuc}. There
one can observe that small bumps begin to develop in the edges of some
of the larger holes.  We now consider the formation of the branched
structures in more detail. In particular, we investigate the
dependence of these `fingered' structures on the parameters of the
model.

\subsection{Influence of mobilities on the fingering}
\label{secNumFing}

\begin{figure}[tbh]
	\begin{minipage}[h]{0.5\linewidth}
		\includegraphics[width=0.46\linewidth]{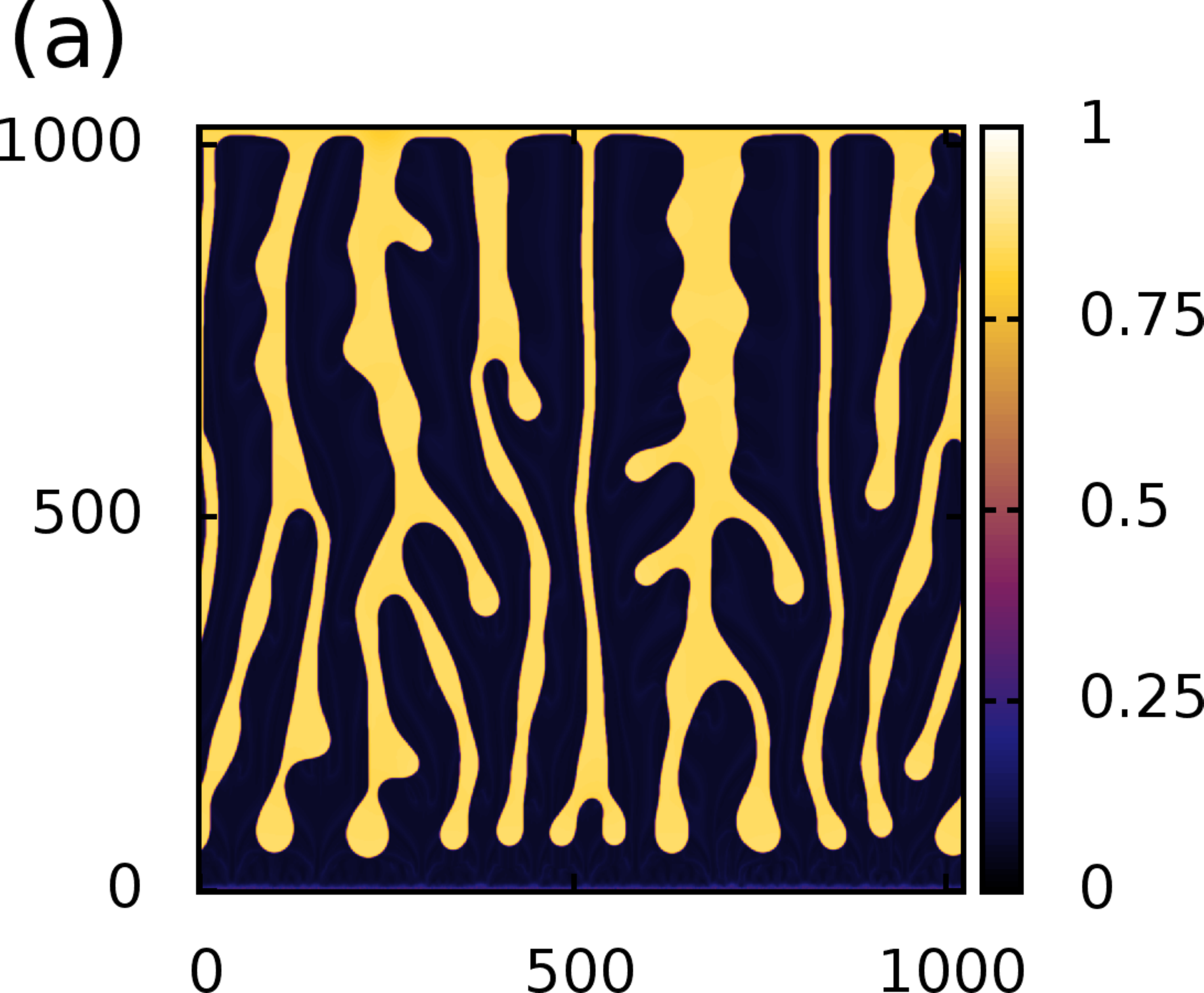}
		\hspace{2mm}
		\includegraphics[width=0.46\linewidth]{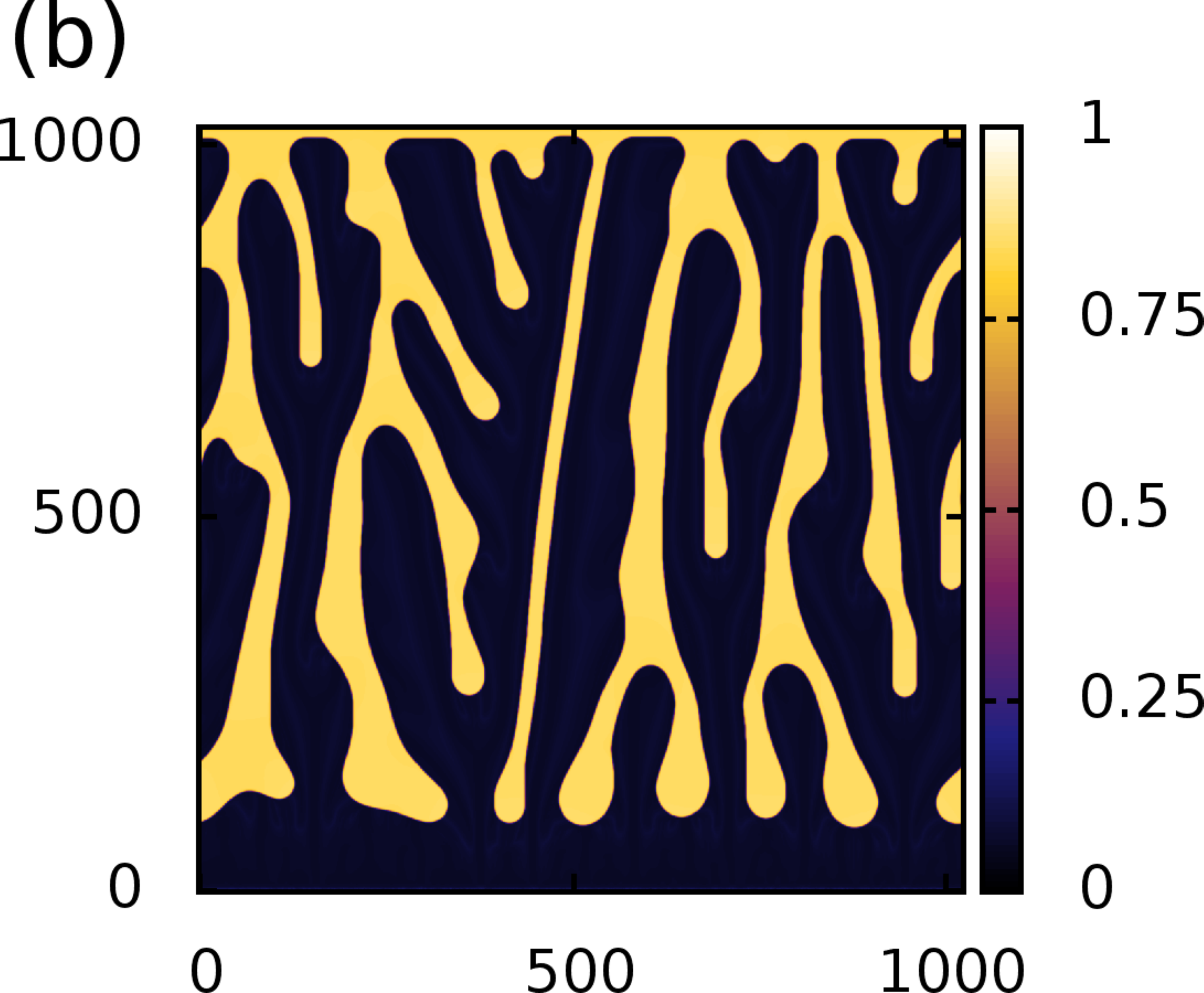}\\
		\includegraphics[width=0.46\linewidth]{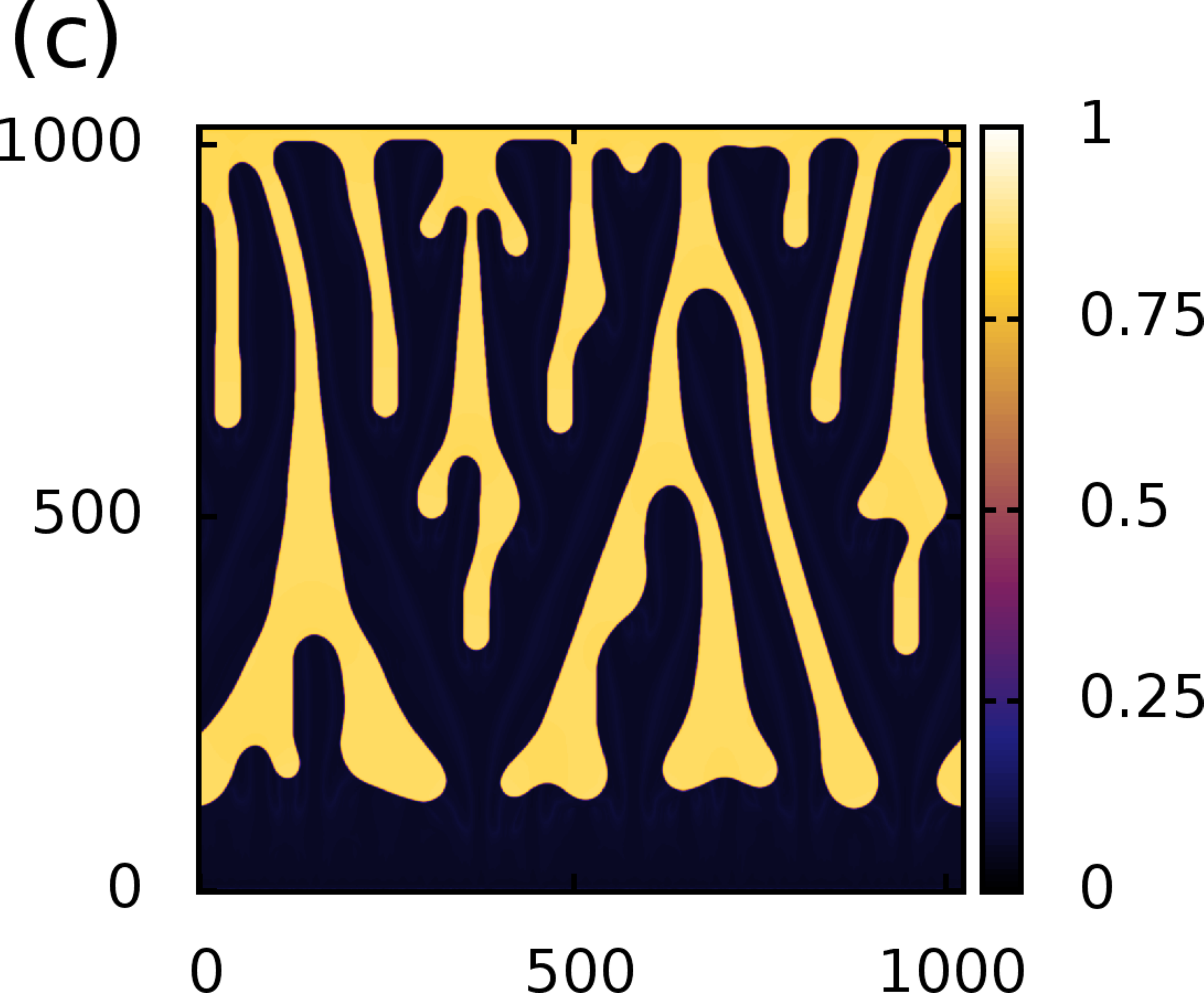}
		\hspace{2mm}
		\includegraphics[width=0.46\linewidth]{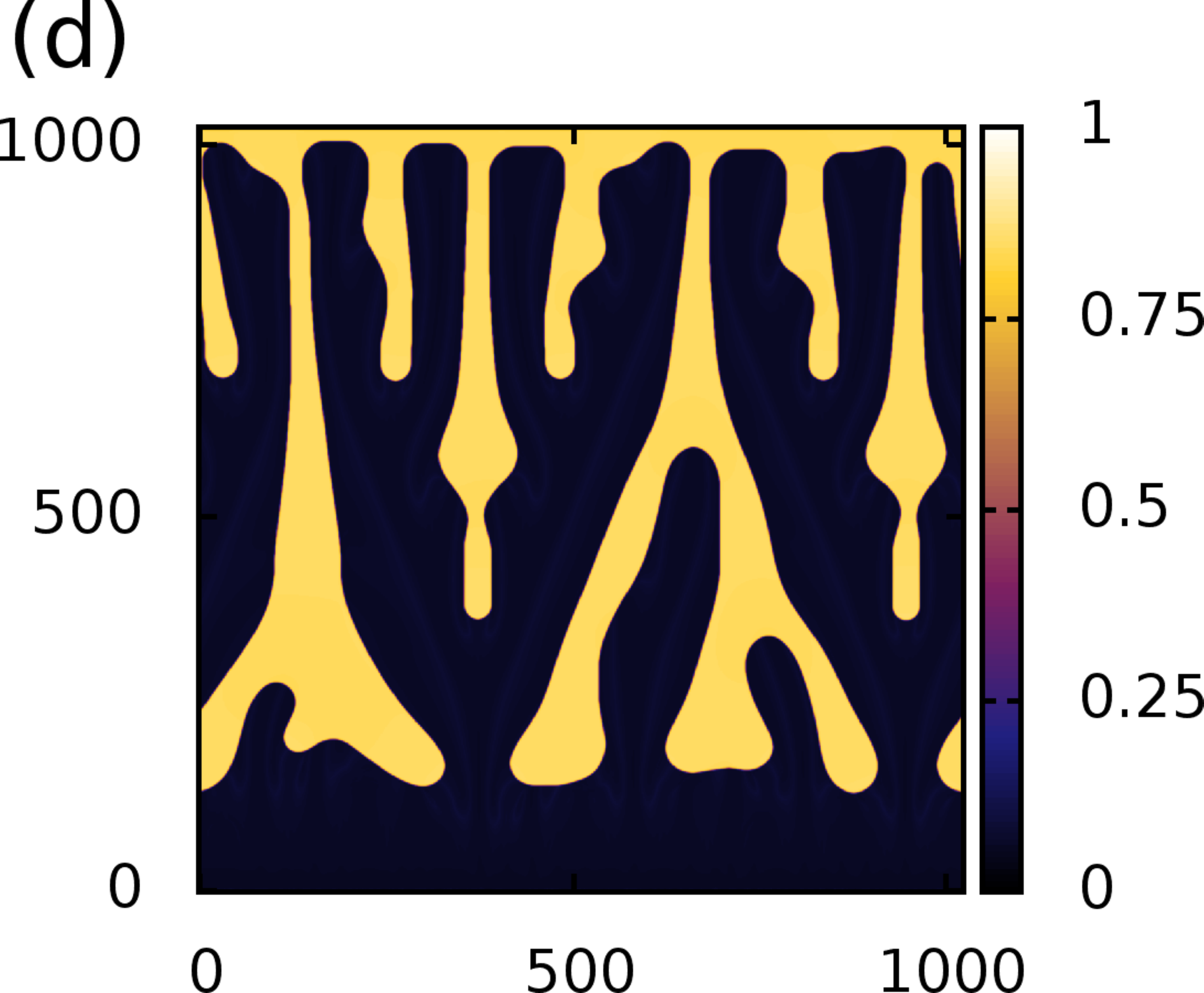}
	\end{minipage}
	\begin{minipage}[h]{0.49\linewidth}
		\includegraphics[width=\linewidth]{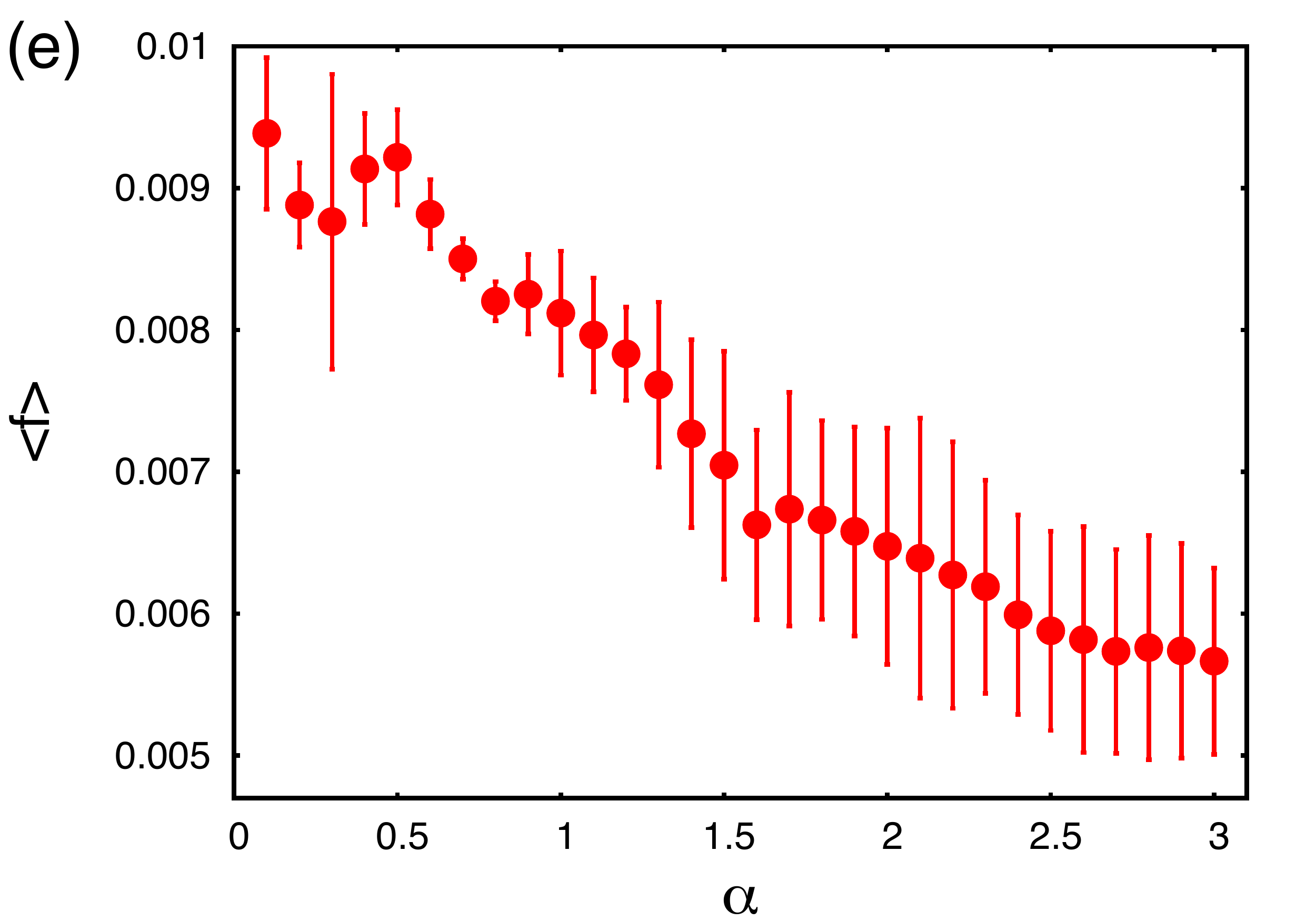}
	\end{minipage}
	\caption{Nano-particle density profiles for calculations with (a) 
		      $\alpha = 0.1$, (b) $\alpha = 1$, (c) $\alpha = 2$ and
		      (d) $\alpha = 3$.  In (e) we display a plot showing the 
		      dependence of the number of fingers on the parameter 
		      value $\alpha$.  The parameter values are: $k_BT=1$, 
		      $\epsilon_l = 1.4$, $\epsilon_n =0.6$, 
		      $\epsilon_{nl} = 0.8$, $M_l^c=0$, $M_l^{nc}=1$, 
		      $\beta\mu=-3.8$, $\Delta x = 1$ and $\lambda = 0.1$.}
	\label{figVaryingAlpha}
\end{figure}

To make a detailed investigation of the branched finger structures it is important to maximise the 
distance a front can recede.  This allows us to obtain better statistics which is important due to the 
fact that solving the DDFT in the fingering regime on a large grid can be time consuming.  To 
achieve this objective we create a straight dewetting front along the bottom edge of the system 
(i.e.~we set the two density values to $\rho_l = \rho_n = 10^{-6}$ for the first five horizontal lines).  
We also set no-flux boundary conditions at the top and the bottom, to prevent a dewetting front 
forming at the top.  The periodic boundary conditions on the left and the right side of the system 
domain remain.  This set-up also allows for easier analysis of the branched structures since we 
begin with an initially straight front.  We define a measure for the average number of fingers 
$\langle f \rangle$ to be the average number of branched structures per unit length in the final 
density profile, after the dewetting front has reached the top of the system.  To calculate this 
quantity we implemented an algorithm which counts the number of transitions between a high 
density of nano-particles and a low density of nano-particles on each horizontal line of the system.  
We then determine the number of fingers on a given horizontal line by dividing this value by two.  
We set a minimum and maximum line for a given set of final density profiles and calculate the 
average number of fingers between these two lines.  This value is then divided by the size of the 
system to give a value that is independent of the system size.  We have investigated how this 
measure is affected by the different parameters of the system.

We begin by discussing the effect that varying the parameter $\alpha$
has on $\langle f \rangle$ (recall that $\alpha$ determines the
mobility of the nano-particles in the liquid film).  We use the same parameter values as above, with $\beta \mu=-3.8$ and 
$\alpha$ varying between $0.1$ and $3$.  Fig.~\ref{figVaryingAlpha} shows final nano-particle 
density profiles for several values of  $\alpha$ and  also a plot of $\langle f \rangle$ versus $\alpha$, which is calculated 
from the average of five runs.  We see that the value of $\alpha$ has a significant influence on the 
average finger number.  Increasing the mobility of the nano-particles results in fewer fingers being 
developed, which is in (qualitative) agreement with the experiments
\cite{PVSM08} and the KMC model results \cite{VTPV08}.  The 
mobility of the nano-particles directly influences the speed of the receding front, so evaporation is 
much slower when $\alpha$ is small.  When $\alpha$ is very small ($\alpha \lsim 0.002$) we find 
that the two density fields become practically decoupled, with the liquid evaporating at high speed 
leaving the nano-particles behind as a homogeneous film of the initial density $\rho_n^{av}$.

\begin{figure}[tbh]
	\begin{minipage}[h]{0.5\linewidth}
		\includegraphics[width=0.46\linewidth]{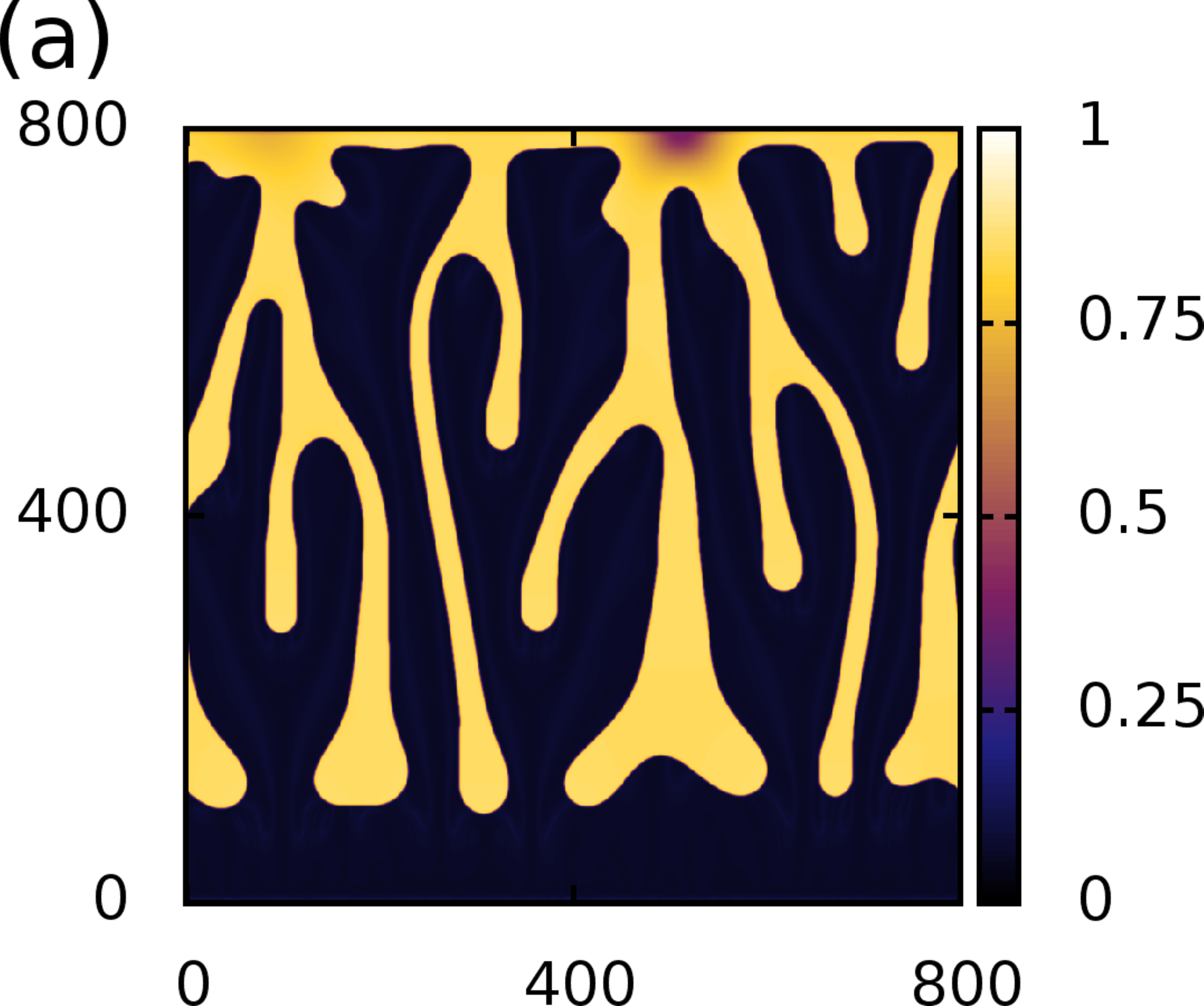}
		\hspace{2mm}
		\includegraphics[width=0.46\linewidth]{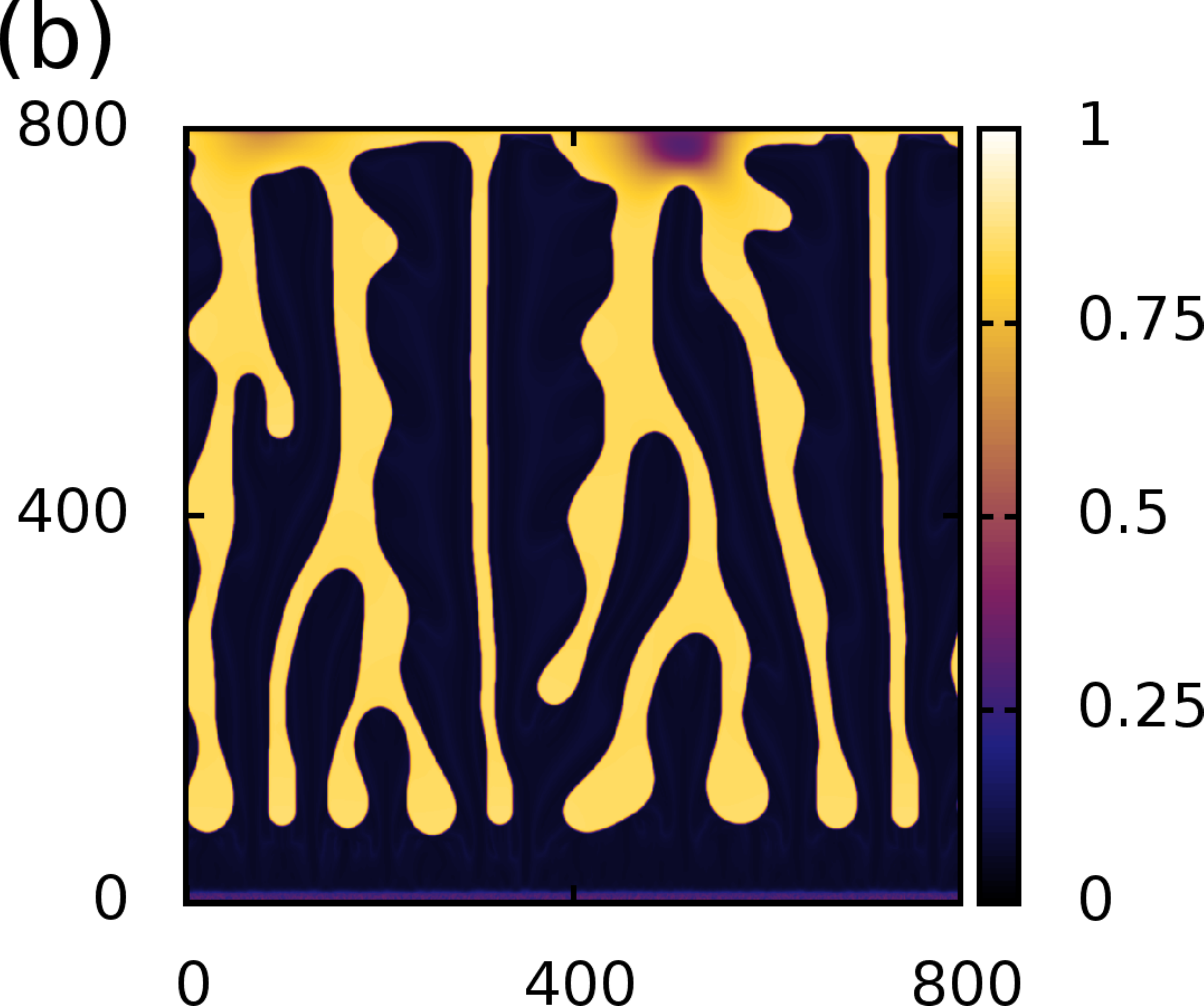}\\
		\includegraphics[width=0.46\linewidth]{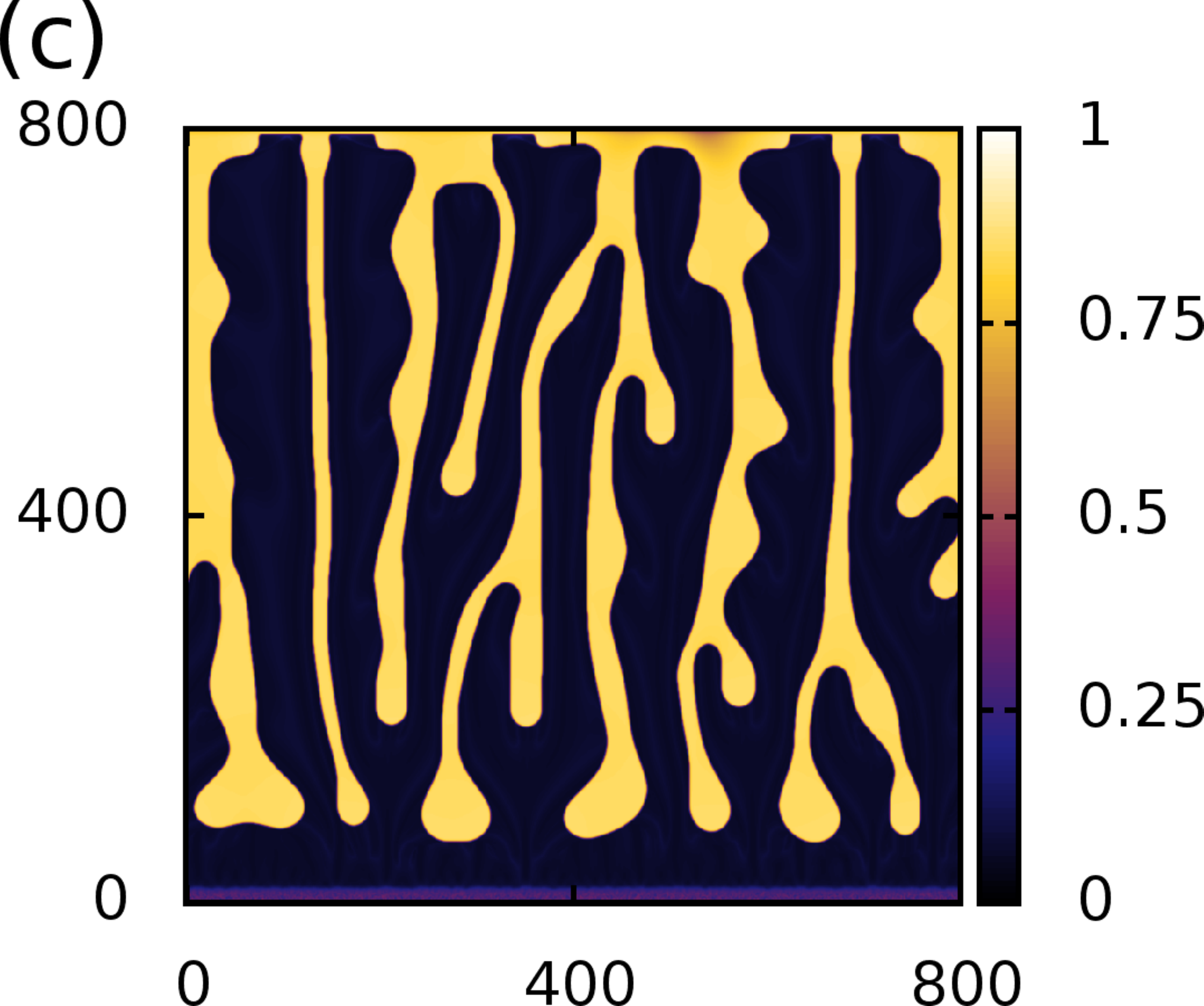}
		\hspace{2mm}
		\includegraphics[width=0.46\linewidth]{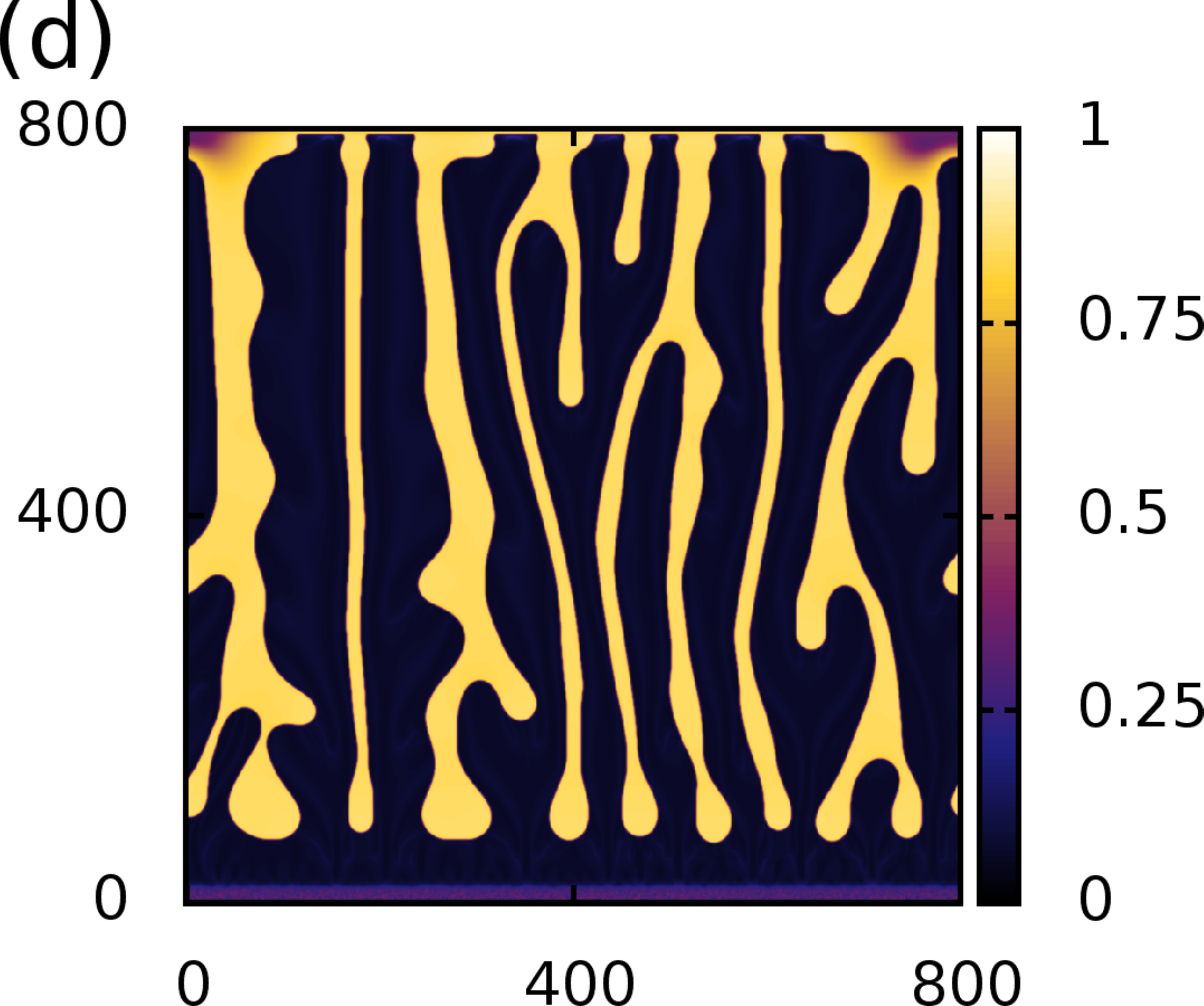}\\
	\end{minipage}
	\begin{minipage}[h]{0.49\linewidth}
		\includegraphics[width=\linewidth]{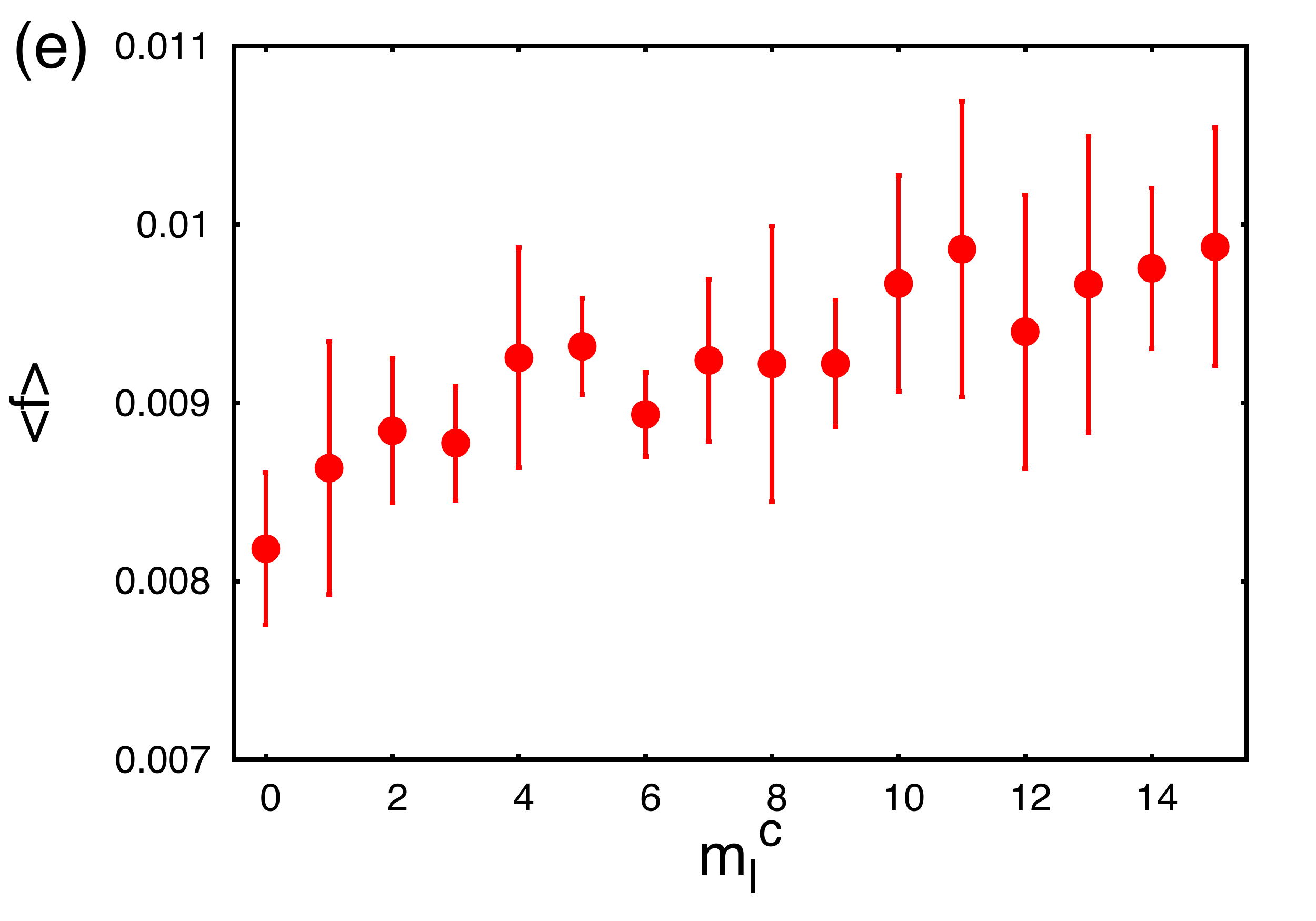}
	\end{minipage}	 
	\caption{Nano-particle density profiles for (a) $M_l^c = 0$, (b) 
	 	      $M_l^c = 5$, (c) $M_l^c = 10$ and (d) $M_l^c = 15$.  In (e) 
		      we display a plot showing how the number of fingers 
		      $\langle f \rangle$ depends on the value of $M_l^c$.  The 
		      parameter values are: $k_BT=1$, $\epsilon_l = 1.4$, 
		      $\epsilon_n =0.6$, $\epsilon_{nl} = 0.8$, 
		      $M_l^{nc}=1$, $\alpha=1$, $\beta \mu=-3.8$ and
		      $\lambda = 0.1$.}
	\label{figLiqCon}
\end{figure}

The effect of liquid diffusion over the surface has also been investigated by varying the liquid 
conserved mobility $M_l^c$.  Using the same parameter values as above, 
and setting $\alpha=1$, we display in  Fig.~\ref{figLiqCon} final nano-particle density profiles for varying $M_l^c$ from 0
to 15 and also the average finger number $\langle f \rangle$ 
versus $M_l^c$ averaged over five runs.  We see that the diffusive mobility of the liquid $M_l^c$ 
does affect the average number of fingers but to a much smaller extent
than the mobility of the nano-particles.  The average finger number generally increases 
as $M_l^c$ is increased. 

\subsection{Influence of liquid-particle demixing on the fingering}
\label{secNumEnn}

We now discuss the effect of possible liquid-particle phase
  separation on the front instability. Such a phase separation may
  occur near the front even for nano-particle concentrations inside
  the liquid film that are far smaller than the binodal value for
  liquid-particle phase separation. This occurs because as a dewetting front recedes it
collects nano-particles (as previously discussed) and therefore
increases the value of $\rho_n$ near the front.  For certain parameter
values we find that if $\rho_n$ increases above a certain threshold
value, then liquid-particle phase separation occurs in the front
region. The liquid separates into two liquid phases, a mobile
  one poor in colloids and a less mobile one rich in colloids.
  To investigate the resulting effects we set
  the interaction energies to $\epsilon_l=1.7$, $\epsilon_{nl}=1$ and
  vary $\epsilon_n$ from $0$ to $1.2$.  Eq.~(\ref{eqLiqLiqCond})
  indicates that we should observe liquid-particle phase separation
  for $\epsilon_n > 0.3$.  We set the average nano-particle density
to be low, $\rho_n^{av}=0.1$, so that there is no liquid-particle
phase separation in the bulk of the fluid film. For the value of the
chemical potential that we use $\mu = -4$, the fluid is linearly stable for all
values of $\epsilon_n$ and leads to a relatively fast dewetting front.

\begin{figure}[tbh]
	\begin{minipage}[h]{0.5\linewidth}
		\includegraphics[width=0.46\linewidth]{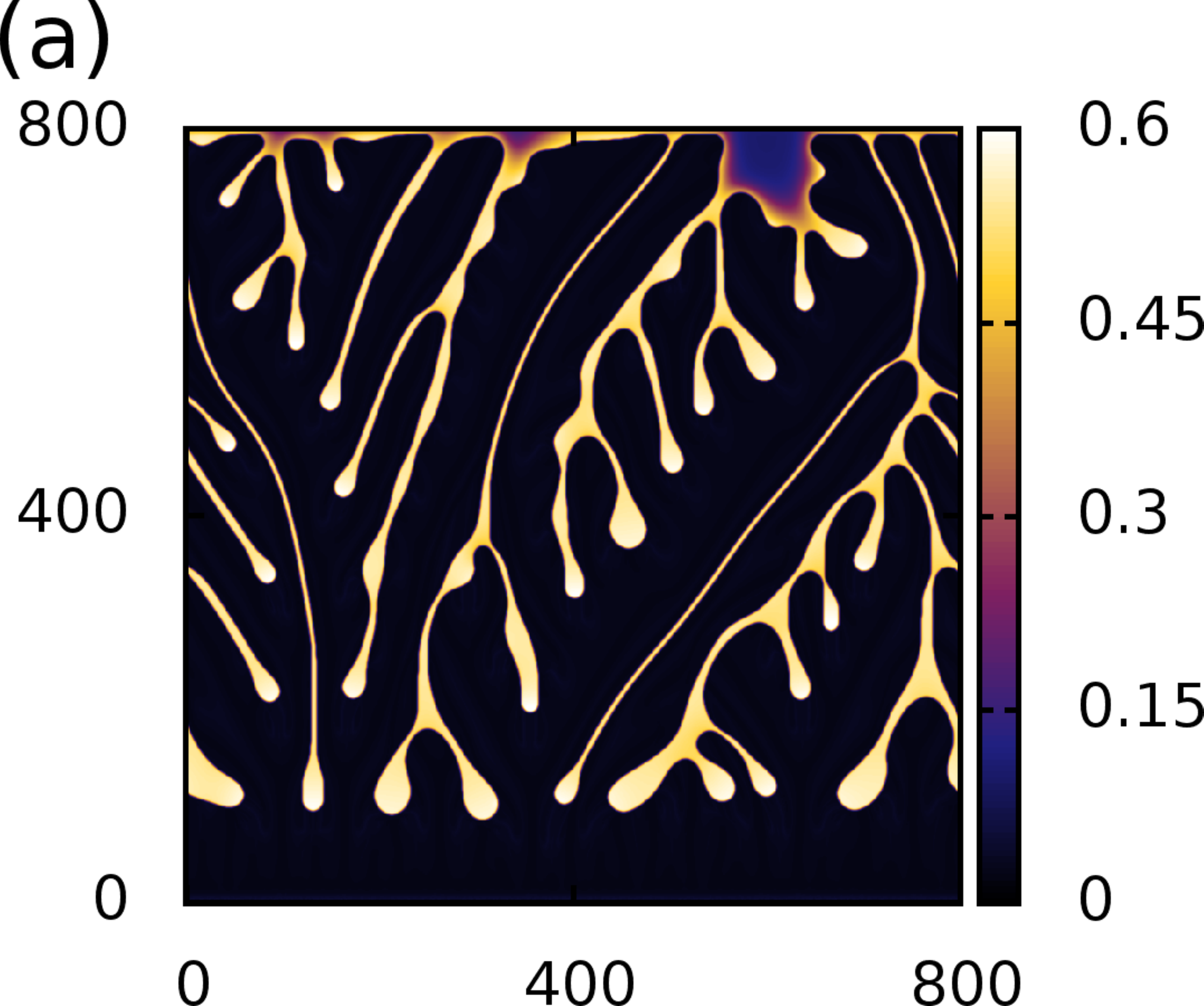}
		\hspace{2mm}
		\includegraphics[width=0.46\linewidth]{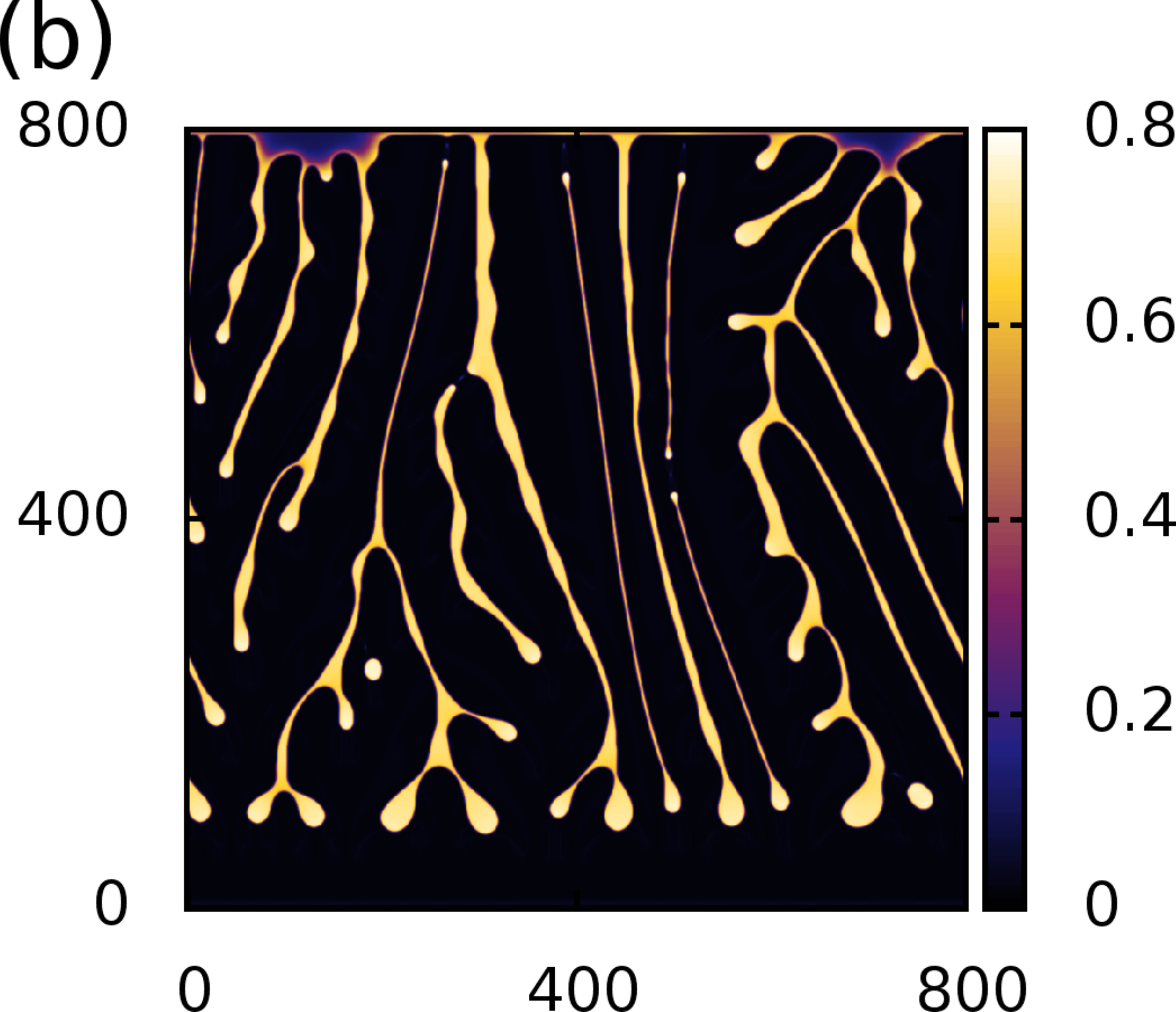}\\
		\includegraphics[width=0.46\linewidth]{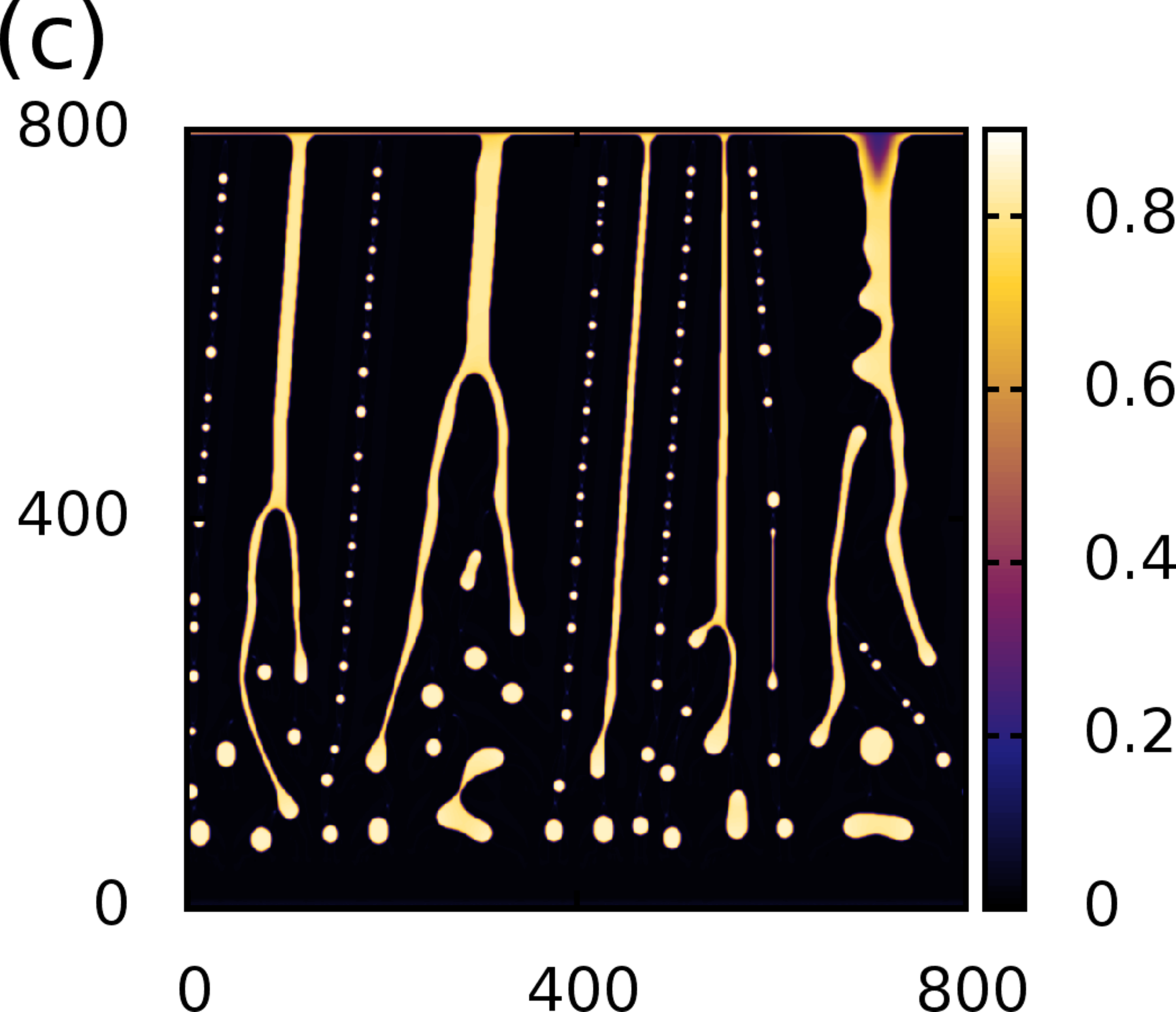}
		\hspace{2mm}
		\includegraphics[width=0.46\linewidth]{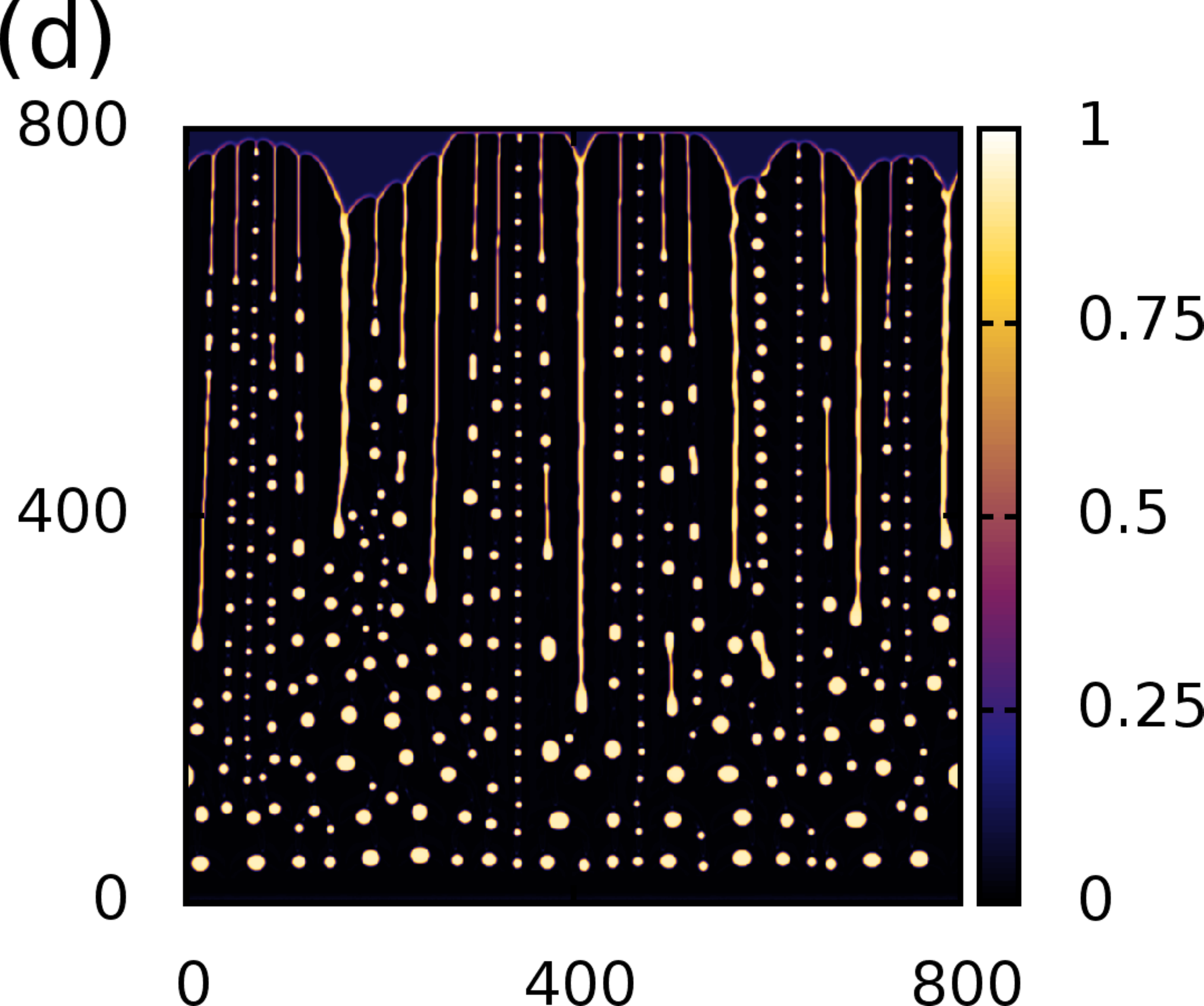}
	\end{minipage}
	\begin{minipage}[h]{0.49\linewidth}
		\includegraphics[width=\linewidth]{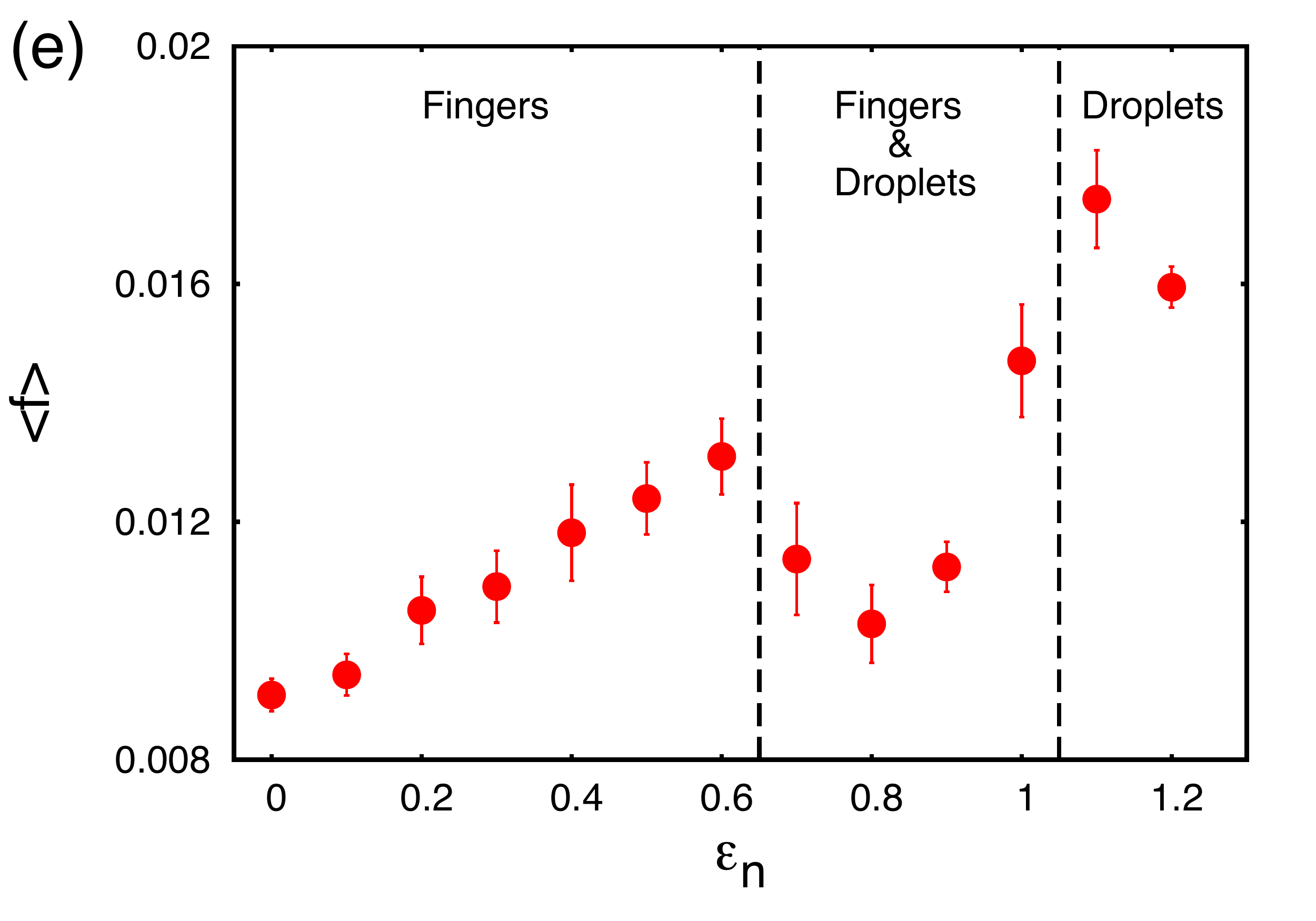}
	\end{minipage}
	\caption{Nano-particle density profiles for (a) $\epsilon_n = 0.3$, (b)
		      $\epsilon_n = 0.6$, (c) $\epsilon_n = 0.8$ and (d) 
		      $\epsilon_n = 1.1$.  In (e) we display a plot 
		      showing how the mean number of fingers $\langle f \rangle$ 
		      depends on the parameter $\epsilon_n$.  When 
		      $\epsilon_n>0.6$ we begin to observe droplets being 
		      deposited along with the branched structures
		      (as shown in (c)).  The parameter values for these calculations 
		      are: $k_BT=1$, $\epsilon_l = 1.7$, $\epsilon_{nl} = 1.0$, 
		      $M_l^c=0$, $M_l^{nc}=1$, $\alpha=0.5$, $\mu = -4$, 
		      $\rho_n^{av}=0.1$ and $\lambda = 0.025$.}
	\label{figLiqLiqSim}
\end{figure}

In Fig.~\ref{figLiqLiqSim}(a)-(d) we display typical final
nano-particle density profiles for various $\epsilon_n$ and the average finger number
$\langle f \rangle$ versus $\epsilon_n$, averaged over five runs.  As
$\epsilon_n$ is increased from $\epsilon_n = 0$, we initially see a
linear increase in the average number of fingers $\langle f \rangle$.
This reaches a peak at $\epsilon_n \approx 0.6$, after which we begin
to see the development of droplets.  When $0.6 \lsim \epsilon_n \lsim
0.9$ we observe separation between regions of high density of
nano-particles and low density of nano-particles occurring locally at
the dewetting front.  The areas with a lower density of nano-particles
form very thin fingers which quickly rupture into a series of
droplets, whereas the areas with a higher density form thicker fingers
which are much more stable as shown in Fig.~\ref{figLiqLiqSim}(c).
%
The sections of the front with a lower density of nano-particles
recede faster than the rest of the front.  This results in a `doublon'
pattern which has been observed in many different systems, e.g.~ in the thin
film directional solidification of a nonfaceted cubic crystal
\cite{AFI95}.  As we increase $\epsilon_n$ further, for $\epsilon_n >
1$ we observe that the dynamics at the front remains similar,
however, now all the finger structures are very thin and therefore
quickly break up into droplets.  One also notices that the
  tendency to form side branches decreases with increasing
  $\epsilon_n$ whereas the orientation of the branches becomes
  increasingly perpendicular to the receding front.  We also observe
an increase in the density of the nano-particles within the
fingers/droplets as $\epsilon_n$ is increased.  This is due to the
increased attraction between the nano-particles.  These results agree
qualitatively with the KMC model \cite{VTPV08}.  The KMC results also
show an initial increase in the number of fingers followed by a
transition from fingers to droplets as the interaction energy between
the nano-particles $\epsilon_{n}$ is increased.  Fig.~17 of
Ref.~\cite{VTPV08} displays a plot of $\langle f \rangle$ versus
$\epsilon_{n}$ which shows a similar trend as the DDFT results
displayed in Fig.~\ref{figLiqLiqSim}(e). However, as it is a
  discrete stochastic model, the details of the transition in the way
  the branching occurs are less discernible than in the present DDFT
  model.


\section{Concluding remarks}
\label{secCon}

We have presented a DDFT based model for the evaporative dewetting
of an ultrathin film of a colloidal suspension. We have derived an
expression for the free energy of the system using a mean-field
approximation for a coarse-grained
Hamiltonian model (\ref{eqHam}) for the system. We have also
derived dynamical equations which describe the diffusive dynamics
of the solvent and of the colloids as well as the evaporation of the
solvent.  We have considered the equilibrium phase behaviour of the
pure solvent and of the two-component fluid and identified parameter
ranges where unstable, metastable and stable phases exist.  We then
solved the coupled dynamical equations numerically to investigate the
different dynamical pathways of the phase transition and the resulting
self-organised patterns of the nano-particles.

The model successfully describes the various self-organised structures
found in experiments \cite{PVSM08} and is in qualitative agreement
with the discrete stochastic KMC model \cite{VTPV08}.  Our numerical
results show how nano-particle network structures can form either from
a spinodal processes (Fig.~\ref{figSpinDe}) or through the nucleation
and growth of holes (Fig.~\ref{figNuc}).  We have also observed how
branched structures develop from a fingering instability of the
receding dewetting front (Fig.~\ref{figFingering}).

The transverse front instability results from a build-up of the
nanoparticles close to the front as the solvent evaporates, when
diffusion is too slow to disperse them. This slows down the front and
renders it unstable.  As a result, density fluctuations along the
front grow into an evolving fingering pattern.  This transverse
front instability can be considered to be a self-optimisation process
which maintains the mean front velocity constant \cite{VTPV08} (see
also the discussion of this in the context of a similar front
instability occurring in the dewetting of non-volatile polymer films
\cite{ReSh01}).  One may also say that the constant average front
velocity is maintained by depositing some of the nano-particles onto the
dry substrate creating the branched structures.  Experimental
observations show that the branched structures found in the ultra-thin
film behind the mesoscopic dewetting front are initiated from random
nucleation sites.  The holes which are nucleated then grow, initially
creating circular dewetting fronts.  We subsequently observe that the
fingering instabilities and the development of branched structures
form on the circular interfaces.  Fig.~\ref{figFingering} shows how
these circular branched patterns develop from a single nucleation
point; our numerical results bare a striking resemblance to the
experimental AFM images of this phenomena \cite{PVSM08}.

We have studied the branched structures in greater detail
using a planar geometry, i.e., by creating initially straight dewetting
fronts. We
have considered how the different mobilities affect the fingering.
The nano-particle mobility in liquid films has a significant effect on
the average number of fingers in the branched structure
(Fig.~\ref{figVaryingAlpha}).  The finger number decreases rapidly
with increasing mobility in agreement with earlier KMC results
\cite{VTPV08}.  This behaviour can be attributed to the lower-mobility
of the nano-particles that hinders re-distribution by diffusion and
also reduces the speed of the dewetting front.  For the system to
attain a higher front speed it must deposit nano-particles onto the
surface at a greater rate.  Therefore, if the mobility of the
nano-particles is low this leads to the creation of more fingers
because in this case the average distance a nano-particle has to
travel to reach a finger is smaller.  Increasing the mobility for the conserved
diffusive dynamics of the liquid has the opposite effect on the
average number of fingers (Fig.~\ref{figLiqCon}).  The complex
relationship between the diffusion of the liquid and the average
finger number is not yet fully understood.  Our hypothesis is that
increasing the mobility of the liquid results in an effective increase
in the speed of the dewetting dynamics of the liquid for fixed
nano-particle mobility. Thus, the mobility of the nano-particles
becomes lower in comparison. The increased finger number then
results from the increased mobility contrast, in agreement with the
general instability mechanism laid out above.

The basic front instability as described above is a purely dynamic
effect and does not depend on particle-liquid and particle-particle
attractive interactions that favour demixing of the liquid and the
nanoparticles. However, beside this regime (that we call the `transport
regime') we have investigated how interactions that favour
demixing influence the instability (Fig.~\ref{figLiqLiqSim}).  In
general, when increasing the interaction energy between the
nano-particles one increases the tendency towards liquid-particle
demixing. However, this has no practical effect as long as the
nano-particle concentration is low, so that it is outside of the two-phase
region. This is normally the case for our initial densities. However,
in the course of the evaporative dewetting the density increases close
to the receding front. Increasing the interaction energy between the
nano-particles causes demixing to occur close to the front (but
not in the bulk film).  The demixing makes the fingering instability
stronger (we call this the `demixing regime'). At first, one finds a
linear increase in the average finger number with increasing
interaction energy. At higher values of $\epsilon_n$, when the localised phase
separation sets in, the fingers become straight with less side branches,
before finally lines of drops are emitted directly at the front.  In this
regime, the mean number of fingers is determined by the dynamics
\textit{and} the energetics of the system.

The results we have obtained with our DDFT model confirm that jamming
of discrete particles (as already discussed in Ref.~\cite{TVAR09}) is not a
necessary factor for the fingering instability to occur. Our
model is a continuum model with a diffusion constant that is
independent of the nanoparticle concentration. The present
two-dimensional DDFT model has several advantages over the
two-dimensional KMC model \cite{RRGB03,VTPV08}: In particular, the
early instability stages are more easy to discern without the
background noise of the KMC. Furthermore, the underlying free energy
may be employed to analyse the equilibrium phase behaviour in detail,
in a similar manner to Ref.~\cite{WoSc06}. Many standard tools for the
analysis of partial differential equations can be applied to the
coupled evolution equations, such as, e.g., the linear stability
analysis of the homogeneous films. In the future, one may perform a
linear stability analysis for the receding straight front and also
investigate steady state solutions as has been done for evaporating films of pure
liquids \cite{Thiele10}. There are many details that would merit
further investigations such as, for example, the doublon structure
mentioned in Section~\ref{secNumEnn} and its relation to such
structures formed in directional solidification \cite{UtBo05}.

The present DDFT model does not include the effect of surface forces,
i.e., wettability effects (substrate-film interactions). Therefore, a
non-volatile liquid film would not dewet the substrate. This implies
that an important avenue for future improvement is to incorporate
wettability effects into the model. This could be done by making a
mean-field approximation to derive an expression for the free energy
for a fully three-dimensional KMC model \cite{SHR05,YoRa06} (after
incorporating substrate-particle and substrate-liquid
interactions). The resulting three-dimensional DDFT could then either
be used directly or be averaged perpendicularly to the substrate
employing e.g., a long-wave approximation.  Another possible option
consists of combining a mesoscopic hydrodynamic approach, e.g., a thin
film evolution equation (see \cite{ODB97,Bonn09,Thiele10}) with
elements of DDFT. For a brief discussion of a similar approach see
Ref.~\cite{Thie11b}.

As a final remark, we recall that in the present work we have
only considered dewetting from homogeneous substrates. However,
it is straightforward to include surface heterogeneities in our model
via the external potentials $\phi_i \vcr$ in Eq.\ \eqref{eqFreeEngCont1}.
As future work, it would be interesting to study the influence of surface patterning
on the finger formation displayed by the present system.

\ack
This work was supported by the EU via the ITN MULTIFLOW
(PITN-GA-2008-214919). MJR also gratefully acknowledges support
from EPSRC and AJA thanks RCUK for support.

\section*{References}

\bibliographystyle{prsty}
\bibliography{RAT11}
	
\end{document}